\documentclass[12pt]{article}
\makeatletter \@addtoreset{equation}{section} \makeatother
\renewcommand{\theequation}{\thesection.\arabic{equation}}
\newcommand{\ba}{\begin{array}}
\newcommand{\ea}{\end{array}}
\newcommand{\beq}{\begin{equation}}
\newcommand{\eeq}{\end{equation}}
\newcommand{\bea}{\begin{eqnarray}}
\newcommand{\eea}{\end{eqnarray}}

\def\bce{\begin{center}}
\def\ece{\end{center}}

\def\nonu{\nonumber}

\def\pa{\partial}

\def\be{\beta}

\def\ep{\epsilon}

\def\eps6{{\displaystyle \mathop{\epsilon}^{6}}{}}
\def\g6{{\displaystyle \mathop{g}^{6}}{}}

\def\nab6{{\displaystyle \mathop{\nabla}^{6}}{}}

\def\0{{\sst{(0)}}}
\def\1{{\sst{(1)}}}
\def\2{{\sst{(2)}}}
\def\3{{\sst{(3)}}}
\def\4{{\sst{(4)}}}
\def\5{{\sst{(5)}}}
\def\6{{\sst{(6)}}}
\def\7{{\sst{(7)}}}
\def\8{{\sst{(8)}}}

\def\ba{\begin{array}}
\def\ea{\end{array}}
\def\beq{\begin{equation}}
\def\eeq{\end{equation}}
\def\be{\begin{equation}}
\def\ee{\end{equation}}

\def\eps{\epsilon}

\def\ba{\begin{array}}
\def\ea{\end{array}}
\def\beq{\begin{equation}}
\def\eeq{\end{equation}}
\def\be{\begin{equation}}
\def\ee{\end{equation}}

\def\eps{\epsilon}

\def\eps6{{\displaystyle \mathop{\epsilon}^{6}}{}}

\def\nab6{{\displaystyle \mathop{\nabla}^{6}}{}}

\newcommand{\bean}{\begin{eqnarray*}}
\newcommand{\eean}{\end{eqnarray*}}

\begin{document}
\thispagestyle{empty} \addtocounter{page}{-1}
   \begin{flushright}
\end{flushright}

\vspace*{1.3cm}
  
\centerline{ \Large \bf   
Higher Spin Currents in Wolf Space for Generic $N$ } 
\vspace*{1.5cm}
\centerline{{\bf Changhyun Ahn } and {\bf Hyunsu Kim}
} 
\vspace*{1.0cm} 
\centerline{\it 
Department of Physics, Kyungpook National University, Taegu
702-701, Korea} 
\vspace*{0.8cm} 
\centerline{\tt ahn@knu.ac.kr, \qquad kimhyun@knu.ac.kr 
} 
\vskip2cm

\centerline{\bf Abstract}
\vspace*{0.5cm}

We obtain the $16$ higher spin currents with spins 
$(1,\frac{3}{2},\frac{3}{2},2),(\frac{3}{2}, 2,2, \frac{5}{2}),
(\frac{3}{2},2,2, \frac{5}{2})$ and $(2,\frac{5}{2},\frac{5}{2},3)$
in the ${\cal N}=4$ superconformal Wolf space coset 
$\frac{SU(N+2)}{SU(N) \times SU(2) \times U(1)}$.
The antisymmetric 
second rank tensor
occurs in the quadratic spin-$\frac{1}{2}$ Kac-Moody currents
of the  higher spin-$1$ current. 
Each higher spin-$\frac{3}{2}$ current contains  
the above antisymmetric second rank tensor 
and three symmetric (and traceless)  second rank tensors (i.e. three 
antisymmetric almost complex
structures contracted by the above antisymmetric  tensor) 
in the product of spin-$\frac{1}{2}$ and spin-$1$ 
Kac-Moody currents respectively.  
Moreover, the remaining higher spin currents of spins $2, \frac{5}{2}, 3$
contain the combinations of the (symmetric) metric, 
the three almost complex structures,
the antisymmetric tensor or the three symmetric tensors in the
multiple product of the above Kac-Moody currents  
as well as the composite currents from the large ${\cal N}=4$ nonlinear 
superconformal algebra.

\baselineskip=18pt
\newpage
\renewcommand{\theequation}
{\arabic{section}\mbox{.}\arabic{equation}}

\section{Introduction}

The three-point function in a two-dimensional conformal field theory 
is an important ingredient which determines the higher point functions
via factorization. 
Another reason to study the three-point function is the fact that 
the three-point function in two-dimensions 
can provide some information on 
the three-point function in the three-dimensional bulk theory.  
In fact, the three-point function shares the same formula in the duality 
\cite{GG1011,GG1205,GG1207}
between the higher spin theory on the $AdS_3$ space
and the $W_N$ minimal model coset conformal field theory in two-dimensions.
The eigenvalue equations (before calculating the three-point function) 
for the zero modes of 
the higher spin currents  in the coset 
model match with those of the higher spin fields in the asymptotic
symmetry algebra of the higher spin theory on the $AdS_3$ space.  
The simplest three-point function contains two scalar primaries with one 
higher spin current 
which is a polynomial combination of the spin-$1$
WZW affine currents acting on the coset primaries. 
Also one can  study the three-point function in the `supersymmetric' 
duality.
In \cite{GG1305}, 
the large ${\cal N}=4$ higher spin theory on $AdS_3$ 
based on the higher spin algebra
is dual to
the 't Hooft limit of the two dimensional 
large ${\cal N}=4$ superconformal coset theory on Wolf space 
\cite{Wolf,Alek,Salamon}.
Therefore, in order to reveal this duality precisely from three-point function 
viewpoint, it is necessary, as a first step, to 
obtain the higher spin currents in Wolf space 
for general $N$ (and general level $k$) 
which is the main motivation 
of this paper. 

Let us describe the large ${\cal N}=4$ superconformal coset theory
in two dimensions.
The general coset with $U(M)$ Chan-Paton factor
and corresponding Virasoro central charge are described by \cite{CHR1306}
\bea
\frac{G}{H}=\frac{SU(N+M)_{k} \times SO(2N M)_1}{SU(N)_{k+M} 
 \times U(1)_{N M(N+M)(N+M+k)}}, \qquad
c = \frac{3N M k }{(N+M+k)}+
\frac{(k+N)(M^2-1)}{(N+M+k)}.
\label{goverh}
\eea

For $M=1$, the coset model (\ref{goverh})  divided by 
$SU(M=1)_{k+N}$ is dual to 
the ${\cal N}=2$ higher spin supergravity from the results in \cite{CHR1111}
within the context of Kazama-Suzuki model \cite{KS1,KS2}.
In the appropriate 't Hooft limit, the central charge behaves as $N$.
For $M=1$, there is no contribution from the second 
term of the central charge.
See also the relevant works in 
\cite{CG1203,HP1203,Ahn1206,CG1207,Ahn1208,Hikida1212}.

For $M=2$, the above coset (\ref{goverh}) is dual to
the ${\cal N}=4$ higher spin supergravity \cite{GG1305}.
The central charge behaves as $N$ in the 't Hooft limit.
Note that the Virasoro central charge is the sum of 
Wolf space central charge $c_{\mbox{W}}=\frac{3 N M k}{(N + M +k)}=\frac{6 N k}{(
N+k+2)}$ 
and the $SU(M=2)_{k+N}$ central charge.
The contributions from the latter
appear in each factor group of the coset except  $SU(N)_{k+M}$ group.
If one divides the coset by $SU(M=2)_{k+N}$ further, then 
one has the Wolf space central charge $c_{\mbox{W}}$.
Once the four spin-$\frac{3}{2}$ currents
with an appropriate normalization constant
are fixed, then the third-order pole provides the above 
Wolf space central charge $c_{\mbox{W}}$
(or the highest order-pole determines the normalization constant)
and the Virasoro central charge $c$ in (\ref{goverh}) 
is determined automatically 
from the large ${\cal N}=4$ nonlinear 
superconformal algebra in these four spin-$\frac{3}{2}$ currents. 
The decomposition of central charge in (\ref{goverh})
implies that the two stress energy tensors from each part
are orthogonal to each other and they commute with each other. 
We will come to this issue including the stress energy tensor 
and the four spin-$\frac{3}{2}$ currents
in section $2$ in detail.
From the supersymmetric coset in the abstract to this bosonic 
coset, one realizes that the $2N M (= 4N)$ free fermions in $SO(2N M)_1$ group
survive after
factoring out the four spin-$\frac{1}{2}$ fermion currents 
and one spin-$1$ current living in the large ${\cal N}=4$ linear 
superconformal algebra.
The original number of free fermions was given by $(2N M + M^2=4N+4)$ which
is the same as the bosonic degree of freedoms.

For general $M (>2)$,
the property of the higher spin algebra and the corresponding asymptotic
symmetry algebra was described in \cite{CPV1408}. See also the work of
\cite{CV1312}.

See also the relevant works \cite{BCG1404,GG1406} on the large ${\cal N}=4$ 
holography \cite{GG1305}. 
For the particular level at 
the Kazama-Suzuki model, 
the ${\cal N}=3$ (enhanced) supersymmetry 
is observed in \cite{CHR1406}.  
This behavior also appeared in previous examples of 
\cite{Ahn1211,Ahn1305,BCGG1305}.

Before we are going to consider the higher spin currents,
let us review on the large ${\cal N}=4$ nonlinear superconformal 
algebra \cite{GS,cqg1989,npb1989,GK}. 
In the work of Van Proeyen \cite{cqg1989}, 
the realization of large ${\cal N}=4$ nonlinear superconformal algebra
was obtained.
The four fermionic spin-$\frac{3}{2}$ currents corresponding to
${\cal N}=4$ supersymmetries  
can be associated with the unit matrix and the three complex structures.
The former plays the role of ${\cal N}=1$ supersymmetry while 
the latter plays the role of the remaining, additional 
${\cal N}=3$ supersymmetries.
The three components of the spin-$1$ currents are related to 
the three complex structures in the bosonic Kac-Moody currents.
The three components of other spin-$1$ currents 
are associated with the three complex structures in the product of two
fermionic 
Kac-Moody currents.
The spin-$2$ stress energy tensor can be written in terms of 
two commuting parts,
the stress energy tensor corresponding to Wolf space
and the stress energy tensor from the Sugawara construction in the two kinds 
of above spin-$1$ currents 
(the above 
$SU(2)_{k+N}$ group describes these two currents where the level is given by
$k$ and $N$ respectively) 
where the index of generators 
appears nontrivially: although each part 
differs for each supersymmetry, 
the sum of these two parts leads to the final spin-$2$ stress energy tensor.
In other words, the four different cases  
for each part provides an unique expression for the spin-$2$ 
stress energy tensor
of large ${\cal N}=4$ nonlinear superconformal algebra. 

Furthermore, in the Gates and Ketov's work \cite{GK},
the above three complex structures 
represent almost quaternionic (tri-hermitian) structure on the 
Wolf space (quaternionic symmetric space) coset.
They have considered the nonsymmetric coset space also where
the four spin-$\frac{3}{2}$ currents contain the rank three tensor 
with coset indices and it turned out that one of the spin-$1$ currents 
contain the quadratic terms in the fermionic Kac-moody currents.
By using the results about the coset realizations of the 
large ${\cal N}=4$ nonlinear superconformal algebra over the so-called 
Freudenthal triple systems \cite{Gunaydin}, 
they identified the above $11$ generators 
with the ${\cal N}=4$ WZW generators 
and the quaternionic structure on the Wolf space can be written 
in terms of the symplectic structure of the associated 
Freudenthal triple systems \footnote{
For the large ${\cal N}=4$ linear superconformal algebra 
\cite{STVplb,npb1988,Schoutensnpb,Ivanov1,Ivanov2,Ivanov3,Ivanov4},
Sevrin and Theodoridis \cite{ST}
have determined the $14$ currents of the 
large ${\cal N}=4$ linear superconformal algebra 
from the ${\cal N}=1$ super Kac-Moody currents in ${\cal N}=1$ superspace.
The various tensors appearing in these seven ${\cal N}=1$ super fields
are satisfied the nontrivial relations.   
In particular, the three tensors appearing in the quadratic in the 
${\cal N}=1$ super Kac-Moody currents  of ${\cal N}=1$ super field of
superspin one play the role of almost complex structures.  
Of course, the usual ${\cal N}=1$ Sugawara stress energy tensor
is given. 
In component approach, in \cite{Saulina},
the complete expressions for the $16$ currents of
large ${\cal N}=4$ linear superconformal algebra were written in terms 
of WZW Kac-Moody currents.
For the spin-$\frac{3}{2}$ currents, there are cubic terms in the fermionic 
WZW Kac-Moody currents with four tensors of rank three.
One of them is given by a structure constant and three of them 
are given by the above almost complex structure contracted with a structure
constant. The spin-$1$ currents contain the quadratic fermion Kac-Moody 
currents with almost complex structure contracted two 
structure constants. Moreover, other spin-$1$ current has 
the quadratic fermion Kac-Moody 
currents with two almost complex structures contracted two 
structure constants. }.

Beyond the large ${\cal N}=4$ nonlinear (or linear) superconformal algebra, 
the existence of higher spin currents 
was observed in \cite{GG1305}.
The lowest ${\cal N}=4$ higher spin multiplet
contains, one spin-$1$ current, four spin-$\frac{3}{2}$ currents,
six spin-$2$ currents, four spin-$\frac{5}{2}$ currents 
and one spin-$3$ current.
They transform as $({\bf 1},{\bf 1})$, 
$({\bf 2}, {\bf 2})$, $({\bf 1}, {\bf 3}) \oplus
({\bf 3},{\bf 1})$, $({\bf 2},{\bf 2})$ and $({\bf 1},{\bf 1})$ under the
$SU(2) \times SU(2)$ respectively.
The explicit realization for an extension of 
large ${\cal N}=4$ nonlinear superconformal algebra in 
the ${\cal N}=4$ Wolf space coset theory has been studied in 
\cite{Ahn1311,Ahn1408} for $N=3$. 
This construction includes the previous descriptions in \cite{cqg1989,GK}.
As we vary $N$ starting from $N=3$, 
the operator product expansions (OPEs) from \cite{Ahn1311,Ahn1408}
reveal the $N$-dependence in their structure constants.
From the lessons in \cite{Ahn1311,Ahn1408}, one can determine 
all the $16$ higher spin currents for general $N$ once one obtains 
the lowest spin-$1$ current completely with the help of
the spin-$\frac{3}{2}$ currents of large ${\cal N}=4$ nonlinear superconformal
algebra. The metric on the $SU(N+2)$ group, the three complex structures and
structure constants played an important role in the construction of 
large ${\cal N}=4$ nonlinear superconformal algebra and they 
will appear in the construction of higher spin currents also.  

The higher spin-$1$ current consists of 
the linear bosonic Kac-Moody current and the quadratic 
fermionic Kac-Moody currents.
One should determine the two coefficient tensors appearing in this 
higher spin-$1$ current, along the lines of \cite{Ahn1311,Ahn1408},  
by requiring that this lowest higher spin-$1$ current
transforms as a primary current under the Virasoro stress energy tensor in 
(\ref{goverh}) and it does not have any singular terms in the OPE with 
two $SU(2)$ spin-$1$ currents of large ${\cal N}=4$ nonlinear superconformal
algebra.

In this paper,
we describe the important aspects in the large ${\cal N}=4$ nonlinear 
superconformal realization considered in \cite{cqg1989,GK}. 
First of all, one has four spin-$\frac{3}{2}$ currents for $N=3$.
We would like to write these four spin-$\frac{3}{2}$ currents for general
$N$ because the OPEs between these currents give all the information on 
the remaining seven currents, six spin-$1$ currents and one spin-$2$ current. 
All the $16$ OPEs have the explicit singular terms in terms of 
spin-$1$ and spin-$\frac{1}{2}$ Kac-Moody currents.  
As observed in \cite{cqg1989}, one observes the three complex structures
$h^i_{\bar{a} \bar{b}}$
in front of the above spin-$\frac{3}{2}$ currents
in terms of $4 N \times 4 N$ matrices where the $4 N$ is the number of coset 
indices. 
These complex structures also appear in the spin-$1$ currents.
One can read off the spin-$2$ stress energy tensor from the first-order pole 
in the some OPEs among the above $16$ OPEs.

As emphasized before, the higher spin-$1$ current 
contains 
the antisymmetric second rank tensor $d^0_{\bar{a} \bar{b}}$ 
with the Wolf space coset indices
in front of the quadratic fermionic Kac-Moody currents. 
This antisymmetric second rank tensor is a new quantity
and satisfies the nontrivial identities, 
$1)$ the square of this tensor gives a metric on the Wolf space coset
and 
$2)$ there exists an identity for the product between this tensor
with the structure constant.  
By construction, the four higher spin-$\frac{3}{2}$ currents 
contain the product of the three complex structures 
(including the unit matrix) and the above antisymmetric
second rank tensor.
The higher spin-$2$ currents
contain the combinations of the (symmetric) metric, 
the three almost complex structures,
the antisymmetric tensor or the three symmetric tensors in the
multiple product of the above Kac-Moody currents  
as well as the composite currents from the large ${\cal N}=4$ nonlinear 
superconformal algebra.
Moreover, the remaining higher 
spin currents of spins $\frac{5}{2}, 3$
contain more higher rank tensors.
For the higher spin currents with spins $s=1, \frac{3}{2}$ and $2$,
the explicit manifest $SU(2) \times SU(2)$ representations 
are given.

In section $2$, we review 
the large ${\cal N}=4$ nonlinear superconformal algebra.
The starting point is to begin with the four spin-$\frac{3}{2}$ currents 
which determine all the remaining currents (six spin-$1$ currents and 
one spin-$2$ current) in the large ${\cal N}=4$
nonlinear algebra.

In section $3$,
the $16$ higher spin currents in terms of bosonic and fermionic 
Kac-Moody currents 
are constructed.
Some higher spin currents appeared in \cite{BCG1404} are identified.
The various tensors appearing in the higher spin currents 
are constructed.
Because these higher spin currents are constructed in terms of 
Kac-Moody currents, the zero modes can be obtained.

In section $4$, we summarize the results of this paper and
future directions are given.

In Appendices $A$-$H$, some detailed calculations are described.

The mathematica package by Thielemans \cite{Thielemans} 
is used for low $N$ cases with $N=3,5,7,9$. 
For these values, one obtains all the higher spin currents which are 
not present in this paper explicitly. 
They can be obtained from the results of section $3$ by substituting
these values. 

\section{ The large $\mathcal N = 4$ nonlinear superconformal algebra in 
the Wolf space coset }

In this section, we review the large ${\cal N}=4$ nonlinear 
superconformal algebra in the Wolf space coset.
The $11$ currents are described in terms of the ${\cal N}=1$ 
Kac-Moody currents.

\subsection{ The $\mathcal N =1$ Kac-Moody current algebra in component 
approach }

Let us consider the group 
$SU(N+2)$ in the Wolf space coset where 
$N$ is odd and the dimension of $SU(N+2)$ is even.
The generators are given in Appendix $A$.
They satisfy the usual commutation relation
$\left[ T_a, T_b \right] = f_{a b}^{\;\;\;\; c} T_c $ where 
the indices run over $a, b, \cdots = 1, 2, \cdots, (N+2)^2-1$. 
The normalization is as follows:
$
g_{ab} = \frac{1}{2 c_G} f_{ac}^{\,\,\,\,\,\, d} f_{bd}^{\,\,\,\,\,\, c}
$
where $c_G$ is the dual Coxeter number of the group $c_{SU(N+2)}=(N+2)$.
Let us denote the inverse of the metric as 
$
g^{ab} \equiv g_{ab}^{-1}
$.

The operator product expansions between the spin-$1$ and the spin-$\frac{1}{2}$
currents  are described as  \cite{KT1985}
\bea
V^a(z) \, V^b(w) & = & \frac{1}{(z-w)^2} \, k \, g^{ab}
-\frac{1}{(z-w)} \, f^{ab}_{\,\,\,\,\,\,c} \, V^c(w) 
+\cdots,
\nonu \\
Q^a(z) \, Q^b(w) & = & -\frac{1}{(z-w)} \, (k+N+2) \, g^{ab} + \cdots,
\nonu \\
V^a(z) \, Q^b(w) & = & + \cdots.
\label{opevq}
\eea
Here $k$ is the level and a positive integer.
Note that there is no singular term in the OPE between 
the spin-$1$ current $V^a(z)$ and the spin-$\frac{1}{2}$ current 
$Q^b(w)$.
One can also obtain these OPEs in the ${\cal N}=1$ superspace.
Furthermore, the ${\cal N}=2$ superspace description where the 
spin-$1$ currents have the additional quadratic fermionic terms 
can be 
obtained by introducing some nonlinear constraints as in 
\cite{HS}. 
Note that the first OPE in (\ref{opevq}) has three independent 
OPEs if one uses the notations (with bar and unbarred indices) 
in \cite{Ahn1311,Ahn1408}. 

\subsection{ The $11$ currents 
of $\mathcal N =4$ nonlinear superconformal algebra using 
the Kac-Moody currents}

As in the abstract, the Wolf space coset in the `supersymmetric'
version is as follows:
\bea
\mbox{Wolf}= \frac{G}{H} = 
\frac{SU(N+2)}{SU(N) \times SU(2) \times U(1)}.
\label{coset}
\eea
Let us denote the indices belonging to
$G$ and Wolf space coset $\frac{G}{H}$
as 
\bea 
G \quad \mbox{indices} &:& a,b,c,\cdots =1, 2, \cdots, (N+2)^2-1,
\nonu \\
\frac{G}{H} \quad \mbox{indices} &:& \bar{a},\bar{b},\bar{c},\cdots
=1, 2, \cdots, 4N.
\label{abnotation}
\eea
One can also introduce the subgroup $H$ indices but
does not present them because they do not appear in later expressions
\footnote{In \cite{Ahn1311}, one introduced the complex basis where
the indices are given by barred and unbarred ones. In this paper, 
one uses the different notations. The indices $a, b, \cdots$ are given by
$A$ and $A^{\ast}$ while 
the indices $\bar{a}, \bar{b}, \cdots$ are given by
$\bar{A}$ and $\bar{A^{\ast}}$ in the complex basis. See also Appendix $A$.}. 
For given $(N+2) \times (N+2)$ matrix, 
one can associate the $4N$ coset indices
as follows:
\bea
\left(\begin{array}{r|rrrrr|r}
  & {*} & {*} & \cdots & {*} & {*} & \\ \hline
 {*} &&&&&& {*} \\
 {*} &&&&&& {*} \\
 \vdots &&&&&&  \vdots \\
 {*} &&&&&& {*} \\
 {*} &&&&&& {*} \\ \hline
 & {*} & {*} & \cdots & {*} & {*} & \\ 
\end{array}\right)_{(N+2) \times (N+2)}.
\label{matrix}
\eea
For example, the Wolf space coset generators are given in Appendix $A$.
The $N \times N$ matrix corresponding to the subgroup $SU(N)$
in (\ref{coset}) is located at the inside of (\ref{matrix}).

Let us start with the four fermionic spin-$\frac{3}{2}$ currents.
From 
the $N=3$ result in \cite{Ahn1311},  
one can write the spin-$\frac{3}{2}$ currents 
$G^{\mu}(z)$ where an index $\mu$ is in the $SO(4)$ vector 
representation as follows 
\footnote{One can assume more general ansatz 
which has cubic fermionic terms \cite{GK}.
But in the Wolf space coset, the rank three tensor in front of 
these terms vanishes. }:
\bea
G^{\mu}(z) &=& A(k,N) \, h^{\mu}_{\bar{a} \bar{b}} \,
Q^{\bar{a}} \, V^{\bar{b}}(z), \qquad \mu=0,1,2,3,
\label{Gansatz}
\eea
where $h^{0}_{\bar{a} \bar{b}} \equiv g_{\bar{a} \bar{b}}$ and
$h^{i}_{\bar{a} \bar{b}}$ ($i=1,2,3$) are 
numerical constants and the $A(k,N)$ is normalization factor
to be determined.
Note that the indices appearing in (\ref{Gansatz})
belong to the Wolf space coset.
There is no difference in the ordering of two different currents 
$Q^{\bar{a}}(z)$ and $V^{\bar{b}}(z)$
because they 
do not have any singular terms as mentioned before. 
By calculating the OPE
$G^{\mu}(z) \, G^{\nu}(w)$ 
which should satisfy the $\mathcal N = 4$ nonlinear superconformal 
algebra (\ref{N4scalgebra}) (or (\ref{N4scalgebraExplic})),
one should obtain the other currents in the right hand side of the OPE
as well as the normalization constant in (\ref{Gansatz}).

Let us compute the OPE $G^{\mu}(z) \, G^{\nu}(w)$ using the 
basic OPEs in (\ref{opevq}). See also the relevant works in 
\cite{BBSS1,BBSS2,BS}. 
It turns out that 
\bea
G^{\mu}(z) \, G^{\nu}(w) &=&
\frac{1}{(z-w)^3} \, A^2 \, \left[ - k(k+N+2) \, h^{\mu}_{\bar{a} \bar{b}}
 \, h^{\nu \bar{a} \bar{b}} \right]
\nonu \\
&+& \frac{1}{(z-w)^2} \, A^2 \, \left[  (k+N+2) 
\, h^{\mu}_{\bar{a} \bar{b}} \, h^{\nu \bar{a}}_{\,\,\,\,\,\, \bar{d}}
\, f^{\bar{b} \bar{d}}_{\,\,\,\,\,\, e} \, V^e
+k \, h^{\mu}_{\bar{a} \bar{b}} \,
h^{\nu \bar{b}}_{\bar{c}} \, Q^{\bar{a}} \, Q^{\bar{c}} \right] (w)
\nonu \\
&+& \frac{1}{(z-w)} \, A^2 \left[ - (k+N+2) \, h^{\mu}_{\bar{a} \bar{b}} 
\, h^{\nu \bar{a}}_{\,\,\,\,\,\, \bar{d}} \, V^{\bar{b}} \, V^{\bar{d}} 
+ k \, h^{\mu}_{\bar{a} \bar{b}} \, h^{\nu \bar{b}}_{\bar{c}} \, \pa 
\, Q^{\bar{a}} \, Q^{\bar{c}}  \right.
\nonu \\
&-&  h^{\mu}_{\bar{a} \bar{b}} \, h^{\nu}_{\bar{c} \bar{d}}
\left. f^{\bar{b} \bar{d}}_{\,\,\,\,\,\, e} \, Q^{\bar{a}} \, Q^{\bar{c}} \,
V^{e} \right] (w) + \cdots.
\label{GGope}
\eea
Note that the spin-$1$ current contracted with the structure constant
in (\ref{GGope})
contains the $SU(N+2)$ group index.
The normalization constant 
$A(k,N)$ is  determined by the Wolf space central charge term
and one takes $\mu=\nu=0$ and  
$
G^{0}(z) \, G^{0}(w) |_{\frac{1}{(z-w)^3}} = \frac{2}{3}  c_{\mbox{Wolf}}
$
where the Wolf space coset central charge is
given by $c_{\mbox{Wolf}}  =  \frac{6 k N}{(2+k+N)}$ in (\ref{goverh}).
This implies that $A(k,N)=\frac{i}{(k+N+2)}$ in (\ref{Gansatz})
where we used $g_{\bar{a} \bar{b}}\, g^{ \bar{a} \bar{b}} = 4N$.
Therefore,  the four spin-$\frac{3}{2}$ currents are 
given by
\bea
G^{0}(z)  & = &  \frac{i}{(k+N+2)}  \, Q_{\bar{a}} \, V^{\bar{a}}(z),
\nonu \\
G^{i}(z) & = &  \frac{i}{(k+N+2)} 
\, h^{i}_{\bar{a} \bar{b}} \, Q^{\bar{a}} \, V^{\bar{b}}(z),
\quad (i=1,2,3).
\label{expressionG}
\eea
These equations correspond to the equation $(3.18)$ of \cite{GK}
\footnote{For $N=3$, one has the following results with (\ref{basischange}) 
from \cite{Ahn1311}:
\bea
G^0(z) & = &
\frac{i}{(5+k)} 
\left(\sum_{(\bar{A},\bar{A}^*)=(\bar{1},\bar{1}^*)}^{(\bar{3},\bar{3}^*)} Q^{\bar{A}} 
V^{\bar{A}^*} +\sum_{(\bar{A}^*,\bar{A})=(\bar{1}^*,\bar{1})}^{(\bar{3}^*,\bar{3})} 
Q^{\bar{A}^*} V^{\bar{A}}
+ \sum_{(\bar{A},\bar{A}^*)=(\bar{4},\bar{4}^*)}^{(\bar{6},\bar{6}^*)} Q^{\bar{A}} 
V^{\bar{A}^*} +
\sum_{(\bar{A}^*,\bar{A})=(\bar{4}^*,\bar{4})}^{(\bar{6}^*,\bar{6})} Q^{\bar{A}^*} 
V^{\bar{A}} \right)(z),
\nonu \\
G^1(z) & = &
\frac{i}{(5+k)} 
\left(\sum_{(\bar{A},\bar{B})=(\bar{4},\bar{1})}^{(\bar{6},\bar{3})} Q^{\bar{A}} V^{\bar{B}} +
\sum_{(\bar{A}^*,\bar{B}^*)=(\bar{4}^*,\bar{1}^*)}^{(\bar{6}^*,\bar{3}^*)} 
Q^{\bar{A}^*} V^{\bar{B}^*}
- \sum_{(\bar{A},\bar{B})=(\bar{1},\bar{4})}^{(\bar{3},\bar{6})} Q^{\bar{A}} V^{\bar{B}} -
\sum_{(\bar{A}^*,\bar{B}^*)=(\bar{1}^*,\bar{4}^*)}^{(\bar{3}^*,\bar{6}^*)} Q^{\bar{A}^*} 
V^{\bar{B}^*} \right)(z),
\nonu \\
G^2(z) & = &
\frac{1}{(5+k)} 
\left(-\sum_{(\bar{A},\bar{B})=(\bar{4},\bar{1})}^{(\bar{6},\bar{3})} Q^{\bar{A}} V^{\bar{B}} +
\sum_{(\bar{A}^*,\bar{B}^*)=(\bar{4}^*,\bar{1}^*)}^{(\bar{6}^*,\bar{3}^*)} 
Q^{\bar{A}^*} V^{\bar{B}^*}
+ \sum_{(\bar{A},\bar{B})=(\bar{1},\bar{4})}^{(\bar{3},\bar{6})} Q^{\bar{A}} V^{\bar{B}} -
\sum_{(\bar{A}^*,\bar{B}^*)=(\bar{1}^*,\bar{4}^*)}^{(\bar{3}^*,\bar{6}^*)} Q^{\bar{A}^*} 
V^{\bar{B}^*} \right)(z),
\nonu \\
G^3(z) & = &
\frac{1}{(5+k)} 
\left(-\sum_{(\bar{A},\bar{A}^*)=(\bar{4},\bar{4}^*)}^{(\bar{6},\bar{6}^*)} Q^{\bar{A}} 
V^{\bar{A}^*} +
\sum_{(\bar{A}^*,\bar{A})=(\bar{4}^*,\bar{4})}^{(\bar{6}^*,\bar{6})} Q^{\bar{A}^*} V^{\bar{A}}
- \sum_{(\bar{A},\bar{A}^*)=(\bar{1},\bar{1}^*)}^{(\bar{3},\bar{3}^*)} Q^{\bar{A}} V^{\bar{A}^*} +
\sum_{(\bar{A}^*,\bar{A})=(\bar{1}^*,\bar{1})}^{(\bar{3}^*,\bar{3})} Q^{\bar{A}^*} 
V^{\bar{A}} \right)(z),
\nonu
\eea
which is consistent with the ones in (\ref{expressionG})
with the help of (\ref{fourh}).
Here the indices $10, 11, 12$ (and their complex conjugated ones)
in \cite{Ahn1311} are replaced by $4, 5, 6$ (and their conjugated ones)
and the indices $7, 8$ (and their complex conjugated ones)
are replaced by $11, 12$ (their complex conjugated ones).
The remaining indices 
can be identified each other.
The Wolf space coset indices in this case are given by $(\bar{1},
\bar{2},\bar{3})$ and 
$(\bar{4},\bar{5},\bar{6})$ (and their complex conjugated ones).
}.

Let us describe the OPE structure (\ref{GGope}) in detail.
Let us focus on the quadratic spin-$1$ currents
which appear in the first term in the first-order pole in (\ref{GGope}).
One has similar term for the OPE $G^{\nu}(z) \, G^{\mu}(w)$.
The sum of these two terms should be equal to the twice of
stress energy term 
(\ref{N4scalgebra}) which contains also quadratic spin-$1$ currents 
as follows:
\bea
h^{\mu}_{\bar{a} \bar{b} } \, h^{\nu \bar{a} }_{\,\,\,\,\,\, \bar{d} }
+h^{\nu}_{\bar{a} \bar{b} } \, h^{\mu \bar{a} }_{\,\,\,\,\,\, \bar{d} }
&=& 
2 \, \delta^{\mu \nu} \, g_{\bar{b} \bar{d}}, \qquad (\mu,\nu=0,1,2,3).
\label{hcondition}
\eea
This equation (\ref{hcondition}) corresponds to 
the equation $(3.6)$ of \cite{GK}. 
One can obtain two properties from (\ref{hcondition}) as follows:
\bea
h^{i}_{\bar{a} \bar{b}}  & = &  - h^{i}_{\bar{b} \bar{a}},
\nonu \\
h^{i}_{\bar{a} \bar{b} } \, h^{i \bar{a} }_{\,\,\,\,\,\, \bar{c} } & = &  g_{\bar{b} \bar{c}},
\label{twopro}
\eea
where $i=1,2,3$ and there is no sum over the index $i$ in  the second
equation of (\ref{twopro}). 
The first one in (\ref{twopro}) corresponds to the case 
$\mu = i$ and $\nu=0$
while the second one in (\ref{twopro})
corresponds to the case $\mu=\nu=i$ in (\ref{hcondition}).
The complex structure is antisymmetric in the Wolf space coset indices
according to the first equation of (\ref{twopro}).
The conditions in (\ref{twopro}) imply that the 
$h^i$ are almost complex
structures and correspond to the equations $(3.8)$ and $(3.9)$ of \cite{GK}.
If one defines the third complex structure as
\bea
h^{3}_{\bar{a} \bar{b}}
&\equiv& h^{1}_{\bar{a} \bar{c}}  \, h^{2 \bar{c} }_{\,\,\,\,\,\, \bar{b}}, 
\label{h3}
\eea
then the 
three almost complex structures $(h^1, h^2, h^3)$ 
satisfy the following relation
\bea
h^{i}_{\bar{a} \bar{c} } \, h^{j \bar{c} }_{\,\,\,\,\,\, \bar{b} }
&=& 
\ep^{ijk} \, h^{k}_{\bar{a} \bar{b} } -  \delta^{ ij } \, g_{\bar{a} \bar{b}}. 
\label{hidentity}
\eea
This relation corresponds to the quaternionic algebra 
$(2.24)$ of \cite{Saulina}. See also \cite{MS9907} where the relation
(\ref{hidentity}) was described.

For the  identities from 
the cubic terms in (\ref{GGope}),  one presents in Appendix $E$.

Now one should determine other remaining currents.
We can identify six spin-$1$ currents $A^{\pm i}(z)$ 
of $SU(2)_{\hat{k}^+} \times SU(2)_{\hat{k}^-}$ 
using the last two equations in (\ref{N4scalgebraExplic})
as follows \footnote{For $N=3$, the results from \cite{Ahn1311}
with (\ref{basischange})
lead to the following expressions
\bea
A^{+1}(z) & = & -\frac{1}{2} \left( -V^9 +V^{9^*} \right)(z), \qquad
A^{+2}(z)   =    \frac{i}{2} \left( V^9 +V^{9^*}\right)(z), \nonu \\
A^{+3}(z) & = & -\frac{1}{12} i (\sqrt{3}+3 i) V^{11}(z)  -
\frac{i (3 \sqrt{5}+i \sqrt{3}) }{12 \sqrt{2}} V^{12}(z) -
\frac{1}{12} (3+i \sqrt{3}) V^{11^*}(z) - \frac{(\sqrt{3}+3 i \sqrt{5})}{
12 \sqrt{2}} V^{12^*}(z),
\nonu \\
A^{-1}(z) & = & \frac{1}{2(k+5)} \left(
\sum_{(\bar{A},\bar{B})=(\bar{1},\bar{4})}^{(\bar{3},\bar{6})} Q^{\bar{A}} Q^{\bar{B}}
+ \sum_{(\bar{A}^*,\bar{B}^*)=(\bar{1}^*,\bar{4}^*)}^{(\bar{3}^*,\bar{6}^*)} 
Q^{\bar{A}^*} Q^{\bar{B}^*} \right)(z),
\nonu \\
A^{-2}(z) & = & \frac{i}{2(k+5)} \left(
\sum_{(\bar{A},\bar{B})=(\bar{1},\bar{4})}^{(\bar{3},\bar{6})} Q^{\bar{A}} Q^{\bar{B}}
- \sum_{(\bar{A}^*,\bar{B}^*)=(\bar{1}^*,\bar{4}^*)}^{(\bar{3}^*,\bar{6}^*)} 
Q^{\bar{A}^*} Q^{\bar{B}^*} \right)(z),
\nonu \\
A^{-3}(z) & = & -\frac{i}{2(k+5)} \left(
\sum_{(\bar{A},\bar{A}^*)=(\bar{4},\bar{4}^*)}^{(\bar{6},\bar{6}^*)} Q^{\bar{A}} Q^{\bar{A}^*}
+ \sum_{(\bar{A},\bar{A}^*)=(\bar{1},\bar{1}^*)}^{(\bar{3},\bar{3}^*)} Q^{\bar{A}} 
Q^{\bar{A}^*} \right)(z),
\nonu 
\eea
which is consistent with the ones in (\ref{sixspin1def}).
For the $A^{+i}(z)$ current, it is not obvious to see the consistency 
because there exists a structure constant in (\ref{sixspin1def}). However,
one can check by looking at the explicit values of the
nonzero structure constants in \cite{Ahn1311}. }:
\bea
A^{+i}(z)  & = &
-\frac{1}{4N} \, f^{\bar{a} \bar{b}}_{\,\,\,\,\,\, c} \, h^i_{\bar{a} \bar{b}} \, V^c(z), 
\nonu \\ 
A^{-i}(z) & = & 
-\frac{1}{4(k+N+2)} \, h^i_{\bar{a} \bar{b}} \, Q^{\bar{a}} \, Q^{\bar{b}}(z), 
\quad  (i=1,2,3).
\label{sixspin1def}
\eea
These equations (\ref{sixspin1def}) 
correspond to the equation $(3.18)$ of \cite{GK}.
They satisfy the following OPEs
\footnote{Actually, we have checked that for the upper case, using the currents
for low values of $N= 3,5,7,9$, this equation holds and for the lower
case, one identifies this equation using the property of
complex structures.}
\bea 
A^{\pm i}(z) \, A^{\pm j}(w) &=& 
-\frac{1}{(z-w)^2} \, \frac{1}{2} \, \delta^{ij} \, \hat{k}^{\pm} 
+ \frac{1}{(z-w)}  
\, \ep^{ijk} \, A^{\pm k} (w) + \cdots,
\label{opeaa}
\eea
where $i,j=1,2,3$ and $\hat{k}^{+}=k$ is the level for $SU(2)_{\hat{k}^{+}}$ while
$\hat{k}^{-}=N$ is the level for $SU(2)_{\hat{k}^{-}}$.
They commute with each other because the spin-$1$ currents and the 
spin-$\frac{1}{2}$ currents do not have any singular terms from 
(\ref{opevq}).
One can easily construct the sum of these two currents 
$(A^{+ i} +A^{- i})(z)$ which has the level $(\hat{k}^{+} +\hat{k}^{-})=(k+N)$
for $SU(2)_{\hat{k}^{+} +\hat{k}^{-}}$
\footnote{One can read off the 
three almost complex structures of $SU(N+2=5)$  
from the spin-$1$ currents in (\ref{sixspin1def}) 
(or the $\hat{ B_i }(z)$   in \cite{Ahn1311})
and they are $12 \times 12$  matrices as follows:
\bea
&&h^1_{\bar{a} \bar{b}} = 
\left(
\begin{array}{cccc}
0 & - 1  & 0 & 0 \\
1 & 0 & 0 & 0 \\
0 & 0 & 0 & -1 \\
0 & 0 & 1 & 0 \\
\end{array}
\right), \quad
h^2_{\bar{a} \bar{b}}  = 
\left(
\begin{array}{cccc}
0 & - i   & 0 & 0 \\
i  & 0 & 0 & 0 \\
0 & 0 & 0 & i  \\
0 & 0 & -i  & 0 \\
\end{array}
\right), \quad
h^3_{\bar{a} \bar{b}}  = h^1_{\bar{a} \bar{c}} \, h^{2 \bar{c}}_{ \,\,\,\,\,\, \bar{b}},
\nonu
\eea
where each element is $3 \times 3$ matrix. 
The $h^3_{\bar{a} \bar{b}}$ was defined in (\ref{h3})
and three complex structures satisfy (\ref{hidentity}).
In Appendix $B$, 
the $N$-generalization for the complex structures is given.}. 

There are many ways to extract the correct stress energy tensor.
For example,  one can identify the stress energy tensor from 
the first-order pole term of $G^0(z) \, G^0(w)$. 
After substituting $\mu=\nu=0$ in both sides of (\ref{GGope})
and subtracting the correct nonlinear terms coming from the two
kinds of spin-$1$ currents, one arrives at the following result
\bea
L(z) &=& 
\frac{1}{2(k+N+2)^2} \left[ (k+N+2) \, V_{\bar{a}} \, V^{\bar{a}} 
+k \, Q_{\bar{a}} \, \pa \, Q^{\bar{a}} 
+f_{\bar{a} \bar{b} c} \, Q^{\bar{a}} \, Q^{\bar{b}} \, V^c  \right] (z)
\nonu \\
&-& \frac{1}{(k+N+2)} \sum_{i=1}^3 \left( A^{+i}+A^{-i}  \right)^2 (z).
\label{Lexpression}
\eea
The first line in (\ref{Lexpression}) corresponds to 
the right hand side of (\ref{GGope})
and the second line corresponds to 
the above nonlinear terms which are the usual 
Sugawara construction for the 
above $SU(2)_{\hat{k}^{+} +\hat{k}^{-}}$ currents.
The above $10$ currents 
$G^{\mu}(w)$ and  $A^{\pm i}(w)$ 
are primary under this stress energy tensor $L(z)$ and furthermore
one can check the following OPE from the OPEs in (\ref{opevq})
\footnote{Also in this case one checked this OPE for low values 
for $N=3,5,7,9$.
Furthermore, one has the second line of (\ref{Lexpression}) as 
\bea
\sum_{i=1}^3 \left( A^{+i}+A^{-i}  \right)^2 (z)
&=&
\sum_{i=1}^3 \left( \frac{1}{16N^2} \,
f^{\bar{a} \bar{b}}_{\,\,\,\,\,\, c} \, h^i_{\bar{a} \bar{b}} \, 
f^{\bar{d} \bar{e}}_{\,\,\,\,\,\, f} \, h^i_{\bar{d} \bar{e}} \, V^c
\, V^f 
+\frac{1}{8N(k+N+2)} \,
f^{\bar{a} \bar{b}}_{\,\,\,\,\,\, c} \, h^i_{\bar{a} \bar{b}} \, 
h^i_{\bar{d} \bar{e}} \, Q^{\bar{d}} \, Q^{\bar{e}}
\, V^c 
\right. \nonu \\
& + & \left. \frac{1}{16(k+N+2)^2} h^i_{\bar{a} \bar{b}} h^i_{\bar{c} \bar{d}}
Q^{\bar{a}} Q^{\bar{b}} Q^{\bar{c}} Q^{\bar{d}}\right)  (z)
- \frac{3}{4(k+N+2)} Q_{\bar{a}} \pa Q^{\bar{a}}   (z).
\nonu
\eea
Note that the spin-$1$ Kac-Moody current and the spin-$\frac{1}{2}$
Kac-Moody current commute with each other and there is no ordering problem 
between them.
For the expression of $(A^{-i} \, A^{-i})(z)$, 
one should be careful about 
the ordering \cite{BS}.
If one simplifies (\ref{Lexpression}) further,
then all the expressions can be combined with the ones in the first line of 
(\ref{Lexpression}) except the quartic fermionic term and the quadratic 
spin-$1$ currents. }
\bea
L(z) \, L(w) &=& \frac{1}{(z-w)^4} \, \frac{\hat{c}}{2} +
\frac{1}{(z-w)^2} \, 2 L(w) +\frac{1}{(z-w)} \, \pa L(w) +\cdots,
\label{opell}
\eea 
where the central charge 
$\hat{c} =\frac{3(k+N+2k N)}{(k+N+2)}$ is equal to the one 
in (\ref{goverh}).

Summarizing this subsection, the 
large ${\cal N}=4$ nonlinear superconformal algebra 
in terms of the Wolf space coset currents
is characterized by (\ref{GGope}), where the 
right hand side should be written in terms of 
the $11$ currents, (\ref{opeaa}) and (\ref{opell}). 
Of course, as above, the $10$ currents are primary.
We are left with the OPEs between the spin-$1$ currents $A^{\pm i}(z)$ 
and the 
spin-$\frac{3}{2}$ currents $G^{\mu}(w)$. 
They appear in Appendix $C$. The role of three almost complex structures
was very crucial. Again the $11$ currents are given by 
(\ref{expressionG}), (\ref{sixspin1def}) and (\ref{Lexpression}). Due to 
the nonlinear terms in (\ref{GGope}), the spin-$2$ stress energy 
tensor takes the form of very nontrivial expression. 
This is the reason why some of the literatures do not have the correct 
and explicit form for the stress energy tensor.
We will explain this 
feature in Appendix $C$ in detail
 \footnote{
One has the explicit relations between the 
$11$ currents in \cite{Ahn1311} and the ones in this paper 
as follows:
\bea
 \hat{ B_i }(z) & = &  A^{- i}(z)  , \quad (i=1,2,3),
\nonu \\
 \hat{ A_1 }(z) & = & 
 -A^{+ 1}(z)   ,  \quad   \hat{ A_2 }(z)=A^{+ 2}(z) , 
\quad  \hat{ A_3 }(z)=-A^{+ 3}(z), 
\nonu \\
 \hat{G}_{11} (z) & = & \frac{1}{\sqrt{2}} (G^1- i G^2) (z) , \qquad
\hat{G}_{12} (z) = -\frac{1}{\sqrt{2}} (G^3- i G^0) (z),
\nonu \\
 \hat{G}_{22} (z) &  = & \frac{1}{\sqrt{2}} (G^1+ i G^2)(z), \qquad
\hat{G}_{21} (z) = -\frac{1}{\sqrt{2}} (G^3+ i G^0) (z),
\nonu \\
\hat{T} (z) & = &  L (z).
\label{basischange}
\eea
One can calculate the first order pole of the OPE 
$\hat{G}_{11}(z) \, \hat{G}_{22}(w)$ using (\ref{basischange})
and (\ref{GGope})
and read off the following quantity
\bea
\hat{T}(z)
&=&
\frac{1}{2(k+N+2)^2} \left[ (k+N+2)\,(g-i h^3)_{\bar{a} \bar{b}} \,
 V^{\bar{a} } \, V^{\bar{b} } 
+k  (g +i h^3)_{\bar{a} \bar{b}} \,  Q^{\bar{a} }\,  \pa \, Q^{ \bar{b} } \right.
\nonu \\
&+&\left. \frac{1}{2} \, (h^1-i h^2)_{\bar{a} \bar{b}} 
\, (h^1+i h^2)_{\bar{c} \bar{d}} \, f^{\bar{b} \bar{d} }_{\,\,\,\,\,\, e} 
\, Q^{\bar{a} } \, Q^{ \bar{c}} \, V^{e}
\right] (z)
- \frac{1}{(k+N+2)}
\left[  i  \pa \, \left( N \hat{A}_3 - k \hat{B}_3 \right)
\right. \nonu \\
& + & \left.   
\frac{1}{2} \, \left( \hat{A}_{+} \hat{A}_{-}+\hat{A}_{-} \hat{A}_{+} \right)
+ 
\hat{A}_{3}\, \hat{A}_{3} + 2 \hat{A}_3 \, \hat{B}_3 +
\frac{1}{2} \left( \hat{B}_{+} \, \hat{B}_{-}+\hat{B}_{-} \, \hat{B}_{+} \right)
+\hat{B}_{3} \, \hat{B}_{3}    \right](z), 
\nonu
\eea
which corresponds to the equation $(4.31)$ of \cite{GP1403}.
In this calculation the properties of complex structures are used
and $\hat{A}_{\pm}(z) \equiv \hat{A}_1 \pm i \hat{A}_2(z)$ and 
$\hat{B}_{\pm}(z) \equiv \hat{B}_1 \pm i \hat{B}_2(z)$.
}. 
\section{The higher spin currents in the Wolf space coset    }

Let us recall that the lowest $16$ higher spin currents 
have the following ${\cal N}=2$ multiplets with spin contents
\bea
\left(1, \frac{3}{2}, \frac{3}{2}, 2 \right)
& : & (T^{(1)}, T_{+}^{(\frac{3}{2})}, T_{-}^{(\frac{3}{2})}, T^{(2)}), 
\nonu \\
 \left(\frac{3}{2}, 2, 2, \frac{5}{2} \right) & : & 
(U^{(\frac{3}{2})}, U_{+}^{(2)}, U_{-}^{(2)}, U^{(\frac{5}{2})} ), \nonu \\
\left(\frac{3}{2}, 2, 2, \frac{5}{2} \right) & : & 
(V^{(\frac{3}{2})}, V^{(2)}_{+}, V^{(2)}_{-}, V^{(\frac{5}{2})}),  \nonu \\
\left(2, \frac{5}{2}, \frac{5}{2}, 3 \right) & : &
 (W^{(2)}, W_{+}^{(\frac{5}{2})}, W_{-}^{(\frac{5}{2})}, W^{(3)}).
\label{lowesthigher}
\eea
In this section, we would like to 
construct these higher spin currents explicitly.
As described in the introduction, it is very crucial to obtain 
the lowest higher spin-$1$ current in (\ref{lowesthigher}).
Once this is found, then it is straightforward to 
obtain the remaining $15$ higher spin currents with the help of 
the four spin-$\frac{3}{2}$ currents.

\subsection{ One higher spin-$1$ current }

From the $N=3$ result in \cite{Ahn1311},
one expresses the higher spin-$1$ current as 
\bea
T^{(1)}(z)&=&
-\frac{1}{(k+5)} 
\left[ \frac{5 \left(\sqrt{3}+3 i\right)  }{6} V^{11} -\frac{\left(9 \sqrt{5}-5 i
   \sqrt{3}\right) }{6 \sqrt{2}} V^{12} +\frac{5 \left(\sqrt{3}-3 i\right)
   }{6 }V^{11^*}  \right.
   \label{newspin1su5} \\
   &-& \left. \frac{\left(9 \sqrt{5}+5 i
   \sqrt{3}\right) }{6 \sqrt{2} } V^{12^*} \right] (z)
   + \frac{k}{(k+5)^2} \left[ \sum_{(\bar{A},\bar{A}^*)=(\bar{1},\bar{1}^*)}^{(\bar{3},
\bar{3}^*)} Q^{\bar{A}} Q^{\bar{A}^*} - 
   \sum_{(\bar{A},\bar{A}^*)=(\bar{4},\bar{4}^*)}^{(\bar{6},\bar{6}^*)} Q^{\bar{A}} 
Q^{\bar{A}^*} \right](z).
   \nonu 
\eea
How does one generalize this higher spin-$1$ current for general
$N$?
For the quadratic terms in (\ref{newspin1su5}), one 
can easily  see the generalization by expanding the corresponding
indices to the $N$ and $N$ respectively.
The second summation has an opposite sign.  
However, for the first four terms in (\ref{newspin1su5}),
one cannot see the behavior for the general $N$ easily
because the structure constants in $SU(5)$ are rather complicated.  
Based on (\ref{newspin1su5}), one can try to write the following quantity
\bea
T^{(1)}(z) &=&
-\frac{1}{2(k+N+2)} \, A_a \, V^a (z) 
+  \frac{k}{2(k+N+2)^2} \, d^{0}_{\bar{a} \bar{b}} \, Q^{\bar{a}} \, Q^{\bar{b}} (z),
\label{newspinone}
\eea
where $A_a$ and $d^{0}_{\bar{a} \bar{b}}$ are numerical constants 
(to be determined later) and 
the $d^{0}_{\bar{a} \bar{b}}$ is antisymmetric by construction.
Recall that from the explicit form for the spin-$1$ currents in previous 
section in (\ref{sixspin1def}), one sees the linear $V^a(z)$ term and 
the quadratic $ Q^{\bar{a}} \, Q^{\bar{b}}(z)$ term. 
For the higher spin-$1$ current, 
the two arbitrary coefficient tensors are inserted. 
We also insert the $(N,k)$ dependences in (\ref{newspinone}) from 
$N=3$ result of (\ref{newspin1su5}).

The 
primary condition of the higher spin-$1$ current 
under the stress energy tensor (\ref{Lexpression})
leads to the following OPE
\bea
T^{(1)} (z) \, \hat{T} (w) |_{\frac{1}{(z-w)^2}} &=& T^{(1)} (w). 
\label{newspin1primary}
\eea
In other words, there are no other singular terms in the OPE
$T^{(1)} (z) \, \hat{T} (w)$ except (\ref{newspin1primary}).
The third-order pole is identically zero.
From the condition of (\ref{newspin1primary}),
one has the following relations
\bea
-\frac{1}{2(N+2)} \, A_a \, f_{\bar{b} \bar{c} }^{\,\,\,\,\,\, a} 
 & = &
d^{0}_{\bar{b} \bar{c}}, \nonu \\
A_a \, f^{a \bar{b} }_{\,\,\,\,\,\, c} \, f^{c}_{\,\,\,\, \bar{b} d}
-\frac{k}{(N+2)} \, A_a \, 
f^a_{\,\,\,\, \bar{b} \bar{c}} \, f^{\bar{b} \bar{c}}_{\,\,\,\,\,\, d}
& = &  2 (k+N+2) A_d.
\label{newspin1primaryresult}
\eea

The regular conditions with the six spin-$1$ currents
imply that
\bea
&&T^{(1)} (z) \, A^{\pm i} (w) =
 +\cdots. 
 \label{newspin1regular}
\eea
From the first condition (with upper sign) of (\ref{newspin1regular}),
one has the two relations (corresponding to 
the first-order pole and second-order pole)
\bea
&&A_a  \, f^{\bar{b} \bar{c}}_{\,\,\,\,\,\, d} \, h^i_{\bar{b} \bar{c}}
 \, f^{ad}_{\,\,\,\,\,\, e}=0, \quad \quad
 A_a \, f^{\bar{b} \bar{c} a}  \, h^i_{\bar{b} \bar{c}} = 0. 
\nonu
\eea
From the second condition (with lower sign) of (\ref{newspin1regular}),
the following relations for the first-order and the second-order poles 
hold
\bea
&&d_{\bar{a}}^{ 0 \,\, \bar{c}} \, h^i_{\bar{c} \bar{b}}
=d_{\bar{b}}^{ 0 \,\, \bar{c}} \, h^i_{\bar{c} \bar{a}}, \quad \quad
d^{0 \bar{a} \bar{b}} \, h^i_{\bar{a} \bar{b}}=0.
 \label{2ndregular}
\eea

One has the following condition
\bea
T^{(1)} (z) \, G^{\mu} (w) |_{\frac{1}{(z-w)^2}}
&=&
0.
\label{newspin1Gcondition}
\eea
From this relation (\ref{newspin1Gcondition}), 
the following relations are satisfied
\bea
A^{\bar{a}} \, h^{\mu}_{\bar{a} \bar{b}} =0, 
\quad (\mbox{or} \quad A_{\bar{a}}=0). 
\label{newspin1Gconditionresult}
\eea
When $\mu=0$ in this equation,  one obtains the relation inside of the 
bracket. Using this result, for the $\mu=i$, one has
$A^{\bar{a}} \, h^{i}_{\bar{a} \bar{b}} =0$ automatically.
One can obtain the following relation from 
(\ref{newspin1primaryresult}) and
(\ref{newspin1Gconditionresult}) as follows:
\bea
A_{a} &=& d^{0}_{\bar{b} \bar{c} } \, f^{\bar{b} \bar{c} }_{\,\,\,\,\,\, a}. 
\label{dfrelation}
\eea

One can read off the explicit form  for the second rank tensor 
$d^{0}_{\bar{a} \bar{b}}$ in $SU(5)$ from \cite{Ahn1311} (or (\ref{newspin1su5}))
and generalize to the $SU(N+2)$ case. 
In $SU(5)$ case, 
one has the following $12 \times 12$ matrix representation
\bea
&&d^{0}_{\bar{a} \bar{b}} = 
\left(
\begin{array}{cccc}
0 & 0 & 1 & 0 \\
0 & 0 & 0 & -1 \\
-1 & 0 & 0 & 0 \\
0 & 1 & 0 & 0 \\
\end{array}
\right),
\label{d0expressionexp}
\eea
where each element  is $3 \times 3$ matrix.
See Appendix $B$ for the $N$-generalization of the tensor 
$d^{0}_{\bar{a} \bar{b}}$
and other related tensors.

Furthermore,
the tensor $d^{0}_{\bar{a} \bar{b}}$ satisfies the 
following properties (for $N=3,5,7,9$)
\bea
&&d^{0}_{\bar{a} \bar{c}} \, d^{0 \bar{c}}_{\,\,\,\,\,\, \bar{b}} 
=
g_{\bar{a} \bar{b}}, \quad \quad 
d^{0}_{\bar{a} \bar{b}} \, f^{\bar{b}}_{\,\,\,\, \bar{c} d} =
d^{0}_{\bar{c} \bar{b}} \, f^{\bar{b}}_{\,\,\,\, \bar{a} d}. 
\label{d0property}
\eea
The first equation of (\ref{d0property}) is similar to the second 
equation of (\ref{twopro}) while the second equation of (\ref{d0property})
looks similar to the equation $(3.11)$ of \cite{GK}.
These will be used on (\ref{minusrelation}) later.

Summarizing this subsection, 
one can rewrite $T^{(1)} (z)$ using (\ref{dfrelation}) where the 
coefficient tensor appearing in the linear spin-$1$ current can be 
written in terms of those appearing in the quadratic fermionic term, 
\bea
T^{(1)} (z) &=&
-\frac{1}{2(k+N+2)} \, d^0_{\bar{a} \bar{b}} \,
f^{\bar{a} \bar{b}}_{\,\,\,\,\,\, c}  V^c  (z)
+ \frac{k}{2(k+N+2)^2} \, d^0_{\bar{a} \bar{b}} \, Q^{\bar{a}} \, Q^{\bar{b}} (z).
\label{finalspinone}
\eea
In the first term, there is sum over the index $c$
which can be expanded as the subgroup index and the coset index.
This coset index term vanishes.
Note that the structure constant satisfies 
$f^{\bar{a} \bar{b}}_{\,\,\,\,\,\,\bar{c}} =0$. 
This is true for the symmetric space.
For $N=3$ case, the expression (\ref{newspin1su5})
does not contain the fields $V^a(z)$ where $a=1, \cdots, 6$ (or its
complex conjugated ones).
Compared to the previous spin-$1$ currents of large ${\cal N}=4$ 
nonlinear superconformal algebra, 
the expression in (\ref{finalspinone})
contains the  $ d^0_{\bar{a} \bar{b}}$ tensor. 
Because this higher spin-$1$ current is a singlet $({\bf 1},{\bf 1})$ 
under the 
$SU(2)_{\hat{k}^{+}} \times SU(2)_{\hat{k}^{-}}$, there is no any index 
corresponding to these groups.
In the construction of the remaining higher spin currents, the presence 
of this $ d^0_{\bar{a} \bar{b}}$ tensor occurs in all the OPEs in this paper.

\subsection{ Four higher spin-$\frac{3}{2}$  currents}

Let us describe the next higher spin currents.
Because the four spin-$\frac{3}{2}$ currents (\ref{expressionG})
and the higher spin-$1$ current (\ref{finalspinone}) are given, 
one can calculate the following OPE
\bea
G^{\mu} (z) \, T^{(1)} (w) 
&=&
\frac{1}{(z-w)} \, G'^{\mu} (w) + \cdots.
\label{Gprime}
\eea
The algebraic structure of (\ref{Gprime}) 
was already given in $N=3$ case.
In other words, the higher spin-$1$ current 
generates four higher spin-$\frac{3}{2}$ currents. 
One can calculate $G'^{\mu} (w)$ using the condition (\ref{2ndregular}) and 
it turns out that
\bea
G'^{\mu}(z) &=&
\frac{i}{(k+N+2)} \, 
d^{\mu}_{\bar{a} \bar{b}} \, Q^{\bar{a}} \, V^{\bar{b}} (z),
\label{higherspin3half}
\eea
where the product of $d^{0 \bar{c} }_{ \bar{a} }$ 
and $h^{\mu }_{\bar{c} \bar{b}}$ is defined via
\bea
d^{\mu}_{\bar{a} \bar{b}} \equiv d^{0 \bar{c} }_{ \bar{a} } \, 
h^{\mu }_{\bar{c} \bar{b}}.
\label{dd}
\eea
The $d^i_{\bar{a} \bar{b}} (i=1,2,3)$ are symmetric and 
traceless from the condition (\ref{2ndregular}).
Also the $\mu=0$ case in (\ref{dd}) 
leads to  the $d^{0}_{\bar{a} \bar{b}}$ tensor.
Compared to the previous spin-$\frac{3}{2}$ currents (\ref{Gansatz}), 
the only difference is the presence of the second rank tensor in (\ref{dd}).
The normalization and the field contents are the same.
It is obvious that the higher spin-$\frac{3}{2}$ currents 
transform as $({\bf 2},{\bf 2})$ under the $SU(2) \times SU(2)$ as done in
the four spin-$\frac{3}{2}$ currents in (\ref{Gansatz}). They 
share the common $\mu$ indices.

One can calculate the OPE $G'^{\mu}(z)  \, G'^{\nu}(w)$ 
using the relation (\ref{d0property}) and 
the result is given by the opposite sign of (\ref{GGope})  
\bea
G'^{\mu}(z)  \, G'^{\nu}(w)
&=&
-G^{\mu}(z)  \, G^{\nu}(w).
\label{minusrelation}
\eea
This is also observed in the context of 
asymptotic symmetry algebra in \cite{GP1403}.
In other words, their singular terms have the above relation 
(\ref{minusrelation}).

From (\ref{basischange}) on the spin-$\frac{3}{2}$ currents, 
one can also have similar relations for the higher spin-$\frac{3}{2}$
currents as
\bea
& \hat{G}'_{11} (z) =\frac{1}{\sqrt{2}} (G'^1- i G'^2) (z) , \quad
\hat{G}'_{12} (z) = -\frac{1}{\sqrt{2}} (G'^3- i G'^0) (z),
\nonu \\
& \hat{G}'_{22} (z)=\frac{1}{\sqrt{2}} (G'^1+ i G'^2)(z), \quad
\hat{G}'_{21} (z) = -\frac{1}{\sqrt{2}} (G'^3+ i G'^0) (z).
\label{ggdoublesingle}
\eea
Then by using the defining equation,
one obtains the following OPE in the bispinor notation
\bea
\hat{G}_{m n} (z) \, T^{(1)} (w) 
&=&
\frac{1}{(z-w)} \, \hat{G}'_{m n} (w)+\cdots,
\quad (m,n=1,2).
\label{gtdouble}
\eea
Note that the currents 
$\hat{G}'_{m n} (w)$ is denoted by
$V_{\frac{1}{2}}^{(1) \alpha \beta} (w)$ in \cite{BCG1404}.
The higher spin-$\frac{3}{2}$ current 
$T_{+}^{(\frac{3}{2})}(w)$ was appeared in the following OPE
\cite{Ahn1311} (see also Appendix $G$)
\bea
\hat{G}_{21} (z) \, T^{(1)} (w) =
\frac{1}{(z-w)} \,
\left[ \hat{G}_{21}  + 2 T_{+}^{(\frac{3}{2})} \right](w)+\cdots.
\label{Oope}
\eea
Then one can rewrite the first-order pole in (\ref{Oope}) as
the one in (\ref{gtdouble})
and one has the following simple relation 
\bea
T_{+}^{(\frac{3}{2})} (z) & = & 
\frac{1}{2} \left( \hat{G}'_{21}  - \hat{G}_{21} \right) (z)
\nonu \\
& = &  -\frac{i}{2\sqrt{2}(k+N+2)} 
(d^3 + i d^0-h^3 -i h^0)_{\bar{a} \bar{b}}  
\, Q^{\bar{a}} \, V^{\bar{b}}(z).
\label{t+3half}
\eea
Similarly,
for completeness with the help of Appendix $G$, 
one writes the other three higher spin-$\frac{3}{2}$
currents as follows:
\bea
T_{-}^{(\frac{3}{2})} (z) &=&
\frac{1}{2} \left( \hat{G}'_{12}  + \hat{G}_{12} \right) (z)=
 -\frac{i}{2\sqrt{2}(k+N+2)} 
(d^3 - i d^0+h^3 -i h^0)_{\bar{a} \bar{b}}  
\, Q^{\bar{a}} \, V^{\bar{b}}(z),
\nonu \\
U^{(\frac{3}{2})} (z) & = &
\frac{1}{2} \left( \hat{G}'_{11}  - \hat{G}_{11} \right) (z)
= \frac{i}{2\sqrt{2}(k+N+2)} 
(d^1 - i d^2-h^1 +i h^2)_{\bar{a} \bar{b}}  
\, Q^{\bar{a}} \, V^{\bar{b}}(z),
\nonu \\
V^{(\frac{3}{2})} (z) &=&
\frac{1}{2} \left( \hat{G}'_{22}  + \hat{G}_{22} \right) (z)
\nonu \\
& = &
 \frac{i}{2\sqrt{2}(k+N+2)} 
(d^1 + i d^2+h^1 +i h^2)_{\bar{a} \bar{b}}  
\, Q^{\bar{a}} \, V^{\bar{b}}(z).
\label{spin-3over2relation}
\eea

Summarizing this subsection,
the expression in (\ref{higherspin3half})
contains the  $ d^{\mu}_{\bar{a} \bar{b}}$ tensor while 
the previous spin-$\frac{3}{2}$ currents of large ${\cal N}=4$ 
nonlinear superconformal algebra has  the  $ h^{\mu}_{\bar{a} \bar{b}}$ tensor. 
Or $\hat{G}_{mn}(z)$  and $\hat{G}_{mn}^{\prime}(z)$ in (\ref{gtdouble})
are in the representation $({\bf 2},{\bf 2})$ 
under the 
$SU(2)_{\hat{k}^{+}} \times SU(2)_{\hat{k}^{-}}$.
According to the defining OPE in (\ref{Gprime}),
the role of $h^{\mu}_{\bar{a}\bar{b}}$ and  $d^0_{\bar{a}\bar{b}}$ 
tensors was crucial. The former gives the $SO(4)$ vector representation
for the higher spin-$\frac{3}{2}$ currents
and the latter gives the additional $\mu$-independence on them
\footnote{Note that for $N=3$ the above four higher 
spin-$\frac{3}{2}$ currents were written as
\bea
T_{+}^{(\frac{3}{2})}(z) & = &  -\frac{\sqrt{2}}{(5+k)} 
\sum_{(\bar{A},\bar{A}^*)=(\bar{4},\bar{4}^*)}^{(\bar{6},\bar{6}^*)} 
Q^{\bar{A}} V^{\bar{A}^*}(z), \qquad
T_{-}^{(\frac{3}{2})}(z)  =   -\frac{\sqrt{2}}{(5+k)} 
\sum_{(\bar{A},\bar{A}^*)=(\bar{4},\bar{4}^*)}^{(\bar{6},\bar{6}^*)} Q^{\bar{A}^*} V^{\bar{A}}(z), \nonu \\
U^{(\frac{3}{2})}(z)&  = & - i \frac{\sqrt{2}}{(5+k)} 
\sum_{(\bar{A},\bar{B})=(\bar{4},\bar{1})}^{(\bar{6},\bar{3})} Q^{\bar{A}} V^{\bar{B}}(z), 
\qquad
V^{(\frac{3}{2})}(z)  =   i \frac{\sqrt{2}}{(5+k)} 
\sum_{(\bar{A}^*,\bar{B}^*)=(\bar{4}^*,\bar{1}^*)}^{(\bar{6}^*,\bar{3}^*)} 
Q^{\bar{A}^*} V^{\bar{B}^*}(z).
\nonu 
\eea
From the representation in (\ref{fourh1}) and (\ref{fourg1}),
the nonzero elements for 
$(d^{3}+ i d^{0} - h^{3}
- i h^{0})_{\bar{a} \bar{b}} $ are given by $\bar{4} \, \bar{4}^*$, $
\bar{5} \, \bar{5}^*$ and 
$\bar{6} \, \bar{6}^*$. 
Then one can check that $T_{+}^{(\frac{3}{2})}(z)$ is consistent with
the general expression in (\ref{t+3half}).
One can also check other cases.}. 

\subsection{ Six  higher spin-$2$ currents }

One can generalize the equation $(4.27)$ in \cite{Ahn1311} 
to the corresponding OPE for general $N$ (in fact $N=3,5,7,9$)
as follows:
\bea
\hat{G}_{11}(z) \, T_{-}^{(\frac{3}{2})}(w) & = & 
\frac{1}{(z-w)^2} \, \frac{2N }{(N+k+2)} \left[ i \hat{A}_{+} \right](w) \nonu \\
& + &
\frac{1}{(z-w)} \, \left[ - U_{-}^{(2)} + \frac{1}{2} \pa \{
\hat{G}_{11} \, T_{-}^{(\frac{3}{2})} \}_{-2}
\right](w) +\cdots.
\label{3rdspin2ope}
\eea
From the relation (\ref{spin-3over2relation}), the left hand side of 
the OPE (\ref{3rdspin2ope}) can be written as
\bea
 \hat{G}_{11}(z) \, T_{-}^{(\frac{3}{2})}(w) 
=\frac{1}{2} \hat{G}_{11}(z) \, \hat{G}'_{12} (w)
+\frac{1}{2} \hat{G}_{11}(z) \, \hat{G}_{12}(w).
\label{somerelation}
\eea
Because the OPE $\hat{G}_{11}(z) \, \hat{G}_{12}(w)$ is already known 
for general $N$ (Appendix $D$),
one arrives at the following OPE
\bea
\hat{G}_{11}(z) \, \hat{G}'_{12} (w)
&=&
\frac{1}{(z-w)} \, \left[ - 2 U_{-}^{(2)} -\frac{4}{(N+k+2)} 
  \hat{A}_{+} \hat{B}_3  \right](w) +\cdots.
\label{ggprime1112}
\eea
In other words, there are no second-order pole terms 
in (\ref{ggprime1112}) because they are cancelled each other in 
(\ref{somerelation}).
From (\ref{ggprime1112}), one can write down the higher spin-$2$ current 
as follows: 
\bea
U_{-}^{(2)}(w)
&=& -\frac{1}{2} \{ \hat{G}_{11} \, \hat{G}'_{12} \}_{-1} (w)
 - \frac{2}{(N+k+2)} 
   \hat{A}_{+} \hat{B}_3  (w).
\label{uuu}
\eea
Let us emphasize that 
in (\ref{3rdspin2ope})
one should calculate the second-order pole
to determine $U_{-}^{(2)}(w)$ for the relative $(N,k)$ dependent coefficient
appearing in the descendant field of $\hat{A}_{+}(w)$ 
but in (\ref{uuu}) one has to 
calculate the first-order pole only. 
Then it is straightforward to 
calculate the first term of (\ref{uuu}) from (\ref{expressionG}),
(\ref{basischange}), 
(\ref{higherspin3half}) and (\ref{ggdoublesingle}) and it turns out (for
$N=3,5,7$) that
\bea
U_{-}^{(2)}(w)
&=&
\frac{1}{2 (k+N+2)^2}\left[ (k+N+2) \, (i d^1+ d^2 )_{\bar{a} \bar{b}}  
\, V^{\bar{a}} \, V^{\bar{b}} 
\right. 
\label{finalexp}
 \\
&-& \left. \frac{1}{2} \, ( h^1 - i h^2 )_{\bar{a} \bar{b}}
 \, (i d^0- d^3)_{\bar{c} \bar{d}} \, f^{\bar{b} \bar{d}}_{\,\,\,\,\,\, e}
 \, Q^{\bar{a}} \, Q^{\bar{c}} \, V^{e} \right] (w)
 - \frac{2}{(N+k+2)} 
\left[   \hat{A}_{+} \, \hat{B}_3  \right] (w).
\nonu
\eea
Furthermore one can substitute (\ref{sixspin1def}) with (\ref{basischange}) 
into the last term of (\ref{finalexp})
\bea
U_{-}^{(2)}(w)
& = &
\frac{1}{2 (k+N+2)^2}\left[ (k+N+2) \,
(i d^1+ d^2 )_{\bar{a} \bar{b}}  \, V^{\bar{a}} \, V^{\bar{b}} 
\right. \nonu \\
& - & \left.  
\frac{1}{4N} \, f^{\bar{c} \bar{d}}_{\,\,\,\,\,\, e}  \,
( h^1 - i h^2 )_{\bar{c} \bar{d}}
\, (i d^0- h^3 )_{\bar{a} \bar{b}}
 \, Q^{\bar{a}} \, Q^{\bar{b}} \, V^{e} \right] (w),
\label{u-2final}
\eea
where the identities appearing in 
(\ref{GGprimeidentity}) are used \footnote{
For $N=3$, this current has the following form:
\bea
U_{-}^{(2)}(w)=
-i \frac{2}{(5+k)^2} \sum_{(\bar{A},\bar{A}^*)=(\bar{4},\bar{4}^*)}^{(\bar{6},\bar{6}^*)} 
Q^{\bar{A}} Q^{\bar{A}^*} V^9(w) - i \frac{2}{(5+k)} 
\sum_{(\bar{A},\bar{B})=(\bar{1},\bar{4})}^{(\bar{3},\bar{6})} 
V^{\bar{A}} V^{\bar{B}}(w)-\frac{3i}{(5+k)} \pa 
V^9(w).
\nonu
\eea 
One can see that the quadratic term  can be checked 
by realizing that $(i d^1 + d^2)_{\bar{a} \bar{b}}$ has nonzero term 
in the $\bar{1}\,\bar{4} $, $\bar{2}\, \bar{5}$ and $\bar{3} \, \bar{6}$ 
elements (and $\bar{4}\, \bar{1}$, $\bar{5}\, \bar{2}$ and 
$\bar{6}\, \bar{3}$ elements) from (\ref{fourh1}) and (\ref{fourg1}).
The derivative term in the above can be written in terms of quadratic terms.
The cubic terms can be checked from Appendices (\ref{fourh1}) and 
(\ref{fourg1})
and the explicit values of nonzero structure 
constants in \cite{Ahn1311}.  }.
There are no quartic terms in (\ref{u-2final}) because the spin-$\frac{3}{2}$
current and the higher spin-$\frac{3}{2}$ current do not have any 
cubic terms.

Now one can describe the next higher spin-$2$ current.
As in previous case, the equation $(4.37)$ of \cite{Ahn1311} (or Appendix $G$)
implies the algebraic structure for the higher spin-$2$ current 
$V_{+}^{(2)}(w)$.
The OPE $\hat{G}_{22}(z) \, T_{+}^{(\frac{3}{2})}(w)$
can be written in terms of 
$\hat{G}_{22}(z)$ with 
$\frac{1}{2} (\hat{G}_{21}' - \hat{G}_{21})(w)$
from (\ref{t+3half}).
Then the second-order pole of $(4.37)$ of \cite{Ahn1311}
is canceled out the one of the OPE between 
$\hat{G}_{22}(z)$ and 
$\frac{1}{2}  \hat{G}_{21}(w)$. 
By reading off the first-order pole in the OPE
between $\hat{G}_{22}(z)$ and $\hat{G}_{21}'(w)$, 
one obtains
the following higher spin-$2$ current (for $N=3,5,7$)
\bea
V_{+}^{(2)}(w)
&=&
\frac{1}{2 (k+N+2)^2}\left[ (k+N+2) \, (-i d^1+ d^2)_{\bar{a} \bar{b}}  
\, V^{\bar{a}} \, V^{\bar{b}} 
\right. 
\label{expexp} \\
&-& \left. \frac{1}{2} \, ( i g+h^3 )_{\bar{a} \bar{b}}
 \, ( d^1+i d^2 )_{\bar{c} \bar{d}} \, f^{\bar{b} \bar{d}}_{\,\,\,\,\,\, e}
 \, Q^{\bar{a}} \, Q^{\bar{c}} \, V^{e} \right] (w)
+ \frac{2}{(N+k+2)} 
\left[   \hat{A}_{-} \, \hat{B}_3  \right] (w).
\nonu
\eea
Moreover, further substitution of 
(\ref{sixspin1def}) and (\ref{basischange})
into the last term of (\ref{expexp})
leads to the following simplified expression
\bea
 V_{+}^{(2)}(w)
& = & 
\frac{1}{2 
(k+N+2)^2}\left[ (k+N+2) \, (-i d^1+ d^2 )_{\bar{a} \bar{b}} \,
  V^{\bar{a}} \, V^{\bar{b}} 
\right. \nonu \\
& + & \left.
  \frac{1}{4N} \, f^{\bar{c} \bar{d}}_{\,\,\,\,\,\, e} \, ( h^1+i h^2 )_{\bar{c} \bar{d}}
 \, ( i d^0 -h^3 )_{\bar{a} \bar{b}} 
 \, Q^{\bar{a}} \, Q^{\bar{b}} \, V^{e} \right] (w),
\label{v+2final}
\eea
where
 the identities appearing in 
(\ref{GGprimeidentity}) are used.
Note that the complete expression of (\ref{v+2final})
looks similar to the one in (\ref{u-2final}). 
The only difference appears in the coefficients of $d^1_{\bar{a} \bar{b}}$
and $h^1_{\bar{c} \bar{d}}$. The $U(1)$ charge 
introduced in \cite{Ahn1311} of (\ref{v+2final})
is opposite to the  one of (\ref{u-2final})
\footnote{
For $N=3$, this current has the following form:
\bea
 V_{+}^{(2)}(w) =
-i \frac{2}{(5+k)^2} \sum_{(\bar{A},\bar{A}^*)=(\bar{4},\bar{4}^*)}^{(\bar{6},\bar{6}^*)} 
Q^{\bar{A}} Q^{\bar{A}^*} V^{9^*}(w) - i \frac{2}{(5+k)} 
\sum_{(\bar{A}^*,\bar{B}^*)=(\bar{1}^*,\bar{4}^*)}^{(\bar{3}^*,\bar{6}^*)} V^{\bar{A}^*} 
V^{\bar{B}^*}(w)+\frac{3i}{(5+k)} \pa 
V^{9^*}(w).
\nonu
\eea
The nonzero elements of 
$(-i d^1 + d^2)_{\bar{a} \bar{b}}$ in (\ref{fourh1}) and (\ref{fourg1}) 
are given by  
in the $\bar{1}^* \,\bar{4}^* $, $\bar{2}^* \, \bar{5}^*$ 
and $\bar{3}^* \, \bar{6}^*$ (and $\bar{4}^*\, \bar{1}^*$, $\bar{5}^*,
\bar{2}^*$ and $\bar{6}^*,\bar{3}^*$) and one can check 
the quadratic terms are identified. The derivative term is associated the 
ordering in the quadratic terms. 
One can check the cubic terms by looking at the nonzero structure constants
appearing in \cite{Ahn1311} and Appendices (\ref{fourh1}) and (\ref{fourg1}).}.

Let us consider the two higher spin-$2$ currents with nonzero
$U(1)$ charges.
The equation $(4.23)$ of \cite{Ahn1311} (or Appendix $G$)
implies the algebraic structure for the higher spin-$2$ current 
$U_{+}^{(2)}(w)$.
The OPE $\hat{G}_{11}(z) \, T_{+}^{(\frac{3}{2})}(w)$
can be written in terms of 
$\hat{G}_{11}(z)$ with 
$\frac{1}{2} (\hat{G}_{21}' - \hat{G}_{21})(w)$
from (\ref{t+3half}).
Then the second-order pole of $(4.23)$ of \cite{Ahn1311}
is canceled out the one of the OPE between 
$\hat{G}_{11}(z)$ and 
$\frac{1}{2}  \hat{G}_{21}(w)$. 
By reading off the first-order pole in the OPE
between $\hat{G}_{11}(z)$ and $\hat{G}_{21}'(w)$, 
one obtains
the following higher spin-$2$ current (for $N=3,5,7$)
\bea
U_{+}^{(2)}(w)
&=&
-\frac{1}{2 (k+N+2)^2}\left[
  k \, (  i d^1+d^2 )_{\bar{a} \bar{b}} \, Q^{\bar{a}} \, \pa \, Q^{\bar{b}}  
\right. 
\label{finalu+2}
\\
& - & \left. \frac{1}{2} \, ( h^1 - i h^2 )_{\bar{a} \bar{b}}
 \, (i d^0+ d^3 )_{\bar{c} \bar{d}} \, f^{\bar{b} \bar{d}}_{\,\,\,\,\,\, e}
 \, Q^{\bar{a}} \, Q^{\bar{c}} \, V^{e} \right] (w)
 - \frac{2}{(N+k+2)} 
\left[ \hat{A}_3  \, \hat{B}_{-}  \right] (w).
\nonu
\eea
The last term of (\ref{finalu+2}) can be replaced with the expressions 
in section $2$.

The equation $(4.41)$ of \cite{Ahn1311} (or Appendix $G$)
implies the algebraic structure for the higher spin-$2$ current 
$V_{-}^{(2)}(w)$.
The OPE $\hat{G}_{22}(z) \, T_{-}^{(\frac{3}{2})}(w)$
can be written in terms of 
$\hat{G}_{22}(z)$ with 
$\frac{1}{2} (\hat{G}_{12}' + \hat{G}_{12})(w)$
from (\ref{spin-3over2relation}).
Then the second-order pole of $(4.47)$ of \cite{Ahn1311}
is canceled out the one of the OPE between 
$\hat{G}_{22}(z)$ and 
$\frac{1}{2}  \hat{G}_{12}(w)$. 
By reading off the first-order pole in the OPE
between $\hat{G}_{22}(z)$ and $\hat{G}_{12}'(w)$, 
one obtains
the following higher spin-$2$ current (for $N=3,5,7$)
\bea
V_{-}^{(2)}(w)
&=&
\frac{1}{2 (k+N+2)^2}\left[
  k \, (  i d^1-d^2 )_{\bar{a} \bar{b}} \, Q^{\bar{a}} \, \pa \, Q^{\bar{b}}  
+ \frac{1}{2} \, ( i g -h^3 )_{\bar{a} \bar{b}}
 \, (d^1+i d^2 )_{\bar{c} \bar{d}} \, f^{\bar{b} \bar{d}}_{\,\,\,\,\,\, e}
 \, Q^{\bar{a}} \, Q^{\bar{c}} \, V^{e} \right] (w)
 \nonu \\
  &+& \frac{2}{(N+k+2)} 
\left[ \hat{A}_3 \, \hat{B}_{+}   \right] (w).
\label{finalv-2} 
\eea
One can also 
replace the last term of (\ref{finalv-2}) with the expressions 
in section $2$.

The equation $(4.16)$ of \cite{Ahn1311} (or Appendix $G$)
implies the algebraic structure for the higher spin-$2$ current 
$T^{(2)}(w)$.
The OPE $\hat{G}_{21}(z) \, T_{-}^{(\frac{3}{2})}(w)$
can be written in terms of 
$\hat{G}_{21}(z)$ with 
$\frac{1}{2} (\hat{G}_{12}' + \hat{G}_{12})(w)$
from (\ref{spin-3over2relation}).
Then the third-order pole of $(4.16)$ of \cite{Ahn1311}
is canceled out the one of the OPE between 
$\hat{G}_{21}(z)$ and 
$\frac{1}{2}  \hat{G}_{12}(w)$. 
By reading off the first-order pole in the OPE
between $\hat{G}_{21}(z)$ and $\hat{G}_{12}'(w)$, 
one obtains
the following higher spin-$2$ current (for $N=3,5,7$)
\bea
T^{(2)} (w) &=&
 -\frac{1}{2 (k+N+2)^2}\left[ (k+N+2)
\, (d^0-i d^3 )_{\bar{a} \bar{b}}  \, V^{\bar{a}} \, V^{\bar{b}} 
 + k \, ( d^0 - i d^3 )_{\bar{a} \bar{b}} 
\, Q^{\bar{a}} \, \pa \, Q^{\bar{b}}  \right.
\nonu \\
&-&\left. \frac{1}{2} ( i g- h^3 )_{\bar{a} \bar{b}}
\, (i d^0+ d^3 )_{\bar{c} \bar{d}} \, f^{\bar{b} \bar{d}}_{\,\,\,\,\,\, e}
 \, Q^{\bar{a}} \, Q^{\bar{c}} \, V^{e} \right] (w)
\label{finalt2}
\\
 &+& \left[ \frac{1}{2} \pa T^{(1)}  +
\frac{1}{(N+k+2)} \left( \hat{A}_i \hat{A}_i +\hat{B}_i \hat{B}_i
-2 \hat{A}_3 \hat{B}_3 \right)
+\frac{(k+N)}{(k+N+2kN)} \hat{T} \right] (w).
\nonu
\eea
The last line of (\ref{finalt2}) comes from Appendix $G$.

The equation $(4.48)$ of \cite{Ahn1311} (or Appendix $G$)
implies the algebraic structure for the higher spin-$2$ current 
$W^{(2)}(w)$.
The OPE $\hat{G}_{22}(z) \, U^{(\frac{3}{2})}(w)$
can be written in terms of 
$\hat{G}_{22}(z)$ with 
$\frac{1}{2} (\hat{G}_{11}' - \hat{G}_{11})(w)$
from (\ref{spin-3over2relation}).
Then the third-order pole of $(4.48)$ of \cite{Ahn1311}
is canceled out the one of the OPE between 
$\hat{G}_{22}(z)$ and 
$\frac{1}{2}  \hat{G}_{11}(w)$. 
By reading off the first-order pole in the OPE
between $\hat{G}_{22}(z)$ and $\hat{G}_{11}'(w)$, 
one obtains
the following higher spin-$2$ current (for $N=3,5,7$)
\bea
W^{(2)} (w) &=&
 \frac{1}{2 (k+N+2)^2}\left[ (k+N+2) \,
(d^0-i d^3 )_{\bar{a} \bar{b}}  \, V^{\bar{a}} \, V^{\bar{b}} 
 + k \, ( d^0 + i d^3 )_{\bar{a} \bar{b}} \, Q^{\bar{a}} \, \pa \, Q^{\bar{b}}  \right.
\nonu \\
&+&\left. \frac{1}{2} ( h^1-i h^2 )_{\bar{a} \bar{b}}
 \, ( d^1+i d^2 )_{\bar{c} \bar{d}} \, f^{\bar{b} \bar{d}}_{\,\,\,\,\,\, e}
 \, Q^{\bar{a}} \, Q^{\bar{c}} \, V^{e} \right] (w)
 \nonu \\
 &+& \left[ -\frac{1}{2} \pa T^{(1)}  +
\frac{1}{(N+k+2)} \left( \hat{A}_i \hat{A}_i +\hat{B}_i \hat{B}_i
+2 \hat{A}_3 \, \hat{B}_3 \right)
+ \hat{T} \right] (w).
\label{w2final}
\eea
In this case, the last line of (\ref{w2final}) comes from Appendix $G$
\footnote{ Strictly speaking, the higher spin-$2$ current $W^{(2)}(w)$ 
in \cite{Ahn1311} is not a primary current but a quasi-primary current.
One can find the primary higher spin-$2$ current 
as follows (for $N=3,5,7,9$):
\bea
\tilde{W}^{(2)}  (w)
&\equiv& W^{(2)} (w) - \frac{2kN}{(k+N+2kN)} \hat{T} (w)
\label{tildeW2} 
\\ 
&=&
\left[ \frac{1}{2} \{ \hat{G}_{11} \, \hat{G}'_{22} \}_{-1}
 -\frac{1}{2} \pa T^{(1)} +
\frac{2}{(N+k+2)} \left( \hat{A}_i \hat{A}_i +\hat{B}_i \hat{B}_i
+2 \hat{A}_3 \hat{B}_3 \right) + \frac{(k+N)}{(k+N+2kN)} \hat{T}  \right] (w).
\nonu
\eea
Note that 
the coefficient of $\hat{T}(w)$  in 
$\tilde{W}^{(2)}(w)$ (\ref{tildeW2}) is the same 
as the one of $\hat{T}(w)$ in $T^{(2)}(w)$.
}.

It is useful to write the above higher spin-$2$ currents 
in terms of  the $ V_{1}^{(s)\pm i} $ ($s=1, i=1,2,3$) introduced in 
\cite{BCG1404}. 
Their equation $(2.10)$ implies the following 
conditions
\bea
\hat{T}(z) \, V_{1}^{(1)\pm i} (w) 
&=&\frac{1}{(z-w)^2} \, 2  \, V_{1}^{(1)\pm i} (w) 
+\frac{1}{(z-w)} \, \pa  \, V_{1}^{(1)\pm i} (w)  + \cdots.
\label{eq1}
\eea
The regular conditions in $(2.11)$ of \cite{BCG1404}
can be written as 
\bea
\left(
\begin{array}{c}
\hat{A}_i \\
\hat{B}_i \\
\end{array} \right)
(z)  \, V_{1}^{(1) \mp j} (w) 
= +\cdots. 
\label{eq2}
\eea
This implies that 
there are no singular terms 
between the triplet of spin-$1$ current $\hat{A}_i(z) [\hat{B}_i(z)]$ 
in the first [second] $SU(2)$ factor 
and the triplet of the higher spin-$2$ current $V_1^{(1)-j}(w) [V_1^{(1)+j}(w)]$ 
in the second [first] $SU(2)$ factor.
Furthermore, one has the following OPEs
\bea
\left(
\begin{array}{c}
\hat{A}_i \\
\hat{B}_i \\
\end{array} \right)
(z)  \, V_{1}^{(1) \pm j} (w) 
&=& \frac{1}{(z-w)^2} \, a \, \delta_{ij} \, T^{(1)} (w)
+\frac{1}{(z-w)} \, \ep^{ijk}  \, V_{1}^{(1) \pm k} (w)  + \cdots.
\label{eq3}
\eea
Therefore, the two triplets in the same $SU(2)$ factor 
do have nontrivial OPEs. 

By requiring the above conditions (\ref{eq1}), (\ref{eq2})
and (\ref{eq3}), one obtains their three higher spin-$2$ currents 
as follows (for $N=3,5,7$):
\bea
V_{1}^{(1)+ 1} (z)
&=&
\left[ U^{(2)}_{-} - V^{(2)}_{+} 
+ \frac{4}{(N+k+2)} \, \hat{A}_{1} \, \hat{B}_{3} \right] (z),
\nonu \\
V_{1}^{(1)+ 2} (z)
&=&
\left[ -i (U^{(2)}_{-} + V^{(2)}_{+}) 
+ \frac{4}{(N+k+2)} \, \hat{A}_{2} \, \hat{B}_{3} \right](z),
\label{Eq1}
\\
V_{1}^{(1)+ 3} (z)
&=&
\left[ T^{(2)} - W^{(2)} 
+ \frac{4}{(N+k+2)} \, \hat{A}_{3} \, \hat{B}_{3} 
+\frac{2kN}{(N+k+2kN)} \, \hat{T} \right](z).
\nonu
\eea
Note that the $U(1)$ charge defined in \cite{Ahn1311}
is not preserved in (\ref{Eq1}).
It is not obvious to  see the $SU(2)$ 
index  appearing in 
the left hand side of (\ref{Eq1}) in the right hand side 
explicitly.
The other three higher spin-$2$ currents
are given by (for $N=3,5,7$)
\bea
V_{1}^{(1)- 1} (z)
&=&
\left[ V^{(2)}_{-} - U^{(2)}_{+} 
- \frac{4}{(N+k+2)} \, \hat{A}_{3} \, \hat{B}_{1} \right] (z),
\nonu \\
V_{1}^{(1)- 2} (z)
&=&
\left[ -i (V^{(2)}_{-} + U^{(2)}_{+}) 
- \frac{4}{(N+k+2)} \, \hat{A}_{3} \, \hat{B}_{2} \right] (z),
\label{Eq2} 
\\
V_{1}^{(1)- 3} (z)
&=&
\left[ T^{(2)} + W^{(2)} 
- \frac{2}{(N+k+2)} \left(
\hat{A}_{i} \hat{A}_{i} 
+\hat{B}_{i} \hat{B}_{i} 
\right) 
- \frac{2(N+k+kN)}{(N+k+2kN)} \hat{T} \right]  (z).
\nonu
\eea
Also in this case, the $U(1)$ charge is not conserved.
Because the right hand side of (\ref{Eq1}) and (\ref{Eq2}) are known 
from the previous results, one can simplify them further \footnote{
That is,
\bea
V_{1}^{(1)+ j} (z)
&=&
\frac{i (-1)^{j+1} }{(k+N+2)^2} \,
\left[ (k+N+2) \, d^j_{\bar{a} \bar{b}} \, V^{\bar{a}} \, V^{\bar{b}}
 -\frac{1}{2}   
C^{+j}_{\bar{a} \bar{b} \bar{c} \bar{d}}
 \, f^{{\bar{b} \bar{d}}}_{\,\,\,\,\,\, e}
\, Q^{\bar{a}}  \, Q^{\bar{c}} \, V^e
\right]  (z),
\label{spin2bcg1}
\eea
where the rank four tensor is introduced as follows:
\bea
C^{\pm j}_{\bar{a} \bar{b} \bar{c} \bar{d}}
\equiv h^j_{\bar{a} \bar{b}} \, d^0_{\bar{c} \bar{d}} \pm \frac{1}{2}
\ep^{jkl} \, h^k_{\bar{a} \bar{b}} \, d^l_{\bar{c} \bar{d}}.
\nonu
\eea
Similarly the right hand side of (\ref{Eq2})
can be further simplified as follows:
\bea
V_{1}^{(1)- j} (z)
&=&
 \frac{i}{(k+N+2)^2} \,
\left[ k \, d^j_{\bar{a} \bar{b}} \, Q^{\bar{a}} \, \pa \, Q^{\bar{b}}
-\frac{1}{2} \, C^{-j}_{\bar{a} \bar{b} \bar{c} \bar{d}}
\, f^{{\bar{b} \bar{d}}}_{\,\,\,\,\,\, e}
\, Q^{\bar{a}}  \, Q^{\bar{c}} \, V^e
\right]  (z).
\label{spin2bcg2}
\eea
The constant $a$ in (\ref{eq3}) is determined in next subsection and 
is fixed as $a=i$. 
After this simplification, one 
sees the $SU(2) \times SU(2)$ representations explicitly.
}. 

Summarizing this subsection, the higher spin-$2$ currents
are given by (\ref{u-2final}), (\ref{v+2final}), (\ref{finalu+2}), 
(\ref{finalv-2}), (\ref{finalt2}) and 
(\ref{w2final}). They contain the quadratic and cubic terms 
in the Kac-Moody currents as well as the composite currents 
from the large ${\cal N}=4$ nonlinear superconformal algebra.

\subsection{ Four higher spin-$\frac{5}{2}$ currents }

Now one can move the other higher spin currents.
The $(4.55)$ of \cite{Ahn1311} can be generalized to the 
following OPE
\bea
\hat{G}_{12} (z) \, W^{(2)} (w) 
&=&
\frac{1}{(z-w)^2} \, \left[ \frac{(N+2k+1)}{(N+k+2)} \hat{G}_{12} 
+ \frac{(N-k)}{(N+k+2)} T_{-}^{(\frac{3}{2})} \right](w)
\nonu \\
&+& 
\frac{1}{(z-w)} \, \left[ W_{-}^{(\frac{5}{2})} 
+\frac{1}{3} \pa \{ \hat{G}_{12} \, W^{(2)} \}_{-2}
\right] (w)+ \cdots.
\label{Oope1}
\eea
Then the spin-$\frac{5}{2}$
current (for $N=3,5,7$) can be read off 
\bea
W_{-}^{(\frac{5}{2})}(w)
= \{ \hat{G}_{12} \, W^{(2)} \}_{-1} (w)  - \frac{1}{3} \pa  \{ \hat{G}_{12} \, W^{(2)} \}_{-2}  (w).
\nonu
\eea
The second term
can be obtained from the second-order pole of (\ref{Oope1}).
From the explicit expressions of spin-$\frac{3}{2}$ current
 (\ref{expressionG}) with (\ref{basischange}) 
and the higher spin-$2$ current (\ref{w2final}), 
one obtains the following result
\bea
W_{-}^{(\frac{5}{2})}(w)
&=& \frac{1}{\sqrt{2}(k+N+2)^2} \left[ (k-1) \, (d^0+i d^3)_{\bar{a} \bar{b}} 
\, \pa \, Q^{\bar{a}} \, V^{\bar{b}}
-(k+1)\, (d^0+i d^3)_{\bar{a} \bar{b}}  \, Q^{\bar{a}} \, \pa  \, V^{\bar{b}} \right.
\nonu \\
&+& \left. A_{\bar{a} \bar{d} e}  \, Q^{\bar{a}} \, V^{\bar{d}} \, V^{e}
+\frac{1}{4(k+N+2)} \, B_{\bar{a} \bar{b} \bar{c} \bar{d}} \, Q^{\bar{a}} \, 
Q^{\bar{b}} \, Q^{\bar{c}} \, V^{\bar{d}} 
\right] (w)
 - \frac{1}{(N+k+2)} \left[ \frac{4}{3} \pa \hat{G}_{12}  \right.
 \nonu \\
&+& \left. \frac{1}{3}(2N+k+3) \pa \hat{G}'_{12}   -  2i \,
(\hat{A}_3+\hat{B}_3) \, \hat{G}_{12}
+i  \hat{A}_{+} \, \hat{G}_{22} +i \hat{B}_{+} \, \hat{G}_{11}  \right] (w).
\label{w-5half}
\eea
The second part (starting from $\pa \hat{G}_{12}(w)$ to the end) 
of (\ref{w-5half})
can be obtained by taking the OPEs between 
the spin-$\frac{3}{2}$ current $\hat{G}_{12}(z)$
and the last line of (\ref{w2final}) (and the contributions from the derivative of the second order pole of (\ref{Oope1})). 
All of these are easily determined from the 
basic OPEs in section $2$ (and the OPE in (\ref{gtdouble})).
Here 
we introduce the rank three tensor and the rank four tensor as follows:
\bea
A_{\bar{a}  \bar{d} e}  & \equiv &
\left[ (-ig+h^3)_{\bar{a} \bar{b}}  \, d^3_{\bar{c} \bar{d} }
+\frac{1}{2} \, (d^1+i d^2)_{\bar{a} \bar{b}} \, (h^1-i h^2)_{\bar{c} \bar{d} } 
\right] f^{\bar{b} \bar{c}}_{\,\,\,\,\,\, e},
\nonu \\
B_{\bar{a} \bar{b} \bar{c} \bar{d}} & \equiv & (-i h^1+h^2)_{\bar{a} \bar{b}} \,
(i d^1+d^2)_{\bar{c} \bar{d}} 
+2 \, d^0_{\bar{a} \bar{b}}  \, (g+i h^3)_{\bar{c} \bar{d}}.
\label{abtensors}
\eea
In the rank three tensor, there is a $SU(N+2)$ index.
In the rank four tensor, all the indices are living in the 
Wolf space coset.
They will appear in the construction of 
the higher spin-$3$ current later.

In particular, the quartic term  in the second line 
of (\ref{w-5half}) can be analyzed 
as follows. 
Let us focus on the OPE $\hat{G}_{12}(z)$ and the third term of
the higher spin-$2$ current in 
(\ref{w2final}). Then according to the last equation of
Appendix (\ref{firstfirst}),
one has the quartic term in the right hand side of (\ref{firstfirst}).
By multiplying the correct coefficients, one arrives at the 
final expression in (\ref{w-5half}). 
Note that in (\ref{w-5half}) there is no quintic fermionic term
because the spin-$\frac{3}{2}$ current has no cubic term in the Wolf space 
coset. For the nonsymmetric space, one expects that there should be 
such quintic term. 

Similarly, from Appendix $G$ (the first equation of 
(\ref{spin5halfgenerating})), one can read off the next higher
spin-$\frac{5}{2}$
current and it turns out, from   the spin-$\frac{3}{2}$ current
(\ref{expressionG}) with (\ref{basischange}) 
and the higher spin-$2$ current (\ref{finalexp}), that (for $N=3,5,7$)
\bea
U^{(\frac{5}{2})} (w) &=&
\{ \hat{G}_{21}  \, U^{(2)}_{-} \}_{-1} (w)  - \frac{1}{3} \pa \{ \hat{G}_{21}  \, U^{(2)}_{-} \}_{-2} (w)
\nonu \\
&=&
\frac{i}{\sqrt{2}(k+N+2)^2} \left[ (2k+N) \,
(d^1-i d^2)_{\bar{a} \bar{b}} \, \pa \, Q^{\bar{a}} \, V^{\bar{b}}
+N \, (d^1-i d^2)_{\bar{a} \bar{b}}  \, Q^{\bar{a}} \, \pa \, V^{\bar{b}} \right.
\nonu \\
&+& \left. A^{1}_{\bar{a} \bar{d} e} \, Q^{\bar{a}}  \, V^{\bar{d}}  \, V^{e} 
+\frac{1}{2(k+N+2)} \, d^0_{\bar{a} \bar{b}}  \, (h^1-i h^2)_{\bar{c} \bar{d}} 
\, Q^{\bar{a}} \, Q^{\bar{b}} \, Q^{\bar{c}} \, V^{\bar{d}}
\right] (w)
\label{u5half}
\\
&+& \frac{1}{(N+k+2)} \left[ \frac{1}{3} \pa \hat{G}_{11}- 
\frac{1}{3}(N+2k+1) \pa \hat{G}'_{11}  
-2i  \hat{G}_{11} \, \hat{B}_3
+i  \hat{A}_{+} \, \hat{G}_{21}  \right] (w),
\nonu
\eea
where the tensor is given by
\bea
A^{1}_{\bar{a} \bar{d} e} \equiv 
\left[ (i g +h^3)_{\bar{a} \bar{b}} \, (i d^1+ d^2)_{\bar{c} \bar{d}} 
+\frac{1}{2} \, (h^1-i h^2)_{\bar{a} \bar{b}} \, 
(d^0- i d^3)_{\bar{c} \bar{d}} \right]  f^{\bar{b} \bar{c}}_{\,\,\,\,\,\, e}.
\nonu
\eea
As described before, the 
last line of (\ref{u5half}) can be obtained from 
the OPE between $\hat{G}_{21}(z)$ and the derivative of the 
second order pole in the first equation in (\ref{spin5halfgenerating}). 
The quartic term  in the second line (in the final expression) 
of (\ref{u5half}) can be analyzed 
as follows. 
The OPE $\hat{G}_{21}(z)$ and the second term of
the higher spin-$2$ current in 
(\ref{finalexp}) implies the last equation of
Appendix (\ref{firstfirst}).
One has the quartic term in the right hand side of (\ref{firstfirst}).
By multiplying the correct coefficients, one arrives at the 
final expression in (\ref{u5half}). 
Note that in (\ref{u5half}) there is no quintic fermionic term
as described before. 

One can continue to describe the following higher spin-$\frac{5}{2}$ current
with the help of  the spin-$\frac{3}{2}$ current
(\ref{expressionG}) with (\ref{basischange}) 
and the higher spin-$2$ current (\ref{finalv-2}) and it turns out that
(for $N=3,5,7$)
\bea
V^{(\frac{5}{2})} (w) &=&
\{ \hat{G}_{21}  \, V^{(2)}_{-} \}_{-1} (w)  - \frac{1}{3} \pa \{ \hat{G}_{21}  \, V^{(2)}_{-} \}_{-2} (w)
\nonu \\
&=&
\frac{i}{\sqrt{2}(k+N+2)^2} \left[ k \, (d^1+i d^2)_{\bar{a} \bar{b} } \,
Q^{\bar{a} } \, \pa \, V^{\bar{b}}
- k \, (d^1+i d^2)_{\bar{a} \bar{b} } \, \pa \, Q^{\bar{a} }  \, V^{\bar{b}}   \right.
\nonu \\
&+& \left. A^{2}_{\bar{b} \bar{c} e}  \, Q^{\bar{b} } \, V^{\bar{c} } \, V^{e} 
+ \frac{1}{4(N+k+2)}  \, B^{2}_{\bar{a} \bar{b} \bar{c} \bar{d}} 
\, Q^{\bar{a}} \, Q^{\bar{b}} \, Q^{\bar{c}} \, V^{\bar{d}} 
\right] (w)
\label{v5half}
\\
&-& \frac{1}{(N+k+2)} \left[ \frac{1}{3} \, \pa \hat{G}_{22} +
\frac{1}{3}(2N +k +1) \pa \hat{G}'_{22} 
-2 i \hat{A}_{3 } \, \hat{G}_{22} +i \hat{G}_{21} \, \hat{B}_{+ }  \right] (w),
\nonu
\eea
where the two tensors are 
\bea
A^{2}_{\bar{b} \bar{c} e} & \equiv &
 ( g -i h^3)_{\bar{a} [ \bar{b}} \, ( d^1+ i d^2)_{\bar{c} ] \bar{d}}   \,
f^{\bar{a} \bar{d}}_{\,\,\,\,\,\, e},
\nonu \\
B^{2}_{\bar{a} \bar{b} \bar{c} \bar{d}} & \equiv &
(h^1 +i h^2)_{\bar{a}  \bar{b}} \, (d^0- i d^3)_{\bar{c} \bar{d}} 
+2 i \, h^3_{\bar{a}  \bar{b}} 
\, (d^1+ i d^2)_{\bar{c} \bar{d}}
-2 \, d^0_{\bar{a}  \bar{b}} \, (h^1 + i h^2)_{\bar{c} \bar{d}}.
\nonu
\eea
The second equation of Appendix (\ref{spin5halfgenerating}) is used.
The derivative of the second order pole in the second equation of
(\ref{spin5halfgenerating}) with the spin-$\frac{3}{2}$ current 
gives the last line of (\ref{v5half}).   
Also the cubic and quartic terms in the second line (in the final expression) 
of
(\ref{v5half}) can be obtained using the property of Appendix 
(\ref{firstfirst}). 

Finally the last higher spin-$\frac{5}{2}$-current 
can be obtained similarly.
From the expressions of the spin-$\frac{3}{2}$ current  
(\ref{expressionG}) with (\ref{basischange}) 
and the higher spin-$2$ current (\ref{w2final}),
one arrives at the following result (from the third equation of Appendix 
(\ref{spin5halfgenerating})) (for $N=3,5,7$)
\bea
W_{+}^{(\frac{5}{2})}(w)
&=&
\{ \hat{G}_{21} \, W^{(2)} \}_{-1} (w)- \frac{1}{3} \pa  \{ \hat{G}_{21} \, W^{(2)} \}_{-2} (w)
\nonu \\
&=& \frac{1}{\sqrt{2}(k+N+2)^2} \left[ (N+2k+1)\, 
(d^0-i d^3)_{\bar{a} \bar{b}} \, \pa \, Q^{\bar{a}} \, V^{\bar{b}}
+(N+1) \, (d^0 - i d^3)_{\bar{a} \bar{b}}  \, Q^{\bar{a}} \, \pa  \, 
V^{\bar{b}} \right.
\nonu \\
&+& \left. A^{3}_{\bar{a} \bar{d} e}  \, Q^{\bar{a}} \, V^{\bar{d}} \, V^{e}
+\frac{1}{4(k+N+2)} \, B^{3}_{\bar{a} \bar{b} \bar{c} \bar{d}} 
\, Q^{\bar{a}} \, Q^{\bar{b}} \, Q^{\bar{c}} \, V^{\bar{d}} 
\right] (w)
 - \frac{1}{(N+k+2)} \left[ \frac{4}{3} \pa \hat{G}_{21}  \right.
 \nonu \\
&+& \left. \frac{1}{3}(N+2k+3) \pa \hat{G}'_{21}   +  2i (\hat{A}_3+\hat{B}_3) \hat{G}_{21}
-i  \hat{A}_{-} \, \hat{G}_{11} - i \hat{B}_{-} \, \hat{G}_{22}  \right] (w),
\label{w+5half}
\eea
where we have the following tensors
\bea
A^{3}_{\bar{a}  \bar{d} e} &  \equiv &
\left[ (ig+h^3)_{\bar{a} \bar{b}} \, d^3_{\bar{c} \bar{d} }
+\frac{1}{2} \, (h^1-i h^2)_{\bar{a} \bar{b}} \, (d^1+i d^2)_{\bar{c} \bar{d} } 
\right] f^{\bar{b} \bar{c}}_{\,\,\,\,\,\, e},
\nonu \\
B^{3}_{\bar{a} \bar{b} \bar{c} \bar{d}} & \equiv & ( h^1- i h^2)_{\bar{a} \bar{b}} \,
(d^1+i d^2)_{\bar{c} \bar{d}}
+2 \, d^0_{\bar{a} \bar{b}}  \, (g-i h^3)_{\bar{c} \bar{d}}.
\nonu
\eea

Summarizing this subsection, the higher spin-$\frac{5}{2}$ currents
are given by (\ref{w-5half}), (\ref{u5half}), (\ref{v5half}) and 
(\ref{w+5half}). They contain the quadratic, cubic and quartic terms 
in the Kac-Moody currents as well as the composite currents 
from the large ${\cal N}=4$ nonlinear superconformal algebra
\footnote{
One also obtains the precise relation between the higher 
spin-$\frac{5}{2}$ currents. 
Because the spin-$\frac{3}{2}$ currents are given in (\ref{expressionG}) 
and the six higher spin-$2$ currents are found in (\ref{spin2bcg1})
and (\ref{spin2bcg2}),
one can calculate the fourth and fifth OPEs of Appendix $(A.1)$ 
in \cite{BCG1404}
and it turns out that their four higher spin-$\frac{5}{2}$ currents 
are
\bea
V_{\frac{3}{2}, 11}^{(1)}(z) & = & 2 i \, U^{(\frac{5}{2})}(z) \nonu \\
& + & \frac{4}{(N+k+2)}
\, \left[
\hat{A}_{+} \hat{G}_{21} 
+ \hat{A}_{+} T_{+}^{(\frac{3}{2})}
- \hat{B}_{-} \hat{G}_{12}+ 2 \hat{B}_{-} T_{-}^{(\frac{3}{2})}
+2 \hat{B}_3 U^{(\frac{3}{2})} - \frac{2 i}{3} \pa  U^{(\frac{3}{2})} 
\right](z),
\nonu \\
V_{\frac{3}{2},22}^{(1)}(z) & = & -2 i \, V^{(\frac{5}{2})}(z) \nonu \\
& + & \frac{4}{(N+k+2)}
\, \left[
\hat{A}_{-} \hat{G}_{12} 
-2 \hat{A}_{-} T_{-}^{(\frac{3}{2})}
-2 \hat{A}_{3} V^{(\frac{3}{2})}- \hat{B}_{+} T_{+}^{(\frac{3}{2})}
 + \frac{2 i}{3} \pa  V^{(\frac{3}{2})} 
\right](z),
\nonu \\
V_{\frac{3}{2},21}^{(1)}(z) & = & 2 i \, W_{+}^{(\frac{5}{2})}(z) 
+  \frac{4}{(N+k+2)}
\, \left[
\hat{A}_{-} \hat{G}_{11} 
+ \hat{A}_{-} U^{(\frac{3}{2})}
- 2 \hat{A}_{3} \hat{G}_{21}- 2 \hat{A}_{3} T_{+}^{(\frac{3}{2})}
\right. \nonu \\
& + & \left.
 \hat{B}_{-} \hat{G}_{22} -  \hat{B}_{-} V^{(\frac{3}{2})} 
+2 \hat{B}_3  T_{+}^{(\frac{3}{2})} + \frac{2 i}{3} \pa \hat{G}_{21}
\right](z),
\nonu \\
V_{\frac{3}{2},12}^{(1)}(z) & = & -2 i \, W_{-}^{(\frac{5}{2})}(z) 
+  \frac{4}{(N+k+2)}
\, \left[
\hat{A}_{+} \hat{G}_{22} 
- \hat{A}_{+} V^{(\frac{3}{2})}
- 2 \hat{A}_{3} \hat{G}_{12}+ 2 \hat{A}_{3} T_{-}^{(\frac{3}{2})}
\right. \nonu \\
& + & \left.
 \hat{B}_{+} \hat{G}_{11} + \hat{B}_{+} U^{(\frac{3}{2})} 
-2 \hat{B}_3  T_{-}^{(\frac{3}{2})} - \frac{2 i}{3} \pa \hat{G}_{12}
\right](z).
\nonu
\eea
Note that in this calculation, the constant $a$ appeared in 
the description of (\ref{spin2bcg1})
and (\ref{spin2bcg2}) is determined by $a=i$. It is not obvious 
to reexpress these in $SU(2) \times SU(2)$ manifest way as done 
in the higher spin-$2$ currents. It would be interesting to
obtain these higher spin-$\frac{5}{2}$ currents 
with manifest $({\bf 2},{\bf 2})$ representation 
of $SU(2) \times SU(2)$.}.

\subsection{ One higher spin-$3$ current }

Let us describe the final higher spin-$3$ current.
The equation $(4.59)$ of \cite{Ahn1311} can be
generalized to the following result  
\bea
\hat{G}_{21}(z) \, W_{-}^{(\frac{5}{2})}(w)  & = & 
\frac{1}{(z-w)^3} \, \left[ a_1 \, \hat{A}_3 +
a_2 \, \hat{B}_3 + a_3 \, T^{(1)}
\right](w) 
\nonu \\
& + & \frac{1}{(z-w)^2} \, \left[ 
a_4 \, \hat{T}
+a_5 \, T^{(2)}
+ a_6 \, W^{(2)}
+  a_7 \, ( \hat{A}_1 \, \hat{A}_1 + \hat{A}_2 \, \hat{A}_2 )
+  a_8  \, \hat{A}_3 \, \hat{A}_3  \right.
\nonu \\
& +& \left. a_9 \, \hat{A}_3 \, \hat{B}_3
+ a_{10} \, (\hat{B}_1 \, \hat{B}_1 +   \hat{B}_2 \, \hat{B}_2)
+ a_{11} \,  \hat{B}_3 \, \hat{B}_3 
+ a_{12} \, (T^{(1)} \, \hat{A}_3 
- T^{(1)} \, \hat{B}_3)   
\right](w)
\nonu \\
& + &
\frac{1}{(z-w)} \,
 \left[ \frac{1}{4} \, \pa  \,  \{ \hat{G}_{21} \, W_{-}^{(\frac{5}{2})} \}_{-2} 
+a_{13} \, ( \hat{T} \, \hat{A}_3-
\frac{1}{2} \pa^2 \hat{A}_3 ) 
 + a_{14}
 \, ( \hat{T} \, \hat{B}_3-
\frac{1}{2} \pa^2 \hat{B}_3 )
\right.
\nonu \\
&+&\left. a_{15}
 \, ( \hat{T} \, T^{(1)}-
\frac{1}{2} \pa^2 T^{(1)} ) +   W^{(3)} 
\right](w) +\cdots,
\label{finalopeope}
\eea
where the coefficients are given in (\ref{15coeff}).
Then the higher spin-$3$ current can be written as
\bea
W^{(3)} (w) &=& \{ \hat{G}_{21} \, W_{-}^{(\frac{5}{2})} \}_{-1}(w) 
-\left[ \frac{1}{4} \, \pa  \,  \{ \hat{G}_{21} \, W_{-}^{(\frac{5}{2})} \}_{-2} 
+a_{13} \, ( \hat{T} \, \hat{A}_3-
\frac{1}{2} \pa^2 \hat{A}_3 ) 
\right. 
\nonu \\
& + &  \left. a_{14}
 \, ( \hat{T} \, \hat{B}_3-
\frac{1}{2} \pa^2 \hat{B}_3 ) + a_{15}
 \, ( \hat{T} \, T^{(1)}-
\frac{1}{2} \pa^2 T^{(1)} )  \right](w).
\label{w3exp}
\eea
Note that in order to obtain the 
second term of (\ref{w3exp}), one 
should differentiate the second order pole in (\ref{finalopeope}). 
After calculating the first-order pole in the OPE between 
 $\hat{G}_{21}(z)$ (\ref{expressionG}) with (\ref{basischange}) 
and $ W_{-}^{(\frac{5}{2})}(w)$ (\ref{w-5half}),
the first term of (\ref{w3exp})
is given by (for $N=3,5$)
\bea
\{ \hat{G}_{21} \, W_{-}^{(\frac{5}{2})} \}_{-1}(w) 
&=& 
\frac{i}{2(k+N+2)^3}
\nonu \\
& \times & 
\left[ 2i \,(k+N+2)\, A_{\bar{a} \bar{d} e} \,V^{\bar{a} } \,V^{\bar{d} } \,V^{e}
-4 \,(k+N+2) \,(ik d^0-d^3)_{\bar{a} \bar{b} } \,V^{\bar{a} } \,\pa \,V^{\bar{b} }  
\right.
\nonu \\
&-& i (k-1)(k+N+2) \,d^0_{\bar{a} \bar{b} } \,f^{\bar{a} \bar{b}}_{\,\,\,\,\,\, c} 
\,\pa^2 \,V^c
+\left( (k+1) \,C_{\bar{c} \bar{a} e} +F_{\bar{a} \bar{c} e} \right) \,Q^{\bar{a} } 
\,Q^{\bar{c} }\, \pa \, V^{e}
\nonu \\
&+& \left( (k-1) \,C_{\bar{a} \bar{c} e} + (k+1) \,C_{\bar{c} \bar{a} e} +
k \,D_{\bar{a} \bar{c} e} 
+F_{\bar{a} \bar{c} e} \right) \,Q^{\bar{a} } \,\pa \,Q^{\bar{c} }  \,V^{e}
\nonu \\
&+& E_{\bar{a} \bar{c} e f} \,Q^{\bar{a} }  \,Q^{\bar{c} } \,V^{e} \,V^{f} 
+ ( G+\frac{3}{4} I )_{\bar{a} \bar{c} \bar{d} \bar{f}  } 
\,Q^{\bar{a} }  \,Q^{\bar{c} } \,V^{\bar{d}} \,V^{\bar{f}} 
+2i \,k(k-1) \,d^0_{\bar{a}  \bar{b}} \,\pa \,Q^{\bar{a} } \,\pa \,Q^{\bar{b} }
\nonu \\
&+& \frac{1}{4(k+N+2)}  \,J_{\bar{a}  \bar{b}  \bar{c} \bar{d} h  } 
\,Q^{\bar{a} }  \,Q^{\bar{b} }  \,Q^{\bar{c} } \,Q^{\bar{d} } \, V^{h} 
+\frac{k}{4(k+N+2)} \,K_{\bar{a}  \bar{b} \bar{c} \bar{d}}  
\,Q^{\bar{a} }   \,Q^{\bar{b} }   \,Q^{\bar{c} }   \,\pa \,Q^{\bar{d} } 
\nonu \\
&+&\left. k \, ( -2 (k+1) \, (id^0-d^3) +\frac{1}{2}  H )_{\bar{a}  \bar{c}}   
\, Q^{\bar{a} } \, \pa^2 \, Q^{\bar{c} } \right](w)
\nonu \\
&-& \frac{1}{(N+k+2)} 
\left[ \frac{4}{3} \, \pa \, \{ \hat{G}_{21} \, \hat{G}_{12} \}_{-1} 
+ \frac{(2N+k+3)}{3} \, \pa \, \{ \hat{G}_{21} \, \hat{G}'_{12} \}_{-1}  
-2 \, \hat{G}_{21} \, \hat{G}_{12} \right.
\nonu \\
&-&  \hat{G}_{11} \,\hat{G}_{22}
- \hat{G}_{22} \, \hat{G}_{11}  
- 2i \,( \hat{A}_3 + \hat{B}_3 ) \, \{ \hat{G}_{21} \, \hat{G}_{12} \}_{-1} 
+i \, \hat{A}_{+} \, \{ \hat{G}_{21} \, \hat{G}_{22} \}_{-1} 
\nonu \\
&+& \left. i \hat{B}_{+} \, \{ \hat{G}_{21} \, \hat{G}_{11} \}_{-1} \right] (w),
\label{w3w3w3}
\eea
where the various tensors are 
\bea
C_{\bar{a} \bar{c} e}  & \equiv &
(ig+h^3)_{\bar{a} \bar{b} } \,(d^0+i d^3)_{\bar{c} \bar{d} } \,
f^{\bar{b} \bar{d} }_{\,\,\,\,\,\, e} ,
\nonu \\
E_{\bar{a} \bar{c} e f}  & \equiv &
A_{\bar{a} \bar{d} e} \, f^{\bar{d} \bar{b} }_{\,\,\,\,\,\, f}  \,
(ig-h^3)_{\bar{b} \bar{c} } ,
\nonu \\
F_{\bar{a} \bar{c} e }  & \equiv & 
 E_{\bar{a} \bar{c} b d}  \, f^{b d }_{\,\,\,\,\,\, e},
\nonu \\
D_{\bar{a} \bar{c} e} & \equiv &
\left[ (ig-h^3)_{\bar{a} \bar{b} } \, (d^0-i d^3)_{\bar{c} \bar{d} }
-(d^1 + i d^2)_{\bar{a} \bar{b} } \, (i h^1 + h^2)_{\bar{c} \bar{d} } \right] 
\,f^{\bar{b} \bar{d} }_{\,\,\,\,\,\, e},
\nonu \\
G_{\bar{a} \bar{c} \bar{d} \bar{f} } & \equiv &
 A_{\bar{a} \bar{d} e} \, (ig-h^3)_{\bar{b} \bar{c} } 
\, f^{e \bar{b} }_{\,\,\,\,\,\, \bar{f}},
\nonu \\
H_{\bar{a} \bar{c} } & \equiv & A_{\bar{a} \bar{d} e} \, (ig-h^3)_{\bar{b} \bar{c} } 
\, f^{e \bar{d} \bar{b}}
=-4N  \, (i d^0 - d^3)_{\bar{a} \bar{c} } ,
\nonu \\
I_{\bar{a} \bar{c} \bar{d} \bar{f} } & \equiv & 
(ig+h^3)^{\bar{e} }_{\,\,\,\, \bar{d}  }  \, B_{[ \bar{e} \bar{a} \bar{c} ] \bar{f} },
\nonu \\
J_{\bar{a} \bar{b} \bar{c} \bar{d}  h } & \equiv &
B_{ \bar{a} \bar{b} \bar{c}  \bar{e} }  \, (ig + h^3)_{\bar{d} \bar{f} } 
\, f^{\bar{e} \bar{f} }_{\,\,\,\,\,\, h },
\nonu \\
K_{\bar{a} \bar{b} \bar{c} \bar{d}   } & \equiv &
2i \, (h^1+i h^2)_{\bar{a} \bar{b} }  \, (d^1-i d^2)_{\bar{c} \bar{d} } 
+4 \, d^0_{\bar{a} \bar{b} } \, (i g- h^3)_{\bar{c} \bar{d} }.
\nonu 
\eea
The $ A_{\bar{a} \bar{d} e}$ and $B_{ \bar{a} \bar{b} \bar{c}  \bar{e} }$
tensors are given in (\ref{abtensors}).
The identity containing $H_{\bar{a} \bar{c}}$ is checked by 
low values of $N=3,5,7$.

One can analyze each term in (\ref{w3w3w3}).
For example, the first term
consists of the cubic term.  
Let us consider the OPE $\hat{G}_{21}(z) \, W_{-}^{(\frac{5}{2})}(w)$
where the cubic term is taken from the higher spin-$\frac{5}{2}$ current.
According to the third equation of Appendix (\ref{finalfirst}),
there exists a first term in the right hand side.   
By combining the right coefficient from the spin-$\frac{3}{2}$
current and the higher spin-$\frac{5}{2}$ current, 
one checks
that the final corresponding expression is given by the 
first term in (\ref{w3w3w3}).
The last three lines in (\ref{w3w3w3}) come from 
the OPE between  $\hat{G}_{21}(z)$ and 
the composite fields appearing in the second part of (\ref{w-5half}).
In order to obtain these expressions completely, 
Appendices $D$ and $G$ are needed.  

Summarizing this subsection,
one way to obtain the last higher spin-$3$ current is given
(the OPEs $\hat{G}_{22}(z) \, U^{(\frac{5}{2})}(w)$ or 
$\hat{G}_{11}(z) \, V^{(\frac{5}{2})}(w)$ provide the higher spin-$3$ current).
The highest order is given by the sextic term which 
appears in $\hat{T} \, \hat{B}_3(w)$ or $\hat{T} \, T^{(1)}(w)$.  
There exists a cubic term in the bosonic
spin-$1$ currents as one expected. 

Therefore, the $16$ higher spin currents 
in (\ref{lowesthigher}) 
are obtained in this section.
Because they are written in terms of Kac-Moody currents explicitly,
one observes that their zero mode can be determined and this will be the 
open problem to accomplish further in the context of three-point function.
  
\section{Conclusions and outlook }

We have determined the $16$ higher spin currents with spins 
$(1,\frac{3}{2},\frac{3}{2},2),(\frac{3}{2}, 2,2, \frac{5}{2}),
(\frac{3}{2},2,2, \frac{5}{2})$ and $(2,\frac{5}{2},\frac{5}{2},3)$
in the ${\cal N}=4$ superconformal Wolf space coset 
$\frac{SU(N+2)}{SU(N) \times SU(2) \times U(1)}$ using the 
bosonic and fermionic Kac-Moody currents.

Let us describe the future directions as follows:

$\bullet$ Three-point functions

As a second  step to the construction of 
three point function  
\cite{GH1101,Ahn1111,CHR1211,MZ1211,AK1308}
for the two scalars and one higher spin current,
one should calculate the eigenvalue equations 
for the zero modes  of the higher spin currents.
In \cite{GG1305}, one of the simple representation of 
the minimal representations is 
given by $Q_{-\frac{1}{2}}^{\bar{A}} | 0 >$ where $\bar{A} = 1, 2, 
\cdots, 2N$ corresponding to $(0;f)$ 
representation where the first element is an integrable highest weight representation of $SU(N+2)$ and the second element 
is the one of $SU(N)$.
The above state contains the group index 
for the fundamental representation of $SU(N)$ with two dimensional
$SU(2)$ 
representation.
Then the eigenvalue equation for the zero mode of spin-$2$ current 
acting on the above state can be calculated.
One can do this by considering the zero mode of (\ref{Lexpression})
and performing it into the above state. Or one can obtain the OPE 
between the spin-$2$ stress energy tensor $L(z)$ 
and the spin-$\frac{1}{2}$ current
$Q^{\bar{A}}(w)$ and read off the relevant commutator relation.
According to the result of \cite{GG1305}, the conformal dimension
$h(0;f)$ is given by $h(0;f) =\frac{(k+\frac{3}{2})}{2(N+k+2)}$
which is equal to the one for BPS bound. 
That is, one should have 
$L_0 \, Q_{-\frac{1}{2}}^{\bar{A}} | 0 > = h(0;f) \, 
Q_{-\frac{1}{2}}^{\bar{A}} | 0 >$.
It would be interesting to see this behavior by analyzing the results of this
paper carefully.
Furthermore, there exist other $10$ currents and $16$ higher spin currents. 
It is an open problem to construct the zero mode eigenvalue equations 
and obtain the various three point functions with two scalars
together with each (higher spin) current.   

$\bullet$ An extension of 
large ${\cal N}=4$ linear superconformal algebra

It would be interesting to obtain the higher spin currents in the 
context of an extension of large ${\cal N}=4$ linear superconformal algebra
by adding the four spin-$\frac{1}{2}$ currents and one spin-$1$ current.
As described  in section $2$, 
the spin-$\frac{3}{2}$ currents have the cubic fermionic terms.    
This can be observed in the explicit transformations between 
the spin-$2$, spin-$\frac{3}{2}$ and spin-$1$ currents in the large 
${\cal N}=4$ nonlinear superconformal algebra and those in the 
large ${\cal N}=4$ linear superconformal algebra, initiated by
Goddard and Schwimmer in \cite{GS}. 
Furthermore, from the work of \cite{Saulina}, one expects that
one can obtain the $16$ currents with the various tensors 
in terms of the spin-$1$ and 
spin-$\frac{1}{2}$ Kac-Moody currents in (\ref{opevq}). 
Due to the presence of the extra $(4+1)$ currents, there are more rooms 
for the coset space $\frac{SU(N+2)}{SU(N)}$.
Once the above construction for the large ${\cal N}=4$ linear superconformal 
algebra is done, 
then it is straightforward to determine the higher spin currents
by following the works of \cite{Ahn1311,Ahn1408} and this paper.  
One of the motivations for this direction is to provide 
an extension of small ${\cal N}=4$ superconformal algebra (one spin-$2$ 
current, four spin-$\frac{3}{2}$ currents and three spin-$1$ currents) 
by taking the 
appropriate limits on the levels in the theory. In the end, 
one obtains the OPEs for the extension of small ${\cal N}=4$ superconformal 
algebra which is related to the type IIB 
string compactification of $AdS_3 \times {\bf S}^3
\times {\bf T}^4$.
It is not clear whether the extension of small ${\cal N}=4$ superconformal 
algebra can be obtained 
from the extension of large ${\cal N}=4$ nonlinear superconformal 
algebra because there exist nonlinear terms in the spin-$1$ currents 
of the OPE between the spin-$\frac{3}{2}$ currents.  
Any simple transformations  do not decouple the three spin-$1$ currents 
in these nonlinear terms of the OPEs.

$\bullet$ The application to an orthogonal Wolf space 

One can apply the results of this paper to other type of Wolf space coset.
There exists an orthogonal Wolf space coset by changing the unitary group
to the orthogonal group. The relevant works in the orthogonal group 
are given by \cite{Ahn1106,GV1106,Ahn1202,CHR1209,CGKV1211,AP1301,AP1310}.
At the level of large ${\cal N}=4$ nonlinear superconformal algebra,
there are no much differences between the unitary Wolf space coset theory
and the orthogonal Wolf space coset theory because the role of 
$SU(2) \times SU(2)$ group is replaced with $SO(3) \times SO(3)$ group.   
The main difference between them arises when one tries to determine the 
extension of large ${\cal N}=4$ nonlinear superconformal algebra because
the lowest ${\cal N}=4$ higher spin multiplet contains the higher 
spin-$2$ current 
as its lowest component rather than higher spin-$1$ current.
This higher spin-$2$ current behaves as a singlet under the above 
$SU(2) \times SU(2)$.
From the results of this paper, it is useful to 
recall that the higher spin-$2$ currents, $T^{(2)}(z)$ and $W^{(2)}$,
contain $d^{\mu}_{\bar{a} \bar{b}}$ tensor, $h^{\mu}_{\bar{a} \bar{b}}$ tensor and 
$f^{\bar{a} \bar{b}}_{\,\,\,\,\,\,c}$ structure constant.
In order to determine the tensorial structure in the higher spin-$2$ current,  
the low $N$ value results for the work of \cite{AP1410} 
will be very useful because they will 
give the explicit realizations on the above tensors.
It is an open problem
why the higher spin-$1$ current like the $T^{(1)}(z)$ (and other four 
higher spin-$\frac{3}{2}$ currents) in this paper 
does not exist in the orthogonal Wolf space coset theory.  
Compared to the unitary Wolf space coset, the orthogonal Wolf space 
coset has very simple nonzero structure constants and maybe this 
does not give the right current contents for the low spins ($s=1$ or 
$s=\frac{3}{2}$). 
Other possibility comes from the fact that 
the regular conditions between the above candidate currents and the four
spin-$\frac{1}{2}$ currents (and one spin-$1$ current) from the 
large ${\cal N}=4$ linear superconformal algebra restrict to their structures
strongly. In other words, for low spin cases, there are too many conditions
to solve for a few unknown terms. As the spin increases, 
one can consider more terms by multiplying the Kac-Moody currents 
and for the lowest higher spin-$2$ current
one expects that there exists a unique solution for the regular conditions.     

$\bullet$ The next $16$ higher spin currents for general $N$ 

In order to see the spin dependence on the three-point function, 
sometimes it is not enough to obtain
the three-point functions on two scalars with higher spin current
living in the lowest ${\cal N}=4$ multiplet (\ref{lowesthigher}) only. 
Then one should look at the three-point functions with higher spin current
in the next ${\cal N}=4$ multiplet.
For $N=3$, the higher spin current generating procedure is given 
in \cite{Ahn1408}.  
For the four higher spin-$\frac{5}{2}$ currents, the OPEs between the 
higher spin-$1$ current
and four higher spin-$\frac{5}{2}$ current found in this paper are needed.
Once these new higher spin-$\frac{5}{2}$ currents are found, then the OPEs 
between the spin-$\frac{3}{2}$ currents living in the large ${\cal N}=4$
nonlinear superconformal algebra and them will determine the six higher
spin-$3$ currents and one higher spin-$2$ current. 
Furthermore, the four higher spin-$\frac{7}{2}$ currents 
can be fixed by the OPEs between the spin-$\frac{3}{2}$ currents 
and the above new higher spin-$3$ currents.
Finally, the higher spin-$4$ current can be determined by the OPE 
between the spin-$\frac{3}{2}$ current and the higher
spin-$\frac{7}{2}$ current. 
In order to see the $N$ dependence on the OPEs, one should try to
calculate them for low values for $N$. For the higher spin-$4$ current,
the first-order pole in $(2.14)$ of \cite{Ahn1408} is rather complicated 
and it is nontrivial to obtain all the $N$ dependence in the first-order 
pole in there.
As the spin increases,
one should  simplify 
the multiple products between the 
$d^{\mu}_{\bar{a} \bar{b}}$ tensor and  the $h^{\mu}_{\bar{a} \bar{b}}$ tensor 
(as well as the structure constants) using some relevant identities.
Otherwise, one cannot write down the higher spin currents in simple form.  

$\bullet$ Oscillator formalism for the higher spin currents

Based on the previous work  in \cite{PV9806,PV9812}, 
the results in  the 
$AdS_3$ bulk theory \cite{GG1305} were obtained for the higher spin-$1$ current
and the four higher spin-$\frac{3}{2}$ currents as well as some currents 
in the large ${\cal N}=4$ nonlinear superconformal algebra.
Then it is an open problem to obtain the oscillator formalism for the 
remaining higher spin currents (with general $N$ and $k$) 
by using the higher spin algebra in the 
$AdS_3$ bulk theory.
In this approach, the nontrivial things are to write  down 
all the higher spin fields in the primary basis.

$\bullet$ For the general $M$

So far, the $M$ in (\ref{goverh}) is fixed as $M=2$.
What happens for $M > 2$?
Along the lines of \cite{CPV1408,CV1312}, it is an open problem to
study the coset model for $M >2$. Are there any supersymmetric 
coset models? 
It would be interesting to observe whether the $M=2$ case is special or not. 

\vspace{.7cm}

\centerline{\bf Acknowledgments}

This work was supported by the Mid-career Researcher Program through
the National Research Foundation of Korea (NRF) grant 
funded by the Korean government (MEST) 
(No. 2012-045385/2013-056327/2014-051185).
CA would like to thank the participants of the focus program of 
Asia Pacific Center
for Theoretical Physics (APCTP) 
on
``Liouville, Integrability and Branes (10) Focus Program at Asia-Pacific
Center for Theoretical Physics'',
Sept. 03-14, 2014 for their feedbacks.
CA appreciates APCTP for its hospitality during completion of this work.
CA acknowledges warm hospitality from 
the School of  Liberal Arts (and Institute of Convergence Fundamental
Studies), Seoul National University of Science and Technology.

\newpage

\appendix

\renewcommand{\thesection}{\large \bf \mbox{Appendix~}\Alph{section}}
\renewcommand{\theequation}{\Alph{section}\mbox{.}\arabic{equation}}

\section{ The generators of $SU(N+2)$ in complex basis }

The $SU(N+2)$ generators can be expressed in the complex (or Cartan-Weyl)
basis.
The index $a$ (\ref{abnotation}) is classified with 
$A$ and $A^{\ast}$.
The $[(N+2)^2-1]$ generators of $SU(N+2)$ are divided by 
the two sets of generators $T_A$ and $T_{A^*}$ 
where $A=1,2, \cdots, \frac{(N+2)^2-1}{2}$ with $N$ odd.
Let us describe the $\frac{(N+2)^2-1}{2}$ unstarred  generators $T_A$. 
There are $\frac{N+1}{2}$ diagonal generators and 
the remaining $[\frac{(N+2)^2-1}{2} -\frac{N+1}{2}]$ 
off-diagonal  generators.
For each off-diagonal matrix, 
the nonzero element (which is equal to $1$) occurs once
(at the specific row and column).
Among those off-diagonal matrices, 
the nonzero elements of the 
$\frac{N(N-1)}{2}$ matrices are located at the lower half 
triangle matrices of the 
middle $N\times N$ matrices,
one matrix has nonzero element with $(N+2)$-th row and first column
and the remaining $2N$ matrices have nonzero elements as follows:    
\bea
&&T_1 = \left(\begin{array}{r|rrrrr|r}
 0 &0&&\cdots&&0& 0 \\ \hline
 1 &&&&&& 0  \\
 0 &&&&&&   \\
  \vdots &&&0&&& \vdots  \\
    &&&&&&   \\
 0 &&&&&&0  \\ \hline
 0   &0 &&\cdots &&0& 0 \\ 
\end{array}\right),
\qquad 
T_2 = \left(\begin{array}{r|rrrrr|r}
 0 &0&&\cdots&&0& 0 \\ \hline
 0 &&&&&& 0  \\
 1 &&&&&&   \\
  \vdots &&&0&&& \vdots  \\
    &&&&&&   \\
 0 &&&&&& 0  \\ \hline
 0   &0 &&\cdots &&0& 0 \\ 
\end{array}\right),  \cdots,
\nonu \\
&&T_N= \left(\begin{array}{r|rrrrr|r}
 0 &0&&\cdots&&0& 0 \\ \hline
 0 &&&&&& 0  \\
 0 &&&&&&   \\
  \vdots &&&0&&& \vdots  \\
    &&&&&&   \\
 1 &&&&&&0  \\ \hline
 0   &0 &&\cdots &&0& 0 \\ 
\end{array}\right),
\qquad
T_{N+1}= \left(\begin{array}{r|rrrrr|r}
 0 &0&&\cdots&&0& 0 \\ \hline
 0 &&&&&& 0  \\
 0 &&&&&&   \\
  \vdots &&&0&&& \vdots  \\
    &&&&&&   \\
 0 &&&&&& 0  \\ \hline
 0   &1 &&\cdots &&0& 0 \\ 
\end{array}\right),
\nonu \\
&&T_{N+2}= \left(\begin{array}{r|rrrrr|r}
 0 &0&&\cdots&&0& 0 \\ \hline
 0 &&&&&& 0  \\
 0 &&&&&&   \\
  \vdots &&&0&&& \vdots  \\
    &&&&&&   \\
 0 &&&&&&0  \\ \hline
 0   &0 &1&\cdots &&0& 0 \\ 
\end{array}\right), \cdots,
\qquad
T_{2N}= \left(\begin{array}{r|rrrrr|r}
 0 &0&&\cdots&&0& 0 \\ \hline
 0 &&&&&& 0  \\
 0 &&&&&&   \\
  \vdots &&&0&&& \vdots  \\
    &&&&&&   \\
 0 &&&&&& 0  \\ \hline
 0   &0 &&\cdots &&1& 0 \\ 
\end{array}\right).
\label{Ts}
\eea
That is, the matrix $T_n$ where $ 1\leq n \leq N $ has nonzero element
(numerical value is $1$) 
at $(n+1)$-th row and first column (and other matrix elements vanish) 
and the matrix $T_{N+n}$ 
 has nonzero element 
at $(N+2)$-th row and $(n+1)$-th column (and other matrix elements are
vanishing). 
By interchanging of the role of both row and column 
(or transposing the matrix and taking the complex conjugation),
\bea
T_{A^*} \equiv T_A^{\dagger},
\label{cc}
\eea
one obtains 
the other set of generators.

There exist $(N+1)$ Cartan generators in $SU(N+2)$ denoted by 
$H_1, H_2, \cdots, H_{N+1}$. The Cartan generators are defined by \cite{Georgi}
\bea
[H_m]_{ij}=\frac{1}{\sqrt{2m(m+1)}} \left( \sum_{k=1}^m \delta_{ik} \delta_{jk} - m \delta_{i,m+1} \delta_{j,m+1}
\right).
\label{matrixH}
\eea
How does one construct these explicitly?
The normalization is given in (\ref{matrixH})
and the diagonal element can be obtained as follows:
One can start with $H_1$ by putting  the number $1$ at the first row
and first column, put $-1$ at the next diagonal element and put the zeros
at the remaining diagonal elements.
For the $H_m$ where $ 1\leq m \leq N+1$,
one puts $1$ at
 each diagonal element until $m$-th diagonal element, 
put $-m$ at the next diagonal element and 
put the zeros at the remaining diagonal elements.
Then one obtains the following matrix representation for the Cartan generators
as follows:
\bea
H_1 &=&  \frac{1}{2} \left(\begin{array}{rrrrrrr}
 1 &0&&\cdots&&& 0 \\  
 0 & -1&&&&&  \\
  &&  \ddots&&&&   \\
  \vdots &&&\ddots &&& \vdots  \\
    &&&&\ddots&&   \\
  &&&&&&  \\
 0   & &&\cdots &&& 0 \\ 
\end{array}\right), \qquad
H_2 =  \frac{1}{\sqrt{12}} \left(\begin{array}{rrrrrrr}
 1 &0&&\cdots&&& 0 \\  
 0 & 1&&&&&  \\
  && -2&&&&   \\
  \vdots &&&\ddots &&& \vdots  \\
    &&&&\ddots&&   \\
  &&&&&&  \\
 0   & &&\cdots &&& 0 \\ 
\end{array}\right), \cdots,
\nonu \\
H_{N+1} &=& \frac{1}{\sqrt{2(N+1)(N+2)}} \left(\begin{array}{rrrrrrr}
 1 &0&&\cdots&&& 0 \\  
 0 & 1&&&&&  \\
  && 1&&&&   \\
  \vdots &&&\ddots &&& \vdots  \\
    &&&&\ddots&&   \\
  &&&&&&  \\
 0   & &&\cdots &&& -(N+1) \\ 
\end{array}\right).
\nonu
\eea
Because $N$ is odd, there are even number of Cartan generators. 
Then one can define the $\frac{N+1}{2}$ diagonal generators. 
\bea
&T_{p+1}=i H_1+ H_2,
\nonu \\
&T_{p+2}=i H_3+ H_4,
\nonu \\
&\vdots
\nonu \\
&T_{p+\frac{N+1}{2}}=i H_N + H_{N+1},
\label{Tss}
\eea
where $p=\frac{(N+2)^2-1}{2} -\frac{N+1}{2}$.
Half of these Cartan generators can be obtained from (\ref{cc}). 

The metric is  
\bea
g_{ab} = \mbox{Tr} (T_a T_b)=
\left(
\begin{array}{cc}
0 & 1 \\
1 & 0 \\
\end{array}
\right), 
\quad (a,b=1,2, \cdots, (N+2)^2-1).
\nonu
\eea
This is consistent with the description of subsection $2.1$.
 The nonvanishing metric components are
\bea
 g_{ A A^* }=g_{  A^* A }=1,
\nonu
\eea 
where $A=1,2,\cdots,\frac{(N+2)^2-1}{2} $.

We have checked some OPEs and other relevant quantities for low 
values of $N$. The structure constants for those cases can be obtained from 
the generators in (\ref{Ts}) and (\ref{Tss}). 
One can also obtain the generators 
corresponding to the nonzero elements of the 
$\frac{N(N-1)}{2}$ matrices located at the lower half 
triangle matrices of the 
middle $N\times N$ matrices.
For example, the structure constants containing the group $SU(N+2)$ index
appear in the spin-$1$ current 
(\ref{sixspin1def}) or 
spin-$2$ current (\ref{Lexpression}).

\section{ Eight rank-two tensors in the Wolf space coset  }

The four rank two tensors ($12 \times 12$ matrices) appearing in the 
spin-$\frac{3}{2}$ currents in (\ref{Gansatz})
can be generalized to  the ones of $4N \times 4N $ matrices
as follows:
\bea
&&h^0_{\bar{a} \bar{b}} \equiv  g_{\bar{a} \bar{b}}=
\left(
\begin{array}{cc|cc}
0 & 0 & 1 & 0 \\
0 & 0 & 0 & 1 \\
\hline
1 & 0 & 0 & 0 \\
0 & 1 & 0 & 0 \\
\end{array}
\right), \quad
h^1_{\bar{a} \bar{b}} = 
\left(
\begin{array}{cc|cc}
0 & - 1  & 0 & 0 \\
1 & 0 & 0 & 0 \\
\hline
0 & 0 & 0 & -1 \\
0 & 0 & 1 & 0 \\
\end{array}
\right), \nonu \\
&&h^2_{\bar{a} \bar{b}}  = 
\left(
\begin{array}{cc|cc}
0 & - i   & 0 & 0 \\
i  & 0 & 0 & 0 \\
\hline
0 & 0 & 0 & i  \\
0 & 0 & -i  & 0 \\
\end{array}
\right), \quad
h^3_{\bar{a} \bar{b}}  \equiv h^1_{\bar{a} \bar{c}} \, h^{2 \bar{c}}_{ \,\,\,\,\,\, \bar{b}}=
\left(
\begin{array}{cc|cc}
0 & 0  & i  & 0 \\
0 & 0  & 0 & i  \\
\hline
-i  & 0  & 0 & 0 \\
0 & -i   & 0 & 0 \\
\end{array}
\right),
\label{fourh}
\eea
where each element in (\ref{fourh}) is a $N \times N$ matrix.
One can specify the indices further.
The rows and columns of the 
first $2N \times 2N$ matrices are given by $\bar{A}$ and $\bar{A}$.  
The rows and columns of the 
first right $2N \times 2N$ matrices are given by $\bar{A}$ and 
$\bar{A^{\ast}}$.  
The rows and columns of the 
second left $2N \times 2N$ matrices are given by $\bar{A^{\ast}}$ 
and $\bar{A}$.  
The rows and columns of the 
second right $2N \times 2N$ matrices are given by $\bar{A^{\ast}}$ 
and $\bar{A^{\ast}}$.
One also has
\bea
&& i h^0_{\bar{a} \bar{b}} + h^3_{\bar{a} \bar{b}}=
\left(
\begin{array}{cc|cc}
0 & 0 & 2i & 0 \\
0 & 0 & 0 & 2i \\
\hline
0 & 0 & 0 & 0 \\
0 & 0 & 0 & 0 \\
\end{array}
\right), \quad
i h^0_{\bar{a} \bar{b}}- h^3_{\bar{a} \bar{b}} = 
\left(
\begin{array}{cc|cc}
0 & 0  & 0 & 0 \\
0 & 0 & 0 & 0 \\
\hline
2i & 0 & 0 & 0 \\
0 & 2i & 0 & 0 \\
\end{array}
\right), \nonu \\
&& h^1_{\bar{a} \bar{b}} +i h^2_{\bar{a} \bar{b}}  = 
\left(
\begin{array}{cc|cc}
0 & 0  & 0 & 0 \\
0 & 0 & 0 & 0 \\
\hline
0 & 0 & 0 & -2 \\
0 & 0 & 2 & 0 \\
\end{array}
\right), \quad
h^1_{\bar{a} \bar{b}} -i h^2_{ \bar{a} \bar{b}}=
\left(
\begin{array}{cc|cc}
0 & -2  & 0 & 0 \\
2 & 0  & 0 & 0 \\
\hline
0 & 0  & 0 & 0 \\
0 & 0  & 0 & 0 \\
\end{array}
\right).
\label{fourh1}
\eea
  
The rank two tensors defined in (\ref{dd}) (see also (\ref{d0expressionexp})) 
are generalized to
the following $4N \times 4N$ matrices
\bea
&&d^0_{\bar{a} \bar{b}} = 
\left(
\begin{array}{cc|cc}
0 & 0 & 1 & 0 \\
0 & 0 & 0 & -1 \\
\hline
-1 & 0 & 0 & 0 \\
0 & 1 & 0 & 0 \\
\end{array}
\right), \quad
d^1_{\bar{a} \bar{b}} \equiv d^{0 \bar{c} }_{ \bar{a} } \, h^{1 }_{\bar{c} \bar{b}} =
\left(
\begin{array}{cc|cc}
0 & -1 & 0 & 0 \\
-1 & 0 & 0 & 0 \\
\hline
0 & 0 & 0 & 1 \\
0 & 0 & 1 & 0 \\
\end{array}
\right), 
\label{fourg}
 \\
&&d^2_{\bar{a} \bar{b}} \equiv d^{0 \bar{c} }_{ \bar{a} } \, h^{2 }_{\bar{c} \bar{b}}=
\left(
\begin{array}{cc|cc}
0 & -i  & 0 & 0 \\
-i  & 0 & 0 & 0 \\
\hline
0 & 0 & 0 & -i   \\
0 & 0 & -i  & 0 \\
\end{array}
\right), \quad
d^3_{\bar{a} \bar{b}} \equiv d^{0 \bar{c} }_{ \bar{a} } \, h^{3 }_{\bar{c} \bar{b}}=
\left(
\begin{array}{cc|cc}
0 & 0 & i  & 0 \\
0 & 0 & 0 & -i   \\
\hline
i  & 0 & 0 & 0 \\
0 & -i  & 0 & 0 \\
\end{array}
\right).
\nonu
\eea
In (\ref{fourg}), the above equations (\ref{fourh}) are used.
Compared to (\ref{fourh}), the locations for the row and column appearing 
 the nonzero elements in 
$d^{\mu}_{\bar{a} \bar{b}}$ are the same as the ones of 
$h^{\mu}_{\bar{a} \bar{b}}$. 
One also has
\bea
&& i d^0_{\bar{a} \bar{b}} + d^3_{\bar{a} \bar{b}}=
\left(
\begin{array}{cc|cc}
0 & 0 & 2i & 0 \\
0 & 0 & 0 & -2i \\
\hline
0 & 0 & 0 & 0 \\
0 & 0 & 0 & 0 \\
\end{array}
\right), \quad
i d^0_{\bar{a} \bar{b}}- d^3_{\bar{a} \bar{b}} = 
\left(
\begin{array}{cc|cc}
0 & 0  & 0 & 0 \\
0 & 0 & 0 & 0 \\
\hline
-2i & 0 & 0 & 0 \\
0 & 2i & 0 & 0 \\
\end{array}
\right), \nonu \\
&& d^1_{\bar{a} \bar{b}} +i d^2_{\bar{a} \bar{b}}  = 
\left(
\begin{array}{cc|cc}
0 & 0  & 0 & 0 \\
0 & 0 & 0 & 0 \\
\hline
0 & 0 & 0 & 2 \\
0 & 0 & 2 & 0 \\
\end{array}
\right), \quad
d^1_{\bar{a} \bar{b}} -i d^2_{ \bar{a} \bar{b}}=
\left(
\begin{array}{cc|cc}
0 & -2  & 0 & 0 \\
-2 & 0  & 0 & 0 \\
\hline
0 & 0  & 0 & 0 \\
0 & 0  & 0 & 0 \\
\end{array}
\right).
\label{fourg1}
\eea
It is easy to obtain eight tensors whose nonzero single element appears
by taking a simple linear combination of (\ref{fourh1}) and (\ref{fourg1}).  
For example, 
\bea
 -i h^0_{\bar{a} \bar{b}} - h^3_{\bar{a} \bar{b}} + 
i d^0_{\bar{a} \bar{b}} + d^3_{\bar{a} \bar{b}}
=
\left(
\begin{array}{cc|cc}
0 & 0 & 0 & 0 \\
0 & 0 & 0 & -4i \\
\hline
0 & 0 & 0 & 0 \\
0 & 0 & 0 & 0 \\
\end{array}
\right),
\nonu
\eea
which appears in the higher spin-$\frac{3}{2}$ current in (\ref{t+3half}).

\section{ The large 
$\mathcal N = 4$ nonlinear superconformal algebra}

In this Appendix, we review the work of Van Proeyen \cite{cqg1989}.
The currents of ${\cal N}=4$ nonlinear superconformal algebra 
are given by 
one spin-$2$ current $L(z)$, four spin-$\frac{3}{2}$ currents $G^{\mu}(z)$, 
six 
spin-$1$ currents $A^{\pm i}(z)$.

The spin-$2$ stress energy tensor satisfies the 
following OPE:
\bea
L(z) \, L(w) &=& \frac{1}{(z-w)^4} \, \frac{\hat{c}}{2} +
\frac{1}{(z-w)^2} \, 2 L(w) +\frac{1}{(z-w)} \, \pa L(w) +\cdots,
\nonu 
\eea
where the central charge is given in (\ref{goverh})
\bea
\hat{c} =  \frac{6(k+1)(N+1)}{(k+N+2)} -3 
=\frac{3(k+N+2k N)}{(k+N+2)}, 
\label{chat}
\eea
and the $-3$ in (\ref{chat}) is the contribution from 
the other terms in the modified stress energy tensor \cite{Ahn1311}.
The above $10$ currents are primary currents 
under the $L(z)$ as follows:
\bea 
L(z) \, A^{\pm i}(w) &=& 
\frac{1}{(z-w)^2} \, A^{\pm i}(w)
+ \frac{1}{(z-w)}  \, \pa A^{\pm i} (w) + \cdots,
\nonu \\
L(z) \, G^{\mu} (w) &=& 
\frac{1}{(z-w)^2} \, \frac{3}{2} G^{\mu} (w)
+ \frac{1}{(z-w)}  \, \pa G^{\mu} (w) + \cdots.
\nonu
\eea

The nontrivial OPEs between the four spin-$\frac{3}{2}$ currents
can be summarized by
\bea
G^{\mu}(z) \, G^{\nu}(w) &=& \frac{1}{(z-w)^3} \,
\frac{2}{3} \delta^{\mu \nu} c_{\mbox{Wolf}}
- \frac{1}{(z-w)^2} \, \frac{8}{(k+N+2)} ( N \, \alpha^{+i}_{\mu \nu} \, A^{+}_i      
+k \, \alpha^{-i}_{\mu \nu} \, A^{-}_i  )(w)
\nonu \\
& + & \frac{1}{(z-w)} \, \left[ 2 \delta^{\mu \nu} L
-  \frac{4}{(k+N+2)} \pa ( N \, \alpha^{+i}_{\mu \nu} \, A^{+}_i      
+k \, \alpha^{-i}_{\mu \nu} \, A^{-}_i  ) \right.
\label{N4scalgebra} \\
&-& \left.  \frac{8}{(k+N+2)}  ( \alpha^{+i} A^{+}_i      
- \alpha^{-i} A^{-}_i  )_{\rho (\mu}
( \alpha^{+j} A^{+}_j      
- \alpha^{-j} A^{-}_j  )_{\nu)}^{\,\,\,\,\, \rho} \right](w) +\cdots,
\nonu 
\eea
where the indices run over $\mu, \nu, \rho=0,1,2,3$ and $i,j=1,2,3$. 
The OPE (\ref{N4scalgebra}) is used in section $2$.
The Wolf space coset central charge 
is given in (\ref{goverh})
\bea
c_{\mbox{Wolf}}  & = & 
\frac{6 k N}{(2+k+N)}.
\label{Wolfcentral}
\eea
The
$\alpha^{\pm i} $ is defined by
\bea
\alpha^{\pm i}_{\mu \nu} &=&
\frac{1}{2} \left( \pm \delta_{i \mu} \delta_{ \nu 0} \mp \delta_{i \nu} \delta_{ \mu 0}
+\ep_{i \mu \nu} \right).
\nonu
\eea

The OPE between the six spin-$1$ currents and the four spin-$\frac{3}{2}$
currents (related to the discussion at the end of section $2$)   
is 
\bea
A^{\pm i}(z) \, G^{\mu} (w) &=& 
\frac{1}{(z-w)} \, \alpha^{\pm i}_{\mu \nu} \, G^{\nu} (w)
+\cdots.
\nonu
\eea 
Finally, the OPEs between the spin-$1$ currents 
are given by
\bea 
A^{\pm i}(z) \, A^{\pm j}(w) &=& 
-\frac{1}{(z-w)^2} \, \frac{1}{2} \, \delta^{ij} \, \hat{k}^{\pm} 
+ \frac{1}{(z-w)} \,  \ep^{ijk} A^{\pm k} (w) + \cdots,
\nonu 
\eea
where $\hat{k}^{+}=k$ and $\hat{k}^{-}=N$ are the levels associated with
each group in 
$SU(2)_{ \hat{k}^{+} } \times SU(2)_{\hat{k}^{-} }$ respectively.

Let us present (\ref{N4scalgebra}) explicitly as follows:
\bea
G^{0}(z) \, G^{0}(w) &=&
\frac{1}{(z-w)^3} \, \frac{2}{3}  c_{\mbox{Wolf}} + \frac{1}{(z-w)} \,
\left[ 2L 
+ \frac{2}{(k+N+2)} 
\sum_{i=1}^3 ( A^{+i}+A^{-i}  )^2 \right](w)+ \cdots,
\nonu\\
G^{i}(z) \, G^{i}(w) &=&
\frac{1}{(z-w)^3} \, \frac{2}{3}  c_{\mbox{Wolf}} + \frac{1}{(z-w)} \,
\left[ 2L 
\right. \nonu \\
&+& \left. \frac{2}{(k+N+2)} \sum_{j=1}^3 ( A^{+j}+a(i,j) A^{-j}  
)^2 \right](w)
+ \cdots, \quad (i=1,2,3),
\nonu \\
G^{0}(z) \, G^{i}(w) &=&
\frac{4}{(k+N+2)} \left[ \frac{1}{(z-w)^2}  \, ( N A^{+i} - k A^{-i} ) 
 \right.
\nonu \\
&+&\left. \frac{1}{(z-w)} \, \left( 
 \frac{1}{2} \pa ( N A^{+i} - k A^{-i} )
+\sum_{j,k=1}^3 \ep^{ijk}   A^{+j} A^{-k}  \right)  \right] (w)
+ \cdots, \quad (i=1,2,3)
\nonu \\
G^{i}(z) \, G^{j}(w) &=&
-\frac{4}{(k+N+2)} \left[ \frac{1}{(z-w)^2}   \, \sum_{k=1}^3 \ep^{ijk} 
( N A^{+k} + k A^{-k} ) \right.
 \label{N4scalgebraExplic}
 \\
&+&\left. \frac{1}{(z-w)} \, \left( \frac{1}{2}  \sum_{k=1}^3 \ep^{ijk} \pa 
( N A^{+k} + k A^{-k} )  
-     (  A^{+i} A^{-j} +  A^{-i} A^{+j} ) \right) \right]  (w) 
+ \cdots, 
\nonu
\eea
where the indices are $(i, j) =(1, 2), (2,3)$ or $(3,1)$ 
in the last OPE of (\ref{N4scalgebraExplic}). 
The following index appearing in the second OPE of 
(\ref{N4scalgebraExplic})
occurs in \cite{cqg1989}
\bea
a(i,j) \equiv
\left\{
	\begin{array}{ll}
		+1  & \mbox{if } i=j \\
		-1 & \mbox{if } i \neq  j
	\end{array}.
\right.
\label{index}
\eea

The second equation in  (\ref{N4scalgebraExplic})
has the following decomposition with (\ref{index})
\bea
L(z) &=& 
\frac{1}{2(k+N+2)^2} \left[ (k+N+2) \, V_{\bar{a}} \, V^{\bar{a}} 
+k \, Q_{\bar{a}} \, \pa \, Q^{\bar{a}} 
+h^i_{\bar{a} \bar{b}} \, h^i_{\bar{c} \bar{d}}  
\, f^{\bar{b} \bar{d} }_{\,\,\,\,\,\, e} \, Q^{\bar{a}} \, 
Q^{\bar{c}} \, V^e  \right] (z)
\nonu \\
&-& \frac{1}{(k+N+2)} \sum_{j=1}^3 ( A^{+j}+ a(i,j) A^{-j}  )^2 (z),
\quad (i=1,2,3),
\nonu \\
& \equiv  & L_W^{i} (z) + L_{SU(2)}^{i} (z).
\nonu
\eea
Similarly, one can write $L(z)$ in terms of the sum of
$L_W^{0} (z)$  and $L_{SU(2)}^{0} (z)$ using the first equation of  
(\ref{N4scalgebraExplic}).
With the index $a(0,j) \equiv 1$, one has the following 
decomposition
\bea
L_W^{\mu} (z) & \equiv &
\frac{1}{2(k+N+2)^2} \left[ (k+N+2) \, V_{\bar{a}} \, V^{\bar{a}} 
+k \, Q_{\bar{a}} \, \pa \, Q^{\bar{a}} 
+h^{\mu}_{\bar{a} \bar{b}} \, h^{\mu}_{\bar{c} \bar{d}}  
\, f^{\bar{b} \bar{d} }_{\,\,\,\,\,\, e} \, Q^{\bar{a}} \, Q^{\bar{c}} \, 
V^e  \right] (z),
\nonu \\
L_{SU(2)}^{\mu} (z) & \equiv &
 - \frac{1}{(k+N+2)} 
\sum_{j=1}^3 ( A^{+j}+ a(\mu,j) A^{-j}  )^2 (z), \quad (\mu=0,1,2,3), 
\nonu
\eea
where there is no sum over the index $\mu$.
Therefore, there are four different $L_W^{\mu} (z)$
and four different $L_{SU(2)}^{\mu} (z)$. 
However, the sum of each  $L_W^{\mu} (z)$
and each $L_{SU(2)}^{\mu} (z)$ leads to a single $L(z)$.
In other words,
for each $\mu$ index, 
\bea
L(z) & =  & L_W^{\mu} (z) + L_{SU(2)}^{\mu} (z).
\nonu
\eea

One can also check the following OPEs between 
 $L_W^{\mu} (z)$  and $ L_{SU(2)}^{\mu} (z)$ as follows:
\bea
L_W^{\mu} (z) \, L_W^{\mu} (w) &=& \frac{1}{(z-w)^4}\, \frac{c_{\mbox{Wolf}} }{2} +
\frac{1}{(z-w)^2} \, 2L_W^{\mu} (w) +\frac{1}{(z-w)} \, \pa L_W^{\mu} (w) +\cdots,
\nonu \\
L_{SU(2)}^{\mu} (z) \, L_{SU(2)}^{\mu} (w) &=& \frac{1}{(z-w)^4} \, \frac{c_{\mbox{SU(2)}} }{2} +
\frac{1}{(z-w)^2} \, 2L_{SU(2)}^{\mu} (w) +\frac{1}{(z-w)} \, \pa L_{SU(2)}^{\mu} (w) 
+\cdots,
\nonu \\
L_W^{\mu} (z) \, L_{SU(2)}^{\mu} (w)
&=& +\cdots.
\label{opell1}
\eea 
From (\ref{opell1}), one can check that 
the central charge in (\ref{goverh}) with (\ref{Wolfcentral}) is given by
\bea
\hat{c} = c_{\mbox{Wolf}} +  c_{\mbox{SU(2)}}, \qquad 
c_{\mbox{SU(2)}}=
\frac{3(k+N)}{(k+N+2)}.
\nonu
\eea

Furthermore, each spin-$\frac{3}{2}$ current
$G^{\mu}(w)$
is a primary current under the corresponding  $L_W^{\mu} (z)$
and the sum of $(A^{+i} + a(\mu, i) A^{-i})(w)$
is primary current 
under the  $L_{SU(2)}^{\mu} (z)$ stress energy tensor as follows:
\bea
L_W^{\mu} (z) \, G^{\mu} (w)
&=&
\frac{1}{(z-w)^2} \, \frac{3}{2} G^{\mu} (w)
+\frac{1}{(z-w)}  \, \pa G^{\mu} (w)+ \cdots,
\nonu \\
L_{SU(2)}^{\mu} (z) \, ( A^{+i}+ a(\mu, i) \, A^{-i} )(w)
&=&\frac{1}{(z-w)^2} \,  ( A^{+i}+ a(\mu, i) \, A^{-i} )(w)
\nonu \\
& + & \frac{1}{(z-w)} \, \pa ( A^{+i}+a(\mu, i) \, A^{-i}) (w)+ \cdots.
\nonu 
\eea
Note that the current  $(A^{+i} + a(\mu, i) A^{-i})(w)$
has a level $(k+N)$ which is the sum of $\hat{k}^{+}$ and $\hat{k}^{-}$
because the currents $A^{+i}(z)$ and the currents $A^{-i}(w)$ commute with
each other. Moreover,  $L_{SU(2)}^{\mu} (z) \, G^{\mu} (w) = +\cdots$
and $L_W^{\mu} (z) \, ( A^{+i}+ a(\mu, i) \, A^{-i} )(w) = +\cdots$. 

\section{ The OPEs between the spin-$\frac{3}{2}$ currents for
general $N$ in the bispinor notation}

We present the  OPEs between the spin-$\frac{3}{2}$ currents for
general $N$ in the bispinor notation (the corresponding equation in $N=3
$ appears in $(A.3)$ of \cite{Ahn1311}) as follows:
\bea
\hat{G}_{11}(z) \, \hat{G}_{11}(w) & = & 
\frac{1}{(z-w)} \, \frac{4}{(N+k+2)} \left[ -\hat{A}_{+} \hat{B}_{-} 
\right](w) + \cdots,
\nonu \\
\hat{G}_{11}(z) \, \hat{G}_{12}(w) & = &
\frac{1}{(z-w)^2} \left[ 4 i \, \gamma_A  \hat{A}_{+}  \right](w) +  
\frac{1}{(z-w)} \left[ 2 i \, \gamma_A  \pa \hat{A}_{+} +\frac{4}{(N+k+2)}
\hat{A}_{+} \hat{B}_3 
\right](w) +\cdots,
\nonu \\
\hat{G}_{11}(z) \, \hat{G}_{21}(w) & = &
\frac{1}{(z-w)^2} \left[ -4 i \, \gamma_B  \hat{B}_{-}  \right](w)
+  
\frac{1}{(z-w)} \left[ -2 i \, \gamma_B  \pa  \hat{B}_{-}
  +\frac{4}{(N+k+2)}
\hat{A}_3  \hat{B}_{-}
\right](w) \nonu \\
& + & \cdots,
\nonu \\
\hat{G}_{11}(z) \, \hat{G}_{22}(w) & = &
\frac{1}{(z-w)^3} \, \frac{2}{3}c_{\mbox{Wolf}} +
\frac{1}{(z-w)^2} \left[ 4 i \left( \gamma_A  \hat{A}_3 - \gamma_B  \hat{B}_3 \right) \right](w)
+\frac{1}{(z-w)}  \left[ 2 \hat{T} \right.
\nonu \\
 & + & \left. 2 i \pa \left( \gamma_A  \hat{A}_3 - \gamma_B  \hat{B}_3 \right)+
 \frac{2}{(k+N+2)}\left( 
  \hat{A}_i \, \hat{A}_i +\hat{B}_i \, \hat{B}_i 
+2 \hat{A}_3 \, \hat{B}_3    \right) \right] (w) 
+  \cdots,
\nonu \\
\hat{G}_{12}(z) \, \hat{G}_{12}(w) & = & 
\frac{1}{(z-w)} \, \frac{4}{(N+k+2)} \left[ \hat{A}_{+} 
\hat{B}_{+} \right](w) + \cdots,
\nonu \\
\hat{G}_{12}(z) \, \hat{G}_{21}(w) & = &
\frac{1}{(z-w)^3} \, \frac{2}{3}c_{\mbox{Wolf}} +
\frac{1}{(z-w)^2} \left[ 4 i \left( \gamma_A  \hat{A}_3 + \gamma_B  \hat{B}_3 \right) \right](w)
+\frac{1}{(z-w)} \left[ 2 \hat{T} \right.
\nonu \\
& + &  \left.  2 i \pa \left( \gamma_A  \hat{A}_3 + \gamma_B  \hat{B}_3 \right) +
 \frac{2}{(k+N+2)}\left( 
    \hat{A}_i \, \hat{A}_i +\hat{B}_i \, \hat{B}_i 
-2 \hat{A}_3 \, \hat{B}_3    \right) \right](w) 
+  \cdots,
\nonu \\
\hat{G}_{12}(z) \, \hat{G}_{22}(w) & = &
\frac{1}{(z-w)^2} \left[ -4i\, \gamma_B  \hat{B}_{+} \right](w) + 
\frac{1}{(z-w)} \left[ -2 i\, \gamma_B \pa  \hat{B}_{+} 
 + \frac{4}{(N+k+2)} \hat{A}_3  \hat{B}_{+} 
 \right](w) \nonu \\
& + & \cdots,
\nonu \\
\hat{G}_{21}(z) \, \hat{G}_{21}(w) & = &  
\frac{1}{(z-w)} \frac{4}{(N+k+2)} \left[ \hat{A}_{-}
\hat{B}_{-} 
\right](w) +\cdots,
\nonu \\
\hat{G}_{21}(z) \, \hat{G}_{22}(w) & = &
\frac{1}{(z-w)^2} \left[ 4i\, \gamma_A  \hat{A}_{-} \right](w) +  
\frac{1}{(z-w)} \left[ 2 i\, \gamma_A  \pa  \hat{A}_{-}
  + \frac{4}{(N+k+2)} \hat{A}_{-}
 \hat{B}_3 \right](w) \nonu \\
& + & \cdots,
\nonu \\
\hat{G}_{22}(z) \, \hat{G}_{22}(w) & = &  
\frac{1}{(z-w)} \frac{4}{(N+k+2)} \left[ -\hat{A}_{-} \hat{B}_{+}  
\right](w) +\cdots,
\label{ggopenonlinear}
\eea
where $\gamma_A \equiv \frac{N}{N+k+2}$ and $\gamma_B \equiv 
\frac{k}{N+k+2}$.
One can express these by one single equation 
as done in (\ref{N4scalgebra}) but 
instead of doing this, one presents them in components.
These OPEs (\ref{ggopenonlinear}) can be used when one calculates 
the higher spin-$3$ current relevant to the term of 
the first order pole in the OPE $\hat{G}_{21} (z) \, \pa \hat{G}_{12}(w)$
(and they are used in many other places).

\section{The identities from the cubic terms in the first-order pole of the 
OPE between the spin-$\frac{3}{2}$ currents in (\ref{GGope}) }

There exist the following relations in 
 the cubic terms in the first-order pole of the 
OPE between the spin-$\frac{3}{2}$ currents in (\ref{GGope})
\bea
 2N f^{\bar{a} \bar{b}}_{\,\,\,\,\,\, e}
 \, h^0_{\bar{a} [ \bar{c} } h^1_{\bar{d } ] \bar{b} }
& = & 
f^{\bar{a} \bar{b}}_{\,\,\,\,\,\, e}  \, h^{[2}_{\bar{a} \bar{b} } \, h^{3]}_{\bar{c} \bar{d} },
  \nonu \\
2N f^{\bar{a} \bar{b}}_{\,\,\,\,\,\, e}
 \, h^0_{\bar{a} [ \bar{c} } \, h^2_{\bar{d } ] \bar{b} }
 & = & 
f^{\bar{a} \bar{b}}_{\,\,\,\,\,\, e}  \, h^{[3}_{\bar{a} \bar{b} } \, h^{1]}_{\bar{c} \bar{d} },
 \nonu \\
 2N f^{\bar{a} \bar{b}}_{\,\,\,\,\,\, e}
 \,h^0_{\bar{a} [ \bar{c} } \, h^3_{\bar{d } ] \bar{b} } & = & 
  f^{\bar{a} \bar{b}}_{\,\,\,\,\,\, e}  \, h^{[1}_{\bar{a} \bar{b} } \, h^{2]}_{\bar{c} \bar{d} },
 \nonu \\
2N f^{\bar{a} \bar{b}}_{\,\,\,\,\,\, e}
 \, h^1_{\bar{a} [ \bar{c} } \, h^2_{\bar{d } ] \bar{b} }
&  = & 
 f^{\bar{a} \bar{b}}_{\,\,\,\,\,\, e}  \, h^{(1}_{\bar{a} \bar{b} } \,h^{2)}_{\bar{d} \bar{c} },
 \nonu \\
 2N f^{\bar{a} \bar{b}}_{\,\,\,\,\,\, e}
 \, h^2_{\bar{a} [ \bar{c} } \, h^3_{\bar{d } ] \bar{b} }
& = &
f^{\bar{a} \bar{b}}_{\,\,\,\,\,\, e}  \, h^{(2}_{\bar{a} \bar{b} } \, h^{3)}_{\bar{d} \bar{c} },
 \nonu \\
 2N f^{\bar{a} \bar{b}}_{\,\,\,\,\,\, e}
 \, h^3_{\bar{a} [ \bar{c} } \, h^1_{\bar{d } ] \bar{b} }
 & = &
  f^{\bar{a} \bar{b}}_{\,\,\,\,\,\, e}  \, h^{(3}_{\bar{a} \bar{b} } \, h^{1)}_{\bar{d} \bar{c} },
 \nonu \\
 2N  f^{\bar{a} \bar{b}}_{\,\,\,\,\,\, e}
 \, ( h^{0}_{\bar{a} \bar{c} }  \, h^{0}_{\bar{b} \bar{d} }
 -h^{1}_{\bar{a} \bar{c} }  \, h^{1}_{\bar{b} \bar{d} } ) & = & 
f^{\bar{a} \bar{b}}_{\,\,\,\,\,\, e}  \, ( h^{2}_{\bar{a} \bar{b} } \, h^{2}_{\bar{c} \bar{d} }
+ h^{3}_{\bar{a} \bar{b} } \, h^{3}_{\bar{c} \bar{d} }),
 \nonu \\
 2N  f^{\bar{a} \bar{b}}_{\,\,\,\,\,\, e}
 \, ( h^{0}_{\bar{a} \bar{c} }  \, h^{0}_{\bar{b} \bar{d} }
 -h^{2}_{\bar{a} \bar{c} }  \, h^{2}_{\bar{b} \bar{d} })
& =  &
f^{\bar{a} \bar{b}}_{\,\,\,\,\,\, e}  \, ( h^{3}_{\bar{a} \bar{b} } \, h^{3}_{\bar{c} \bar{d} }
+ h^{1}_{\bar{a} \bar{b} } \, h^{1}_{\bar{c} \bar{d} }),
 \nonu \\
 2N  f^{\bar{a} \bar{b}}_{\,\,\,\,\,\, e}
 \, ( h^{0}_{\bar{a} \bar{c} }  \, h^{0}_{\bar{b} \bar{d} }
 -h^{3}_{\bar{a} \bar{c} }  \, h^{3}_{\bar{b} \bar{d} })
& =  &
 f^{\bar{a} \bar{b}}_{\,\,\,\,\,\, e}  \, ( h^{1}_{\bar{a} \bar{b} } \, h^{1}_{\bar{c} \bar{d} }
+ h^{2}_{\bar{a} \bar{b} } \, h^{2}_{\bar{c} \bar{d} }).
\label{express}
\eea
The left hand side corresponds to the cubic term of the first order pole
in (\ref{GGope}) while the right hand side 
corresponds to the same cubic term of   the first order pole
in (\ref{N4scalgebra}).  

In the bispinor notation,
from the relations of 
(\ref{basischange}) and (\ref{ggopenonlinear}),
the above relations  (\ref{express}) lead to the following equations
\bea
 f^{\bar{a} \bar{b}}_{\,\,\,\,\,\, e} \, ( i g+h^3 )_{\bar{a} [ \bar{c} } 
 \, ( i g-h^3 )_{\bar{d}  ] \bar{b} }
 &=& -\frac{1}{2N} \,  f^{\bar{a} \bar{b}}_{\,\,\,\,\,\, e} 
 \, ( h^1-i h^2 )_{\bar{a} \bar{b} }
 \, ( h^1+i h^2 )_{\bar{c} \bar{d} },  
 \nonu \\
  f^{\bar{a} \bar{b}}_{\,\,\,\,\,\, e} \, ( i g+h^3 )_{\bar{a} [ \bar{c} } 
 \,( i g+h^3 )_{\bar{d}  ] \bar{b} }
 &=& \frac{1}{2N} \,  f^{\bar{a} \bar{b}}_{\,\,\,\,\,\, e} 
 \, (  h^1_{\bar{a} \bar{b} } \, h^1_{\bar{c} \bar{d} }
+ h^2_{\bar{a} \bar{b} } \, h^2_{\bar{c} \bar{d} } )
 -2 f_{   \bar{c} \bar{d}  e    },
 \nonu \\
  f^{\bar{a} \bar{b}}_{\,\,\,\,\,\, e} \,( i g+h^3 )_{\bar{a} [ \bar{c} }
 \, ( h^1 - i h^2 )_{\bar{d}  ] \bar{b} }
 &=& -\frac{1}{2N} \,  f^{\bar{a} \bar{b}}_{\,\,\,\,\,\, e} 
 \, ( h^1-i h^2 )_{\bar{a} \bar{b} }
 \, h^3_{\bar{c} \bar{d} },  
 \nonu \\
  f^{\bar{a} \bar{b}}_{\,\,\,\,\,\, e} \, ( i g+h^3 )_{\bar{a} [ \bar{c} }
 \, ( h^1 + i h^2 )_{\bar{d}  ] \bar{b} }
 &=&  -\frac{1}{2N} \,  f^{\bar{a} \bar{b}}_{\,\,\,\,\,\, e} 
  \, h^3_{\bar{a} \bar{b} } \, ( h^1+ i h^2 )_{\bar{c} \bar{d} },  
  \nonu \\
  f^{\bar{a} \bar{b}}_{\,\,\,\,\,\, e} \, ( i g-h^3 )_{\bar{a} [ \bar{c} }
 \, (  i g+h^3 )_{\bar{d}  ] \bar{b} }
 &=&  -\frac{1}{2N} \,  f^{\bar{a} \bar{b}}_{\,\,\,\,\,\, e} 
  \, ( h^1+ i h^2 )_{\bar{a} \bar{b} } \, ( h^1- i 
h^2 )_{\bar{c} \bar{d} },  
  \nonu \\
  f^{\bar{a} \bar{b}}_{\,\,\,\,\,\, e} \, ( i g-h^3 )_{\bar{a} [ \bar{c} }
 \, (  h^1- i h^2 )_{\bar{d}  ] \bar{b} }
 &=&  \frac{1}{2N}  \, f^{\bar{a} \bar{b}}_{\,\,\,\,\,\, e} 
  \, h^3_{\bar{a} \bar{b} } \, ( h^1- i h^2 )_{\bar{c} \bar{d} },  
  \nonu \\
  f^{\bar{a} \bar{b}}_{\,\,\,\,\,\, e} \, ( i g-h^3 )_{\bar{a} [ \bar{c} }
 \, (  h^1+ i h^2 )_{\bar{d}  ] \bar{b} }
 &=&  \frac{1}{2N} \,  f^{\bar{a} \bar{b}}_{\,\,\,\,\,\, e} 
 \, ( h^1 + i h^2 )_{\bar{a} \bar{b} } \,  h^3_{\bar{c} \bar{d} },  
 \nonu \\
  f^{\bar{a} \bar{b}}_{\,\,\,\,\,\, e} \, ( h^1 - i h^2 )_{\bar{a} [ \bar{c} }
 \, (  h^1- i h^2 )_{\bar{d}  ] \bar{b} }
 &=&  - \frac{1}{2N} \,  f^{\bar{a} \bar{b}}_{\,\,\,\,\,\, e} 
 \, ( h^1- i h^2 )_{\bar{a} \bar{b} }  \, ( h^1- i 
h^2 )_{\bar{c} \bar{d} }, 
 \nonu \\
  f^{\bar{a} \bar{b}}_{\,\,\,\,\,\, e} \, ( h^1 - i h^2  )_{\bar{a} [ \bar{c} }
 \, (  h^1+ i h^2 )_{\bar{d}  ] \bar{b} }
 &=&   \frac{1}{2N} \,  f^{\bar{a} \bar{b}}_{\,\,\,\,\,\, e}
 \, (   h^1_{\bar{a} \bar{b} } \, h^1_{\bar{c} \bar{d} }    
 +h^2_{\bar{a} \bar{b} } \, h^2_{\bar{c} \bar{d} }    
 +2 h^3_{\bar{a} \bar{b} } \, h^3_{\bar{c} \bar{d} }    ) 
 -2 f_{   \bar{c} \bar{d}  e    },
 \nonu \\
  f^{\bar{a} \bar{b}}_{\,\,\,\,\,\, e} \, ( h^1 + i h^2 )_{\bar{a} [ \bar{c} }
 \,(  h^1+ i h^2 )_{\bar{d}  ] \bar{b} }
&=&  - \frac{1}{2N} \,  f^{\bar{a} \bar{b}}_{\,\,\,\,\,\, e} 
 \, ( h^1 +  i h^2 )_{\bar{a} \bar{b} }  \, ( h^1 + i h^2 )_{\bar{c} \bar{d} }.
\nonu 
\eea
From the 
cubic terms in the first-order pole of 
$G^{\mu}(z) \, G'^{\mu}(w)$ (for $N=3,5,7$)
one has the following relation where there is no sum over the 
index $\mu$
\bea
h^{\mu}_{\bar{a} \bar{b} } \, d^{\mu}_{\bar{c} \bar{d} } \,
f^{\bar{b} \bar{d} }_{\,\,\,\,\,\, e} 
=
h^{\mu}_{\bar{c} \bar{b} } \, d^{\mu}_{\bar{a} \bar{d} } \, 
f^{\bar{b} \bar{d} }_{\,\,\,\,\,\, e}  
\quad
(\mu=0,1,2,3).
\label{mixedidentity}
\eea
One can 
prove the identities (\ref{mixedidentity})  
using the second property in  (\ref{d0property}).

From the OPE $G_{mn}(z) G'_{mn}(w)$ (for $N=3,5,7$),
the following identities hold
\bea
 f^{\bar{a} \bar{b}}_{\,\,\,\,\,\, e} \, ( i g+h^3 )_{\bar{a} [ \bar{c} } 
 \, ( d^1 - i d^2 )_{\bar{d}  ] \bar{b} }
 &=& -\frac{i}{2N} \,  f^{\bar{a} \bar{b}}_{\,\,\,\,\,\, e} 
 \, ( h^1-i h^2 )_{\bar{a} \bar{b} }
 \, d^0_{\bar{c} \bar{d} },  
 \nonu \\
 f^{\bar{a} \bar{b}}_{\,\,\,\,\,\, e} \, ( i g-h^3 )_{\bar{a} [ \bar{c} } 
 \,( d^1 + i d^2 )_{\bar{d}  ] \bar{b} }
 &=&- \frac{i}{2N} \,  f^{\bar{a} \bar{b}}_{\,\,\,\,\,\, e} 
 \, ( h^1+i h^2 )_{\bar{a} \bar{b} }
\, d^0_{\bar{c} \bar{d} },  
 \nonu \\ 
  f^{\bar{a} \bar{b}}_{\,\,\,\,\,\, e} \, ( h^1-i h^2 )_{\bar{a} [ \bar{c} } 
 \,( i d^0 - d^3 )_{\bar{d}  ] \bar{b} }
 &=&- \frac{i}{2N}  \, f^{\bar{a} \bar{b}}_{\,\,\,\,\,\, e} 
 \, ( h^1-i h^2 )_{\bar{a} \bar{b} }
 \, d^0_{\bar{c} \bar{d} },  
 \nonu \\
  f^{\bar{a} \bar{b}}_{\,\,\,\,\,\, e} \, ( h^1+i h^2 )_{\bar{a} [ \bar{c} } 
 \, ( i d^0 + d^3 )_{\bar{d}  ] \bar{b} }
 &=& - \frac{i}{2N} \,  f^{\bar{a} \bar{b}}_{\,\,\,\,\,\, e} 
 \, ( h^1  + i h^2 )_{\bar{a} \bar{b} }
 \, d^0_{\bar{c} \bar{d} },  
 \nonu \\
  f^{\bar{a} \bar{b}}_{\,\,\,\,\,\, e} \, ( i g+h^3 )_{\bar{a} [ \bar{c} } 
 \, ( i d^0 -  d^3 )_{\bar{d}  ] \bar{b} }
 &=& 0,
 \nonu \\
  f^{\bar{a} \bar{b}}_{\,\,\,\,\,\, e} \,
( i g-h^3 )_{\bar{a} [ \bar{c} } 
 \, ( i d^0 +  d^3 )_{\bar{d}  ] \bar{b} }
 &=& 0,
 \nonu \\
  f^{\bar{a} \bar{b}}_{\,\,\,\,\,\, e} \, 
( h^1 - i h^2 )_{\bar{a} [ \bar{c} } 
 \, ( d^1 - i d^2 )_{\bar{d}  ] \bar{b} }
 &=& 0,
 \nonu \\
 f^{\bar{a} \bar{b}}_{\,\,\,\,\,\, e} \, ( h^1 + i h^2 )_{\bar{a} [ \bar{c} } 
 \, ( d^1 + i d^2 )_{\bar{d}  ] \bar{b} }
 &=& 0.
 \label{GGprimeidentity}
\eea
These are used in (\ref{u-2final}) and (\ref{v+2final}).

\section{ Some OPEs and identities relevant to 
the higher spin-$\frac{5}{2}$ currents}

For the higher spin-$\frac{5}{2}$ currents,
one should calculate 
the spin-$\frac{3}{2}$ currents and  the spin-$2$ currents.
The former has $Q^{\bar{a}} \, V^{\bar{b} }(z)$ and the latter
contains
the two quadratic terms, 
 $V^{\bar{c} } \, V^{\bar{d}} (w)$
and $ Q^{\bar{c}} \, \pa \, Q^{\bar{d}}  (w)$,
and the cubic term $ Q^{\bar{c}}  \, Q^{\bar{d}} \, 
V^{e} (w)$.
It turns out that the first order poles of the corresponding OPEs,
which will be used in the subsection $3.4$,
are given by 
\bea
Q^{\bar{a}} \, V^{\bar{b} }(z) \, V^{\bar{c} } \, V^{\bar{d}} (w)|_{\frac{1}{(z-w)}}
&=& \left[ k \,( g^{\bar{c} \bar{b} } \, \pa \, Q^{\bar{a}}  \, V^{\bar{d}} 
+ g^{\bar{d} \bar{b} } \, \pa \, Q^{\bar{a}}  \, V^{\bar{c}}  ) 
+f^{\bar{c} \bar{b} }_{\,\,\,\,\,\, e} \, Q^{\bar{a}} \, V^{\bar{d}} \, V^{e}  
+f^{\bar{d} \bar{b} }_{\,\,\,\,\,\, e} \, Q^{\bar{a}} \, V^{\bar{c}} \, V^{e}   \right.
\nonu \\
&+&\left. f^{\bar{c} \bar{b} }_{\,\,\,\,\,\, e} \, f^{\bar{d} e }_{\,\,\,\,\,\, \bar{f}}
\, \pa \, ( Q^{\bar{a}} \, V^{\bar{f}} ) \right] (w),
\nonu \\
Q^{\bar{a}} \, V^{\bar{b} }(z) \,  Q^{\bar{c}} \, \pa \, Q^{\bar{d}}  (w)|_{
\frac{1}{(z-w)}}
&=& (k+N+2) ( 
g^{\bar{d} \bar{a} }  \, Q^{\bar{c}} \, \pa  \, V^{\bar{b}}  -
g^{\bar{c} \bar{a} }  \, \pa \, Q^{\bar{d}}  \, V^{\bar{b}}  ) (w), \nonu \\
Q^{\bar{a}} \, V^{\bar{b} }(z) \,  Q^{\bar{c}}  \, Q^{\bar{d}} \, 
V^{e} (w)|_{\frac{1}{(z-w)}}
&=&\left[ (k+N+2) \, ( 
g^{\bar{d} \bar{a} } \, Q^{\bar{c}}  \, V^{\bar{b}} \, V^{e} -
g^{\bar{c} \bar{a} } \, Q^{\bar{d}} \, V^{\bar{b}} \, V^{e}
)
+k \, g^{e \bar{b} } \, Q^{\bar{c}} \, Q^{\bar{d}} \, \pa \, Q^{\bar{a}} \right.
\nonu \\
&+& \left. f^{e \bar{b} }_{\,\,\,\,\,\, f} \, Q^{\bar{c}} \,Q^{\bar{d}} \,
Q^{\bar{a}} \, V^{f}
\right] (w).
\label{firstfirst}
\eea
Therefore, the final higher spin-$\frac{5}{2}$ currents can be obtained 
by multiplying $h^{\mu}_{\bar{a} \bar{b}}$ and other relevant rank two and
three tensors
(appearing in the spin-$2$ currents) 
in (\ref{firstfirst}).
The spin-$2$ currents have the composite terms from the currents
of large ${\cal N}=4$ nonlinear superconformal algebra. 
One obtains the contributions from the OPEs between the spin-$\frac{3}{2}$
currents and those composite terms without any difficulty because 
they are part of large ${\cal N}=4$ nonlinear superconformal algebra 
(and their descendants).

One presents further simplifications as follows (for $N=3,5,7$):
\bea
 ( ig+h^3 )_{\bar{a} \bar{b}} \, (i d^1 +  d^2 )_{\bar{c} \bar{d}} 
\, f^{\bar{b} \bar{c}}_{\,\,\,\,\,\,e} \, f^{ \bar{d} e}_{\,\,\,\,\,\, \bar{f}} 
\, \pa \, ( Q^{ \bar{a} } V^{ \bar{f} })
& = & 2 N  \, (d^1 - i d^2)_{\bar{a} \bar{b}} \, \pa \,
( Q^{ \bar{a} } V^{ \bar{b} }),
\nonu \\
( ig+h^3 )_{\bar{a} \bar{b}} \, (d^0- i d^3)_{\bar{c} \bar{d}} 
\, f^{\bar{b} \bar{c}}_{\,\,\,\,\,\,e} \, f^{e \bar{d}}_{\,\,\,\,\,\, \bar{f}} 
\, \pa ( Q^{ \bar{a} } V^{ \bar{f} })
& = & 
2 i \, (N+1) \, (d^0 - i d^3)_{\bar{a} \bar{b}} \, \pa \, 
( Q^{ \bar{a} } V^{ \bar{b} } ),
\nonu \\
( ig-h^3 )_{\bar{a} \bar{b}} \, (d^0- i d^3 )_{\bar{c} \bar{d}} 
\, f^{\bar{b} \bar{c}}_{\,\,\,\,\,\,e} \, f^{e \bar{d}}_{\,\,\,\,\,\, \bar{f}} 
\, \pa \, ( Q^{ \bar{a} } V^{ \bar{f} } )
& = & 
2 \, (i d^0 - d^3)_{\bar{a} \bar{b}} \,  \pa \, 
( Q^{ \bar{a} } V^{ \bar{b} } ).
\label{relationabove}
\eea
For example, the first relation in (\ref{relationabove})
comes from the equation (\ref{u5half}).
The spin-$\frac{3}{2}$ 
current $\hat{G}_{21}(z)$ has a term $(h^3 + i g)_{\bar{a} \bar{b}}$
and the higher spin-$2$ current has $(i d^1 + d^2)_{\bar{c} \bar{d}}$ in the 
quadratic term of (\ref{finalexp}).
Moreover, according to the first equation in (\ref{firstfirst}),
the quadratic structure constant term has total derivative term.
Combining these factors leads to the left hand side of the first equation
of (\ref{relationabove}).
The right hand side appears in the final expression of 
the higher spin-$\frac{5}{2}$ current given in (\ref{u5half}).

There are other types of identities as follows (for $N=3,5$):
\bea
&& ( ig + h^3 )_{\bar{a} \bar{b}} \, ( h^1 - i h^2 )_{ \bar{e} \bar{f}}
\, ( i d^0 -  d^3 )_{\bar{c} \bar{g}} \, f^{\bar{f} \bar{g}}_{\,\,\,\,\,\, d}
\, f^{d \bar{b}}_{\,\,\,\,\,\, \bar{h}} \,Q^{\bar{e}} \,Q^{\bar{c}} \,Q^{\bar{a}} 
\, V^{\bar{h}}=
2 \, d^0_{\bar{a} \bar{b}} \, ( h^1 - i  h^2 )_{\bar{c} \bar{d}}  
\, Q^{\bar{a}} \, Q^{\bar{b}} \, Q^{\bar{c}} \, V^{\bar{d}},
\nonu \\
&& (- ig - h^3 )_{\bar{a} \bar{b}} \, ( i g +  h^3 )_{  \bar{f} \bar{e}}
\, ( d^1 +i  d^2 )_{\bar{c} \bar{g}} \, f^{\bar{f} \bar{g}}_{\,\,\,\,\,\, d}
\, f^{d \bar{b}}_{\,\,\,\,\,\, \bar{h}} \, Q^{\bar{e}} \, Q^{\bar{c}} \, Q^{\bar{a}} 
\, V^{\bar{h}}=
\nonu \\
&&
\left[ (h^1+i h^2)_{ \bar{a} \bar{b}}  \, ( d^0 - i  d^3 )_{\bar{c} \bar{d}}  
+2i \, h^3_{ \bar{a} \bar{b}}  \, (d^1+i d^2)_{ \bar{c} \bar{d}}
-2 d^0_{ \bar{a} \bar{b}} \, (h^1+i h^2)_{ \bar{c} \bar{d}} \right]
\, Q^{\bar{a}} \, Q^{\bar{b}} \, Q^{\bar{c}} \, V^{\bar{d}},
\nonu \\
&& ( ig+h^3 )_{\bar{a} \bar{b}} \, ( h^1 - i h^2 )_{ \bar{e} \bar{f} }
\, ( d^1 +i d^2 )_{\bar{c} \bar{g}} \, f^{\bar{f} \bar{g}}_{\,\,\,\,\,\, d}
\, f^{d \bar{b}}_{\,\,\,\,\,\, \bar{h}} \, Q^{\bar{e}} \, Q^{\bar{c}} 
\, Q^{\bar{a}} \, V^{\bar{h}}=
\nonu \\
&&
\left[ ( i h^1 + h^2 )_{\bar{a} \bar{b}}  \,
( d^1 + i d^2 )_{\bar{c} \bar{d}} 
+2 d^0_{\bar{a} \bar{b}} \, ( i g +  h^3 )_{\bar{c} \bar{d}}  \right]
\, Q^{\bar{a}} \, Q^{\bar{b}} \, Q^{\bar{c}} \, V^{\bar{d}},
\nonu \\
&& ( ig-h^3 )_{\bar{a} \bar{b}} \, ( h^1 - i h^2 )_{ \bar{e} \bar{f} }
\, ( d^1 +i d^2 )_{\bar{c} \bar{g}} \, f^{\bar{f} \bar{g}}_{\,\,\,\,\,\, d}
\, f^{d \bar{b}}_{\,\,\,\,\,\, \bar{h}} \, Q^{\bar{e}} \, Q^{\bar{c}} 
\, Q^{\bar{a}} \, V^{\bar{h}}=
\nonu \\
&&
-\left[ (  h^1 +i  h^2 )_{\bar{a} \bar{b}}  \, (i  d^1 + d^2)_{\bar{c} \bar{d}} 
+2 d^0_{\bar{a} \bar{b}} \, ( i g -  h^3 )_{\bar{c} \bar{d}}  \right]
\,Q^{\bar{a}} \,Q^{\bar{b}} \,Q^{\bar{c}} \,V^{\bar{d}}.
\label{relationrel}
\eea
Again the first relation in (\ref{relationrel})
comes from the equation (\ref{u5half}).
The spin-$\frac{3}{2}$ 
current $\hat{G}_{21}(z)$ has a term $(h^3 + i g)_{\bar{a} \bar{b}}$
and the higher spin-$2$ current has $( h^1 - i h^2 )_{ \bar{e} \bar{f}}
\, ( i d^0 -  d^3 )_{\bar{c} \bar{g}} \, f^{\bar{f} \bar{g}}_{\,\,\,\,\,\, d}$ in the 
cubic term of (\ref{finalexp}).
Moreover, according to the third equation in (\ref{firstfirst}),
the last term has a quartic term.
Combining these factors leads to the left hand side of the first equation
of (\ref{relationrel}).
The right hand side appears in the final expression of 
the higher spin-$\frac{5}{2}$ current given in (\ref{u5half}).

\section{Some OPEs generating the higher spin currents (for $N=3,5,7,9$)}

Let us collect some relevant OPEs, $(4.9)$ $(4.13)$, $(4.20)$ 
and $(4.34)$ of \cite{Ahn1311}, 
which generate the higher spin-$\frac{3}{2}$
currents.
One can check they are the same for general $N$
from (\ref{expressionG}), (\ref{basischange}) and (\ref{finalspinone}).
Therefore, one has
\bea 
\hat{G}_{21} (z) \, T^{(1)} (w) &=&
\frac{1}{(z-w)} \left[ \hat{G}_{21}  + 2 T_{+}^{(\frac{3}{2})} \right](w)+\cdots,
\nonu \\
\hat{G}_{12} (z) \, T^{(1)} (w) &=&
\frac{1}{(z-w)} \left[ -\hat{G}_{12}  + 2 T_{-}^{(\frac{3}{2})} \right](w)+\cdots,
\nonu \\
\hat{G}_{11} (z) \, T^{(1)} (w) &=&
\frac{1}{(z-w)} \left[ \hat{G}_{11}  + 2 U^{(\frac{3}{2})} \right](w)+\cdots, 
\nonu \\
\hat{G}_{22} (z) \, T^{(1)} (w) &=&
\frac{1}{(z-w)} \left[- \hat{G}_{22}  + 2 V^{(\frac{3}{2})} \right](w)+\cdots.
\label{spin3halfgenerating}
\eea
These OPEs (\ref{spin3halfgenerating})
are used in subsection $3.2$.
One can also express these OPEs in the $SO(4)$ representation using 
(\ref{basischange}) in order to calculate (\ref{Gprime}). 

Let us describe how  one can generate the next higher spin-$2$ currents.
In (\ref{expressionG}) and (\ref{basischange}), the spin-$\frac{3}{2}$ currents
in the bispinor notation are determined and 
the higher spin-$\frac{3}{2}$ currents are given in (\ref{higherspin3half}).
Then one has
\bea
\hat{G}_{11}(z) \, \hat{G}'_{12} (w)
&=&
\frac{1}{(z-w)} \, \left[ - 2 U_{-}^{(2)} -\frac{4}{(N+k+2)} 
 \hat{A}_{+}  B_3  \right](w) +\cdots,
\nonu \\
\hat{G}_{21}(z) \,  \hat{G}'_{22} (w)
&=&
\frac{1}{(z-w)} \, \left[  2 V_{+}^{(2)} -\frac{4}{(N+k+2)} 
 \hat{A}_{-}  B_3  \right](w) +\cdots,
\nonu \\
\hat{G}_{11}(z) \, \hat{G}'_{21} (w)
&=&
\frac{1}{(z-w)} \, \left[ - 2 U_{+}^{(2)} -\frac{4}{(N+k+2)} 
\hat{A}_3 \hat{B}_{-}  \right](w) +\cdots,
\nonu \\
\hat{G}_{12}(z) \, \hat{G}'_{22} (w)
&=&
\frac{1}{(z-w)} \, \left[ 2 V_{-}^{(2)} -\frac{4}{(N+k+2)} 
 \hat{A}_3 \hat{B}_{+}  \right](w) +\cdots,
\nonu \\
 \hat{G}_{12}(z) \, \hat{G}'_{21} (w)
&=&
\frac{1}{(z-w)^2} \, 2 T^{(1)} (w) +
\frac{1}{(z-w)} \, \left[ -2 T^{(2)} +\pa T^{(1)} +  \frac{2(k+N)}{(k+N+2kN)} 
\hat{T}  \right.
\nonu \\
&+& \left.   \frac{2}{(N+k+2)} \left( \hat{A}_i \hat{A}_i +\hat{B}_i \hat{B}_i
-2 \hat{A}_3 \hat{B}_3 \right)
 \right](w) +\cdots,
 \nonu \\
\hat{G}_{11}(z) \, \hat{G}'_{22} (w)
&=&
\frac{1}{(z-w)^2} \, 2 T^{(1)} (w) +
\frac{1}{(z-w)}\, \left[ 2 W^{(2)} +\pa T^{(1)} -2 \hat{T}  \right.
\nonu \\
&-& \left.   \frac{2}{(N+k+2)} \left( \hat{A}_i \hat{A}_i +\hat{B}_i \hat{B}_i
+2 \hat{A}_3 \hat{B}_3 \right)
 \right](w) +\cdots,
\nonu \\
\hat{G}_{21}(z) \, \hat{G}'_{12} (w)
&=&
\frac{1}{(z-w)^2} \, 2 T^{(1)} (w) +
\frac{1}{(z-w)} \, 
\left[ 2 T^{(2)} +\pa T^{(1)} -  \frac{2(k+N)}{(k+N+2kN)} \hat{T}  \right.
\nonu \\
&-& \left.   \frac{2}{(N+k+2)} \left( \hat{A}_i \hat{A}_i +\hat{B}_i \hat{B}_i
-2 \hat{A}_3 \hat{B}_3 \right)
 \right](w) +\cdots,
 \nonu \\
\hat{G}_{22}(z) \, \hat{G}'_{11} (w)
&=&
\frac{1}{(z-w)^2} \, 2 T^{(1)} (w) +
\frac{1}{(z-w)} \, \left[ -2 W^{(2)} +\pa T^{(1)} + 2 \hat{T}  \right.
\nonu \\
&+& \left.   \frac{2}{(N+k+2)} \left( \hat{A}_i \hat{A}_i +\hat{B}_i \hat{B}_i
+2 \hat{A}_3 \hat{B}_3 \right)
 \right](w) +\cdots.
\label{spin2generating1} 
\eea
These OPEs (\ref{spin2generating1})
correspond to the second equation of Appendix $A$ of \cite{BCG1404}
where the right hand side contains only linear terms.
One can check that 
according to the analysis in the subsection $3.3$ the nonlinear terms 
in (\ref{spin2generating1}) can be absorbed into the redefined 
higher spin-$2$ currents.
One has also following relations concerning on the singular terms
\bea
\hat{G}_{12}(z) \, \hat{G}'_{11} (w)
& = & 
-\hat{G}_{11}(z) \, \hat{G}'_{12} (w), \nonu \\
\hat{G}_{22}(z) \,  \hat{G}'_{21} (w)
& = &
-\hat{G}_{21}(z) \,  \hat{G}'_{22} (w),
\nonu \\
\hat{G}_{21}(z) \, \hat{G}'_{11} (w)
& = & 
-\hat{G}_{11}(z) \, \hat{G}'_{21} (w), 
\nonu \\
\hat{G}_{22}(z) \, \hat{G}'_{12} (w)
& = &
-\hat{G}_{12}(z) \, \hat{G}'_{22} (w).
\label{spin2generating2}
\eea
These OPEs (\ref{spin2generating1}) and (\ref{spin2generating2})
are used in subsection $3.3$.
The OPEs $G_{mn}'(z) \, G'_{pq}(w)$ can be obtained from the 
relation (\ref{minusrelation}) together with (\ref{ggopenonlinear}), 
(\ref{ggdoublesingle}) and (\ref{basischange}). 

The $(4.31)$ $(4.45)$, $(4.52)$ 
and $(4.55)$ of \cite{Ahn1311}, 
which generate the higher spin-$\frac{5}{2}$ current
can be generalized for general $N$
from (\ref{expressionG}), (\ref{basischange}), (\ref{u-2final}), 
(\ref{finalv-2}) 
and (\ref{w2final}).
Therefore, one obtains
\bea
\hat{G}_{21} (z) \, U^{(2)}_{-} (w) 
&=&
\frac{1}{(z-w)^2}\, \left[ \frac{(N+2k)}{(N+k+2)} \hat{G}_{11} + 
\frac{2(N+2k+1)}{(N+k+2)} U^{(\frac{3}{2})} \right](w)
\nonu \\
&+& 
\frac{1}{(z-w)} \, \left[ U^{ (\frac{5}{2}) } 
+\frac{1}{3} \pa \{ \hat{G}_{21}  \, U^{(2)}_{-} \}_{-2} 
\right](w) + \cdots,
\nonu \\
\hat{G}_{21} (z) \, V^{(2)}_{-} (w) 
&=&
\frac{1}{(z-w)^2}  \, \left[ - \frac{(2N+k)}{(N+k+2)} \hat{G}_{22} + 
\frac{2(2N+k+1)}{(N+k+2)} 
V^{(\frac{3}{2})} \right](w)
\nonu \\
&+& 
\frac{1}{(z-w)} \, \left[ V^{ (\frac{5}{2}) } 
+\frac{1}{3} \pa \{  \hat{G}_{21}  \, V^{(2)}_{-}  \}_{-2}
\right] (w)+ \cdots,
\nonu \\
\hat{G}_{21} (z) \, W^{(2)} (w) 
&=&
\frac{1}{(z-w)^2}  \, \left[ \frac{(N+2k+1)}{(N+k+2)} 
\hat{G}_{21} + \frac{(k-N)}{(N+k+2)} T_{+}^{(\frac{3}{2})} \right](w)
\nonu \\
&+& 
\frac{1}{(z-w)} \, \left[ W_{+}^{(\frac{5}{2})} 
+\frac{1}{3} \pa \{ \hat{G}_{21}  \, W^{(2)} \}_{-2}
\right] (w)+ \cdots,
\nonu \\
\hat{G}_{12} (z) \, W^{(2)} (w) 
&=&
\frac{1}{(z-w)^2}  \, \left[ \frac{(N+2k+1)}{(N+k+2)} 
\hat{G}_{12} + \frac{(N-k)}{(N+k+2)} T_{-}^{(\frac{3}{2})} \right](w)
\nonu \\
&+& 
\frac{1}{(z-w)} \, \left[ W_{-}^{(\frac{5}{2})} 
+\frac{1}{3} \pa \{ \hat{G}_{12}  \, W^{(2)} \}_{-2}
\right] (w)+ \cdots.
\label{spin5halfgenerating}
\eea
These OPEs (\ref{spin5halfgenerating})
are used in subsection $3.4$.

\section{The coefficients and some OPEs 
relevant to  the higher spin-$3$ current}

Furthermore, the equation $(4.59)$ of \cite{Ahn1311}
can be generalized to the equation (\ref{finalopeope})
and 
the $(N,k)$ dependent coefficients are given by (for $N=3,5,7,9$)
\bea
a_1 & = & \frac{8 i N (3k+1)}{3(N+k+2)^2},  \nonu \\
a_2 & = & \frac{8 i k (3N+1)}{3(N+k+2)^2}, \nonu \\
a_3 & = & \frac{8  (k-N)}{3(N+k+2)}, 
\nonu\\
a_4 & = & 
-\frac{4\left[ (2N^2-N)+(6N^2+4N-1)k+(6N+4)k^2 \right]}{3(N+k+2)(N+k+2kN)}, 
\nonu \\
a_5 & = & \frac{4(k-N)}{3(N+k+2)}, \nonu \\
a_6 & = & \frac{4 (N+k+1)}{(N+k+2)},
\nonu\\
a_7 & = & - \frac{4 (-N+4k-1) }{ 3(N+k+2)^2 }, \nonu \\
a_8 & = &  - \frac{4 (-4N+4k-1) }{ 3(N+k+2)^2 }, \nonu \\
a_9 & = & \frac{8(-N+k-1)}{3(N+k+2)^2},
\nonu \\
a_{10} & = &  - \frac{4(2N+k-1)}{3(N+k+2)^2}, \nonu \\
a_{11} & = & \frac{4(-2N+2k+1)}{3(N+k+2)^2}, \nonu \\
a_{12} & = &  - \frac{4 i }{(N+k+2)}, 
\nonu \\
a_{13} & = & \frac{8 i N(3k+1)}{(N+k+2) ( 5N+4+(6N+5)k )}, \nonu \\
a_{14} & = &  \frac{8 i k (3N+1)}{(N+k+2) ( 5N+4+(6N+5)k )},
\nonu \\
a_{15} & = & \frac{8(k-N)}{ (5N+4+(6N+5)k) }.
\label{15coeff}
\eea
The numerators of these coefficients 
have a simple linear in $N$ except 
$a_4$.
For the higher spin-$3$ current,
one should calculate 
the spin-$\frac{3}{2}$ current and  the spin-$\frac{5}{2}$ current.
The former has $Q^{\bar{a}} \, V^{\bar{b} }(z)$ and the latter
contains
the two quadratic terms, the cubic term and the quartic term.

It turns out that the first order poles of the corresponding OPEs
are given by 
\bea
Q^{\bar{a}} \, V^{\bar{b} }(z) \, 
\pa Q^{\bar{c}}  \, V^{\bar{d}} (w)|_{\frac{1}{z-w}}
&=&
\left[ (k+N+2) \, (
-g^{\bar{c} \bar{a}} \, V^{\bar{d}} \, \pa \, V^{\bar{b}}
+\frac{1}{2} \, g^{\bar{c} \bar{a}} \, f^{\bar{b}  \bar{d} }_{\,\,\,\,\,\, e}
\, \pa^2 \, V^e )
\right. \nonu \\
& - & \left. 
k \, g^{\bar{d} \bar{b}} \, \pa \, Q^{\bar{c}} \, \pa \, Q^{\bar{a}} 
- f^{\bar{b}  \bar{d} }_{\,\,\,\,\,\, e}  \, Q^{\bar{a}} \, \pa \, Q^{\bar{c}} 
\, V^e \right] (w),
\nonu \\
Q^{\bar{a}} \, V^{\bar{b} }(z) \, 
 Q^{\bar{c}} \, \pa \, V^{\bar{d}} (w)|_{\frac{1}{z-w}}
&=&
\left[ -(k+N+2) \, g^{\bar{c} \bar{a}}  \, V^{\bar{b}} \, \pa \, V^{\bar{d}}
-k \, g^{\bar{d} \bar{b}} \, Q^{\bar{c}} \, \pa^2 \, Q^{\bar{a}}
\right.
\nonu \\
& - & \left. f^{\bar{d}  \bar{b} }_{\,\,\,\,\,\, e}
 \, Q^{\bar{c}} \, \pa \, (  Q^{\bar{a}} \, V^e )  \right] (w),
 \nonu \\
 Q^{\bar{a}} \, V^{\bar{b} }(z) \, 
 Q^{\bar{c}}  \, V^{\bar{d}} \, V^e (w)|_{\frac{1}{z-w}}
 &=&
\left[ -(k+N+2) \, g^{\bar{c} \bar{a}} \, V^{\bar{b}}  \, V^{\bar{d}}  \, V^{e} 
 -k \, g^{\bar{d} \bar{b}} \, Q^{\bar{c}}  \, \pa \, Q^{\bar{a}} \, V^e
 \right. \nonu \\
& - & f^{\bar{d}  \bar{b} }_{\,\,\,\,\,\, g}  
 \, Q^{\bar{c}} \, Q^{\bar{a}} \, V^e \, V^g 
 -  f^{\bar{d}  \bar{b} }_{\,\,\,\,\,\, g} \, f^{e g }_{\,\,\,\,\,\, h}
 \, Q^{\bar{c}} \, \pa \, (Q^{\bar{a}} \, V^h)
 -f^{e  \bar{b} }_{\,\,\,\,\,\, g} 
  \, Q^{\bar{c}} \, Q^{\bar{a}} \, V^{\bar{d}} \, V^{g}
\nonu \\ 
 & - & \left. \frac{1}{2} \, k \, f^{\bar{d}  \bar{b} }_{\,\,\,\,\,\, g}  
  \, g^{e g} \, Q^{\bar{c}} \, \pa^2 \, Q^{\bar{a}} 
 - k \, g^{e \bar{b} } \, Q^{\bar{c}} \, \pa \, Q^{\bar{a}} \, 
V^{\bar{d}} 
  \right] (w),
  \nonu \\
  Q^{\bar{a}} \, V^{\bar{b} }(z) \, 
Q^{\bar{c}}  \, Q^{\bar{d}} \, Q^{\bar{e}} \, V^{f} (w)|_{\frac{1}{z-w}}
&=& 
\left[ (k+N+2) \, ( 
g^{\bar{e} \bar{a}  } \,  Q^{\bar{d}} \,Q^{\bar{c}} \,V^{\bar{b}}\,  V^{f} -
g^{\bar{d} \bar{a}  } \,  Q^{\bar{e}} \,Q^{\bar{c}}\, V^{\bar{b}} \, V^{f} 
\right. \nonu \\
& + & 
g^{\bar{c} \bar{a}  } \,  Q^{\bar{e}} \,Q^{\bar{d}} \,V^{\bar{b}} \, V^{f} ) 
+  f^{f \bar{b} }_{\,\,\,\,\,\, g}
 \,Q^{\bar{e}} \,Q^{\bar{d}} \,Q^{\bar{c}} \,Q^{\bar{a}}\,V^{g}
\nonu \\ 
& + & \left. 
k \,g^{f \bar{b}}  \,Q^{\bar{e}} \,Q^{\bar{d}} \,Q^{\bar{c}} \,\pa \,
Q^{\bar{a}}
 \right] (w).
\label{finalfirst}
\eea
Then, the final higher spin-$3$ current can be obtained 
by multiplying $h^{\mu}_{\bar{a} \bar{b}}$ and other relevant rank two, three
and four  tensors
(appearing in the higher spin-$\frac{5}{2}$ current) 
in (\ref{finalfirst}).
The spin-$3$ current have the composite terms from the currents
of large ${\cal N}=4$ nonlinear superconformal algebra. 
One obtains the contributions from the OPEs between the spin-$\frac{3}{2}$
current and those composite terms without any difficulty as before.



\begin{thebibliography}{99}

\bibitem{GG1011} 
  M.~R.~Gaberdiel and R.~Gopakumar,
  ``An $AdS_3$ Dual for Minimal Model CFTs,''  
Phys.\ Rev.\ D {\bf 83}, 066007 (2011)  
[arXiv:1011.2986 [hep-th]].  

\bibitem{GG1205} 
  M.~R.~Gaberdiel and R.~Gopakumar,
  ``Triality in Minimal Model Holography,''  JHEP {\bf 1207}, 127
  (2012)  
[arXiv:1205.2472 [hep-th]].  

\bibitem{GG1207} 
  M.~R.~Gaberdiel and R.~Gopakumar,
  ``Minimal Model Holography,''
  J.\ Phys.\ A {\bf 46}, 214002 (2013)
  [arXiv:1207.6697 [hep-th]].

\bibitem{GG1305} 
  M.~R.~Gaberdiel and R.~Gopakumar,
  ``Large $\mathcal{N}=4$ Holography,''  JHEP {\bf 1309}, 036 (2013)  [arXiv:1305.4181 [hep-th]].  

\bibitem{Wolf}
J.~A.~Wolf,
``Complex Homogeneous Contact Manifolds and Quaternionic Symmetric Spaces,''
J. \ Math. \ Mech. {\bf 14}, 1033 (1965).

\bibitem{Alek}
D.~V.~Alekseevskii,
``Classification of Quarternionic Spaces with a Transitive Solvable Group of
Motions,''
\ Math. \ USSR \ Izv. {\bf 9}, 297 (1975).

\bibitem{Salamon}
S.~Salamon,
``Quaternionic Kahler Manifolds,''
\ Invent. \ Math. {\bf 67}, 143 (1982).

\bibitem{CHR1306} 
  T.~Creutzig, Y.~Hikida and P.~B.~Ronne,
  ``Extended higher spin holography and Grassmannian models,''
  JHEP {\bf 1311}, 038 (2013)
  [arXiv:1306.0466 [hep-th]].

\bibitem{CHR1111} 
  T.~Creutzig, Y.~Hikida and P.~B.~Ronne,
  ``Higher spin $AdS_3$ supergravity and its dual CFT,''
  JHEP {\bf 1202}, 109 (2012)
  [arXiv:1111.2139 [hep-th]].

\bibitem{KS1} 
  Y.~Kazama and H.~Suzuki,
  ``New N=2 Superconformal Field Theories and Superstring Compactification,''  Nucl.\ Phys.\ B {\bf 321}, 232 (1989).  

\bibitem{KS2} 
  Y.~Kazama and H.~Suzuki,
  ``Characterization of N=2 Superconformal Models Generated by Coset Space Method,''  Phys.\ Lett.\ B {\bf 216}, 112 (1989).  

\bibitem{CG1203} 
  C.~Candu and M.~R.~Gaberdiel,
  ``Supersymmetric holography on $AdS_3$,''
  JHEP {\bf 1309}, 071 (2013)
  [arXiv:1203.1939 [hep-th]].

\bibitem{HP1203} 
  K.~Hanaki and C.~Peng,
  ``Symmetries of Holographic Super-Minimal Models,''
  JHEP {\bf 1308}, 030 (2013)
  [arXiv:1203.5768 [hep-th]].

\bibitem{Ahn1206} 
  C.~Ahn,
  ``The Large N 't Hooft Limit of Kazama-Suzuki Model,''
  JHEP {\bf 1208}, 047 (2012)
  [arXiv:1206.0054 [hep-th]].

\bibitem{CG1207} 
  C.~Candu and M.~R.~Gaberdiel,
  ``Duality in N=2 Minimal Model Holography,''
  JHEP {\bf 1302}, 070 (2013)
  [arXiv:1207.6646 [hep-th]].

\bibitem{Ahn1208} 
  C.~Ahn,
  ``The Operator Product Expansion of the Lowest Higher Spin Current at Finite N,''
  JHEP {\bf 1301}, 041 (2013)
  [arXiv:1208.0058 [hep-th]].

\bibitem{Hikida1212} 
  Y.~Hikida,
  ``Conical defects and $N=2$ higher spin holography,''
  JHEP {\bf 1308}, 127 (2013)
  [arXiv:1212.4124].

\bibitem{CPV1408} 
  C.~Candu, C.~Peng and C.~Vollenweider,
  ``Extended supersymmetry in $AdS_3$ higher spin theories,''
  arXiv:1408.5144 [hep-th].

\bibitem{CV1312} 
  C.~Candu and C.~Vollenweider,
  ``On the coset duals of extended higher spin theories,''
  JHEP {\bf 1404}, 145 (2014)
  [arXiv:1312.5240 [hep-th]].

\bibitem{BCG1404} 
  M.~Beccaria, C.~Candu and M.~R.~Gaberdiel,
  ``The large N = 4 superconformal $W_{\infty}$ algebra,''
  JHEP {\bf 1406}, 117 (2014)
  [arXiv:1404.1694 [hep-th]].

\bibitem{GG1406} 
  M.~R.~Gaberdiel and R.~Gopakumar,
  ``Higher Spins $\&$ Strings,''
  arXiv:1406.6103 [hep-th].

\bibitem{CHR1406} 
  T.~Creutzig, Y.~Hikida and P.~B.~Ronne,
  ``Higher spin $AdS_3$ holography with extended supersymmetry,''
  arXiv:1406.1521 [hep-th].

\bibitem{Ahn1211} 
  C.~Ahn,
  ``The Higher Spin Currents in the N=1 Stringy Coset Minimal Model,''  JHEP {\bf 1304}, 033 (2013)  [arXiv:1211.2589 [hep-th]].  

\bibitem{Ahn1305} 
  C.~Ahn,
  ``Higher Spin Currents with Arbitrary N in the ${\cal N} = 1$ 
Stringy Coset Minimal Model,''  JHEP {\bf 1307}, 141 (2013)  [arXiv:1305.5892 [hep-th]].  

\bibitem{BCGG1305} 
  M.~Beccaria, C.~Candu, M.~R.~Gaberdiel and M.~Groher,
  ``$\mathcal{N}=1$ extension of minimal model holography,''
  arXiv:1305.1048 [hep-th].

\bibitem{GS} 
  P.~Goddard and A.~Schwimmer,
  ``Factoring Out Free Fermions And Superconformal Algebras,''
  Phys.\ Lett.\ B {\bf 214}, 209 (1988).

\bibitem{cqg1989} 
  A.~Van Proeyen,
  ``Realizations Of N=4 Superconformal Algebras On Wolf Spaces,''
  Class.\ Quant.\ Grav.\  {\bf 6}, 1501 (1989).

\bibitem{npb1989} 
  M.~Gunaydin, J.~L.~Petersen, A.~Taormina and A.~Van Proeyen,
  ``On The Unitary Representations Of A Class Of N=4 Superconformal Algebras,''
  Nucl.\ Phys.\ B {\bf 322}, 402 (1989).

\bibitem{GK} 
  S.~J.~Gates, Jr. and S.~V.~Ketov,
  ``No N=4 strings on wolf spaces,''
  Phys.\ Rev.\ D {\bf 52}, 2278 (1995)
  [hep-th/9501140].

\bibitem{Gunaydin} 
  M.~Gunaydin,
  ``N=4 superconformal algebras and gauged WZW models,''
  Phys.\ Rev.\ D {\bf 47}, 3600 (1993)
  [hep-th/9301049].

\bibitem{STVplb} 
  A.~Sevrin, W.~Troost and A.~Van Proeyen,
  ``Superconformal Algebras in Two-Dimensions with N=4,''
  Phys.\ Lett.\ B {\bf 208}, 447 (1988).

\bibitem{npb1988} 
  A.~Sevrin, W.~Troost, A.~Van Proeyen and P.~Spindel,
  ``EXTENDED SUPERSYMMETRIC sigma MODELS ON GROUP MANIFOLDS. 2. CURRENT ALGEBRAS,''
  Nucl.\ Phys.\ B {\bf 311}, 465 (1988).

\bibitem{Schoutensnpb} 
  K.~Schoutens,
  ``O(n) Extended Superconformal Field Theory in Superspace,''  Nucl.\ Phys.\ B {\bf 295}, 634 (1988).  

\bibitem{Ivanov1} 
  E.~A.~Ivanov and S.~O.~Krivonos,
  ``N=4 Superliouville Equation. (in Russian),''  J.\ Phys.\ A {\bf 17}, L671 (1984).  

\bibitem{Ivanov2} 
  E.~A.~Ivanov and S.~O.~Krivonos,
  ``$N=4$ Superextension of the Liouville Equation With Quaternionic Structure,''  Theor.\ Math.\ Phys.\  {\bf 63}, 477 (1985)  [Teor.\ Mat.\ Fiz.\  {\bf 63}, 230 (1985)].  

\bibitem{Ivanov3} 
  E.~A.~Ivanov, S.~O.~Krivonos and V.~M.~Leviant,
  ``A New Class of Superconformal $\sigma$ Models With the {Wess-Zumino} Action,''  Nucl.\ Phys.\ B {\bf 304}, 601 (1988).  

\bibitem{Ivanov4} 
  E.~A.~Ivanov, S.~O.~Krivonos and V.~M.~Leviant,
  ``Quantum N=3, N=4 Superconformal WZW Sigma Models,''  Phys.\ Lett.\ B {\bf 215}, 689 (1988)  [Erratum-ibid.\ B {\bf 221}, 432 (1989)].  

\bibitem{ST} 
  A.~Sevrin and G.~Theodoridis,
  ``N=4 Superconformal Coset Theories,''
  Nucl.\ Phys.\ B {\bf 332}, 380 (1990).

\bibitem{Saulina} 
  N.~Saulina,
  ``Geometric interpretation of the large N=4 index,''
  Nucl.\ Phys.\ B {\bf 706}, 491 (2005)
  [hep-th/0409175].

\bibitem{Ahn1311} 
  C.~Ahn,
  ``Higher Spin Currents in Wolf Space. Part I,''
  JHEP {\bf 1403}, 091 (2014)
  [arXiv:1311.6205 [hep-th]].

\bibitem{Ahn1408} 
  C.~Ahn,
  ``Higher Spin Currents in Wolf Space: Part II,''
  arXiv:1408.0655 [hep-th].

\bibitem{Thielemans} 
  K.~Thielemans,
  ``A Mathematica package for computing operator product expansions,''  Int.\ J.\ Mod.\ Phys.\ C {\bf 2}, 787 (1991).  

\bibitem{KT1985} 
  V.~G.~Kac and I.~T.~Todorov,
  ``Superconformal Current Algebras And Their Unitary Representations,''
  Commun.\ Math.\ Phys.\  {\bf 102}, 337 (1985).

\bibitem{HS} 
  C.~M.~Hull and B.~J.~Spence,
  ``$N=2$ Current Algebra and Coset Models,''  Phys.\ Lett.\ B {\bf 241}, 357 (1990).  

\bibitem{BBSS1} 
  F.~A.~Bais, P.~Bouwknegt, M.~Surridge and K.~Schoutens,
  ``Extensions of the Virasoro Algebra 
Constructed from Kac-Moody Algebras Using Higher Order Casimir
  Invariants,''  
Nucl.\ Phys.\ B {\bf 304}, 348 (1988).  

\bibitem{BBSS2} 
  F.~A.~Bais, P.~Bouwknegt, M.~Surridge and K.~Schoutens,
  ``Coset Construction for Extended Virasoro Algebras,''  
Nucl.\ Phys.\ B {\bf 304}, 371 (1988).  

\bibitem{BS} 
  P.~Bouwknegt and K.~Schoutens,
  ``W symmetry in conformal field theory,''  
Phys.\ Rept.\  {\bf 223}, 183 (1993)  [hep-th/9210010].  

\bibitem{MS9907} 
  J.~Michelson and A.~Strominger,
  ``The Geometry of (super)conformal quantum mechanics,''
  Commun.\ Math.\ Phys.\  {\bf 213}, 1 (2000)
  [hep-th/9907191].

\bibitem{GP1403} 
  M.~R.~Gaberdiel and C.~Peng,
  ``The symmetry of large $\mathcal N= 4$ holography,''
  JHEP {\bf 1405}, 152 (2014)
  [arXiv:1403.2396 [hep-th]].

\bibitem{GH1101} 
  M.~R.~Gaberdiel and T.~Hartman,
  ``Symmetries of Holographic Minimal Models,''
  JHEP {\bf 1105}, 031 (2011)
  [arXiv:1101.2910 [hep-th]].

\bibitem{Ahn1111} 
  C.~Ahn,
  ``The Coset Spin-4 Casimir Operator and Its Three-Point Functions with Scalars,''  JHEP {\bf 1202}, 027 (2012)  [arXiv:1111.0091 [hep-th]].  

\bibitem{CHR1211} 
  T.~Creutzig, Y.~Hikida and P.~B.~Ronne,
  ``Three point functions in higher spin $AdS_3$ supergravity,''
  JHEP {\bf 1301}, 171 (2013)
  [arXiv:1211.2237 [hep-th]].

\bibitem{MZ1211} 
  H.~Moradi and K.~Zoubos,
  ``Three-Point Functions in N=2 Higher-Spin Holography,''
  JHEP {\bf 1304}, 018 (2013)
  [arXiv:1211.2239 [hep-th]].

\bibitem{AK1308} 
  C.~Ahn and H.~Kim,
  ``Spin-5 Casimir operator its three-point functions with two scalars,''
  JHEP {\bf 1401}, 012 (2014)
  [arXiv:1308.1726 [hep-th]].

\bibitem{Ahn1106} 
  C.~Ahn,
  ``The Large N 't Hooft Limit of Coset Minimal Models,''  JHEP {\bf 1110}, 125 (2011)  [arXiv:1106.0351 [hep-th]].  

\bibitem{GV1106} 
  M.~R.~Gaberdiel and C.~Vollenweider,
  ``Minimal Model Holography for SO(2N),''
  JHEP {\bf 1108}, 104 (2011)
  [arXiv:1106.2634 [hep-th]].

\bibitem{Ahn1202} 
  C.~Ahn,
  ``The Primary Spin-4 Casimir Operators in the Holographic SO(N) Coset Minimal Models,''  JHEP {\bf 1205}, 040 (2012)  [arXiv:1202.0074 [hep-th]].  

\bibitem{CHR1209} 
  T.~Creutzig, Y.~Hikida and P.~B.~Ronne,
  ``N=1 supersymmetric higher spin holography on $AdS_3$,''
  JHEP {\bf 1302}, 019 (2013)
  [arXiv:1209.5404 [hep-th]].

\bibitem{CGKV1211} 
  C.~Candu, M.~R.~Gaberdiel, M.~Kelm and C.~Vollenweider,
  ``Even spin minimal model holography,''
  JHEP {\bf 1301}, 185 (2013)
  [arXiv:1211.3113 [hep-th]].

\bibitem{AP1301} 
  C.~Ahn and J.~Paeng,
  ``The OPEs of Spin-4 Casimir Currents in the Holographic $SO(N)$ Coset Minimal Models,''  Class.\ Quant.\ Grav.\  {\bf 30}, 175004 (2013)  [arXiv:1301.0208].  

\bibitem{AP1310} 
  C.~Ahn and J.~Paeng,
  ``Higher Spin Currents in the Holographic $\mathcal{N} = 1$ Coset Minimal Model,''
  JHEP {\bf 1401}, 007 (2014)
  [arXiv:1310.6185 [hep-th]].

\bibitem{AP1410} 
  C.~Ahn and J.~Paeng,
  ``Higher Spin Currents in Orthogonal Wolf Space,''
  arXiv:1410.0080 [hep-th].

\bibitem{PV9806} 
  S.~F.~Prokushkin and M.~A.~Vasiliev,
  ``Higher spin gauge interactions for massive matter fields in 3-D AdS space-time,''
  Nucl.\ Phys.\ B {\bf 545}, 385 (1999)
  [hep-th/9806236].

\bibitem{PV9812} 
  S.~Prokushkin and M.~A.~Vasiliev,
  ``3-d higher spin gauge theories with matter,''
  hep-th/9812242.

\bibitem{Georgi}
  H.~Georgi,
  ``Lie Algebras In Particle Physics.
From Isospin To Unified Theories,'', 2nd Edition(1999),
Front.\ Phys.\  {\bf 54}, 1 (1982).

\end{thebibliography}
\end{document}